\documentclass[12pt,titlepage]{report}
\usepackage{babel,german,a4,macros}
\usepackage{longtable}
\usepackage[dvips]{epsfig}

\begin{document}

\thispagestyle{empty}

\normalsize \hfill
\parbox{50mm}{DESY-THESIS-1999-031\\October 1999}

\vspace*{2.0cm}
\begin{center}
{\Large Francesco Knechtli}
\end{center}
\vskip 1.5cm
\begin{center}
  {\huge The Static Potential in the \\[0.5cm]
             SU(2) Higgs Model}
\end{center}
\vskip 1.5cm
\begin{center}
{\Large
Ph.D. thesis\\[0.5cm]
Humboldt-{Universit\"at} zu Berlin\\[0.5cm]
October 1999}
\end{center}

\newpage

%---------------------
\selectlanguage{english}
\abstract
The static potential in the confinement
``phase'' of the SU(2) Higgs model is studied.
In particular, the observation
of the screening (called {\em string breaking}) of the static quarks
by the dynamical light quarks leading to the formation of two 
static-light mesons was not observed
before my work in non-Abelian gauge theories. The tool that I employ
is lattice gauge simulation. The observable from which the spectrum
of the Hamiltonian in presence of two static quarks can be extracted,
is a matrix correlation whose elements are constructed not only from
string-type states represented by Wilson loops (like in pure gauge
theories). Additional matrix elements representing transitions from 
string-type to meson-type states and the propagation of meson-type
states are taken into account. From this basis of states it is
possible to extract the ground state and first excited state static
potentials employing a variational method.
The crossing of these two energy levels in the string
breaking region is clearly visible and the inadequacy of the Wilson
loops alone can be demonstrated. I also address the question of the
lattice artifacts. For this purpose lines of constant physics in the
confinement ``phase'' of the model have to be constructed. This problem
has only partially been solved. Nevertheless it is possible to
show that the static potentials have remarkable scaling properties
under a variation of the lattice spacing by a factor two and are
almost independent of the quartic Higgs coupling.
%---------------------
\newpage
\selectlanguage{english}
\pagenumbering{roman}
\tableofcontents
\newpage
\pagenumbering{arabic}
\chapter{Introduction \label{intro}}

{\em The success of the quark-constituent picture both for resonances
  and for deep-inelastic electron and neutrino processes makes it
  difficult to believe quarks do not exist. The problem is that quarks
  have not been seen.} \\[0.2cm] \hspace*{\fill} K.G. Wilson, 1974
\\[0.5cm]

The theory of strong interactions plays a pivotal role in particle
physics. It is part of the Standard Model of elementary particles
which successfully describes the constituents of the matter in terms
of quantum gauge field theories. These theories are based on the gauge
principle \cite{Weyl:1929fm}: the fields in the theory have internal 
degrees of freedom associated with a gauge group and it is required 
that local
transformations of these degrees of freedom leave the physics
unchanged. The gauge group of the Standard Model is
$\SUthree_c\times\SUtwo_{\rm L}\times{\rm U(1)}_{\rm Y}$ 
and the degrees of freedom
associated with them are color for SU(3), weak (left-handed) isospin
for SU(2) and hypercharge for U(1). The gauge group together with the
gauge principle dictate the structure and properties of the
interactions. The particle content, described by means of relativistic
local quantum fields, has to be deduced from what nature tells us.

The particles which take part in strong interactions are called hadrons:
Gell-Mann and Zweig \cite{Gell-Mann:1964nj,Zweig:1964} proposed a
model that explained the low energy properties of the hadrons (like mass and
spin) in terms of elementary constituents called {\em quarks}. 
Bjorken \cite{Bjorken:1969dy} studied, within the framework of current
algebra, the electron-nucleon scattering and discovered the {\em scaling}
property of the structure functions for large electron momentum transfer
(deep inelastic scattering). The Bjorken scaling was
experimentally confirmed and could be
understood with the assumption that the electrons scatter off
almost-free pointlike constituents \cite{Bjorken:1969ja} inside the
nucleon, which were called partons \cite{Feynman:1969,Feynman:1969ej}.
Later the partons were identified with the quarks on the basis of
their quantum numbers.
The question at that moment was to find a theory in which particles
are free at high energies. The decisive step was then
made with the proof that non-Abelian gauge field theories exhibit
{\em asymptotic freedom}
\cite{thooft:1973asfree,Gross:1973th,Politzer:1973um}.
The strength of the interaction given by the gauge coupling
becomes weak at shorter distances (or equivalently at high energies)
and this is consistent with the Bjorken scaling.
In order to resolve several
difficulties of the quark model, like the construction of an
antisymmetric wave function for the $\Delta^{++}$ baryon and the
discrepancy between the prediction and experimental data on the total
cross section $e^+e^-\to\mbox{hadrons}$, it was
already suggested that quarks must have a new quantum number called
color and exhibit color symmetry.
Fritzsch, Gell-Mann and Leutwyler \cite{Fritzsch:1973pi}
proposed that the theory describing the quark dynamics is a non-Abelian
gauge theory with gauge group SU(3) associated with the color
symmetry. This theory was named {\em quantum chromodynamics} (QCD):
its ingredients are quarks and gluons, usually called partons. 
The gluons are the vector bosons that mediate the interactions: 
in contrast with an Abelian gauge field theory where 
the vector boson (the photon) 
is gauge neutral, the gluons carry color quantum numbers and
therefore have self-interactions. It is this property which is responsible
for asymptotic freedom.
Due to asymptotic freedom, the short distance
behavior of the partons can be described with a perturbative expansion
in the small value of the gauge coupling. Within the perturbative
approach, QCD found important confirmations as the theory of 
strong interactions,
such as the prediction of a logarithmic deviation from Bjorken scaling 
in structure functions, confirmed experimentally in deep inelastic
lepton-nucleon scatterings.

What is observed in nature are not the partons, but the
hadrons, which are color-neutral objects. The fact that colored partons
cannot be seen isolated led to the conjecture of
{\em color confinement}: the partons are always bound into the hadrons. 
In order to prove this assumption from QCD
one should be able to describe its properties at long distances
corresponding to the size of the hadrons.
Perturbation theory is not applicable because the gauge coupling
is large at this scale.
Wilson \cite{Wilson:1974sk} proposed in 1974 a new approach to gauge
field theories amounting to the discretisation of the
four-dimensional space-time on a Euclidean lattice. 
The quantisation of this theory is naturally performed in
the path integral formalism. The matter fields are treated as classical
variables living on the points of the lattice and the gauge field is
represented by connections (links) between the matter fields on
nearest-neighbor points. The quantum effects in the observables of the
theory are introduced by
evaluating their expectation values expressed as
Feynman path integrals \cite{Feynman:1948aa}.
In the Euclidean lattice formulation,
a quantum field theory looks like a classical 
statistical system. Particle and solid state physics mutually profited
by this relationship \cite{Jegerlehner:cprg}. The concept of
renormalisation of a gauge field theory receives new insights. The
regularisation of the theory on the lattice is associated with an
ultraviolet cut-off, the inverse lattice spacing $a^{-1}$. 
The field theory is changed
in the short distance region while its long distance properties are
preserved. The question one is interested in, is whether there is 
the possibility of
constructing a continuum quantum field theory from the lattice field
theory: that is, is the limit $a\to0$ of the lattice field theories
well defined? To answer this question, we should be able to reproduce
the same physical situation on lattices with different cut-offs $a$
and consider the behavior of dimensionless physical quantities when
$a\to0$. The equations describing the change in the parameters 
of the theory under variation of the lattice spacing $a$ 
are the {\em renormalisation group equations} (RGEs) of the lattice field
theory. One consequence of the RGEs is that the continuum limit of a
lattice regularised asymptotically-free theory is reached when the
lattice bare gauge coupling $g$ is sent to zero.  
To show the correspondence between a lattice field theory and a
statistical system one considers the field propagator on the
lattice. For example, in a statistical system of Ising spins, the
corresponding quantity
is the spin-spin correlation, whose exponential decay is governed by
the correlation length. On the lattice, the correlation length
equals the inverse mass gap. By 
keeping the mass gap fixed at its physical value, the correlation length
expressed in units of $a$ diverges in the continuum limit.
Thus, the continuum limit of a
lattice field theory, if it exists, corresponds to a second order
phase transition in the parameter space of the statistical system.
%%%%%%%%%%%%%%%%%%%%%%%%%%%%%FIGURE%%%%%%%%%%%%%%%%%%%%%%%%%%%%%%%%%%%
\begin{figure}[tb]
\hspace{0cm}
\vspace{-1.0cm}
\centerline{\epsfig{file=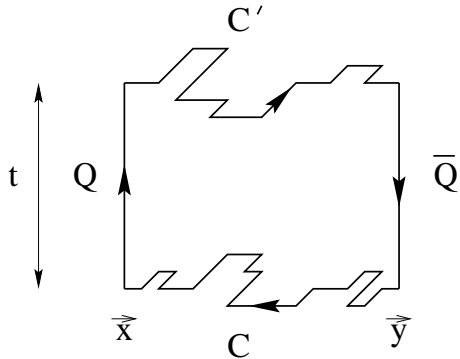,width=6cm}}
\vspace{0.5cm}
\caption{The Wilson loop, here represented on a lattice, 
describes the confinement of static quarks.
\label{f_wilson}}
\end{figure}
%%%%%%%%%%%%%%%%%%%%%%%%%%%%%%%%%%%%%%%%%%%%%%%%%%%%%%%%%%%%%%%%%%%%%%

Wilson \cite{Wilson:1974sk} originally proposed lattice gauge theories 
in order to explain color confinement. To this end, he derived an
expansion valid for strong gauge coupling in which confinement arises
naturally. However, in non-Abelian gauge theories the continuum limit 
is reached when $g\to0$ due to asymptotic freedom. Another method must
be developed to study the confinement in the weak gauge coupling regime.
We consider the system composed of a pair
of infinitely heavy or static quark and anti-quark.
The static quark (anti-quark) is treated as an 
external source in the (complex conjugate of the) 
fundamental representation of the gauge group. 
In pure SU($N$) Yang-Mills gauge theories,
the potential between the static quarks,
called the {\em static potential}, can be
extracted in the path integral formalism from the expectation value of
Wilson loops represented in \fig{f_wilson}. On the lattice,
they are defined as the trace of the product of the gauge links $U$ 
over a closed path composed of two straight time-like lines and arbitrary
space-like paths C and ${\rm C}^{\prime}$ connecting the time-like lines:
\bes\label{genwilloop}
 \langle W_{{\rm C}{\rm C^{\prime}}} \rangle \;=\;
 \langle \tr\prod U \rangle &
 _{\mbox{$\stackrel{\displaystyle\sim}
 {\scriptstyle t\to\infty}$}} &
 \rme^{-t\cdot V_{\rm Q\bar{\rm Q}}(\vec{x}-\vec{y})} \,,
\ees
where $t$ is the time extension of the Wilson loop and
$V_{\rm Q\bar{\rm Q}}(r)$ is the static potential for the separation
$r$ of the static quarks.
The expectation value in \eq{genwilloop} can
be computed by Monte Carlo simulation of the Yang-Mills theory on the
lattice.
The seminal work was done by 
Creutz \cite{Creutz:1980zw} for the gauge group SU(2) and since then,
there have been a number of detailed studies which show a {\em linear
confinement potential} at large distances between the static quarks
close to the continuum limit, both for gauge group SU(2)
\cite{Booth:1993bk,Sommer:1993ce} and SU(3)
\cite{Bali:1992ab,Booth:1992bm}. 

When the Yang-Mills gauge theories are coupled to matter fields in the
fundamental (quark) representation of the gauge group,
the potential between a
pair of static quarks is expected to flatten at large distances:
the ground state of the system is better interpreted in terms
of two weakly interacting static-light mesons which are
bound states of a static and a dynamical quark.
The dynamical quarks are pair-created in the strong
gauge field binding the static quarks. This phenomenon is
called {\em string breaking} or {\em screening} of the static
charges. The name ``string'' refers to the gauge 
field configuration which confines the static quarks and leads to the 
linear confinement in pure gauge theories. 

In recent attempts in QCD with two flavors of dynamical quarks, 
this string breaking effect was not visible
\cite{Glassner:1996xi,Gusken:1997sa,Allton:1998gi,
Burkhalter:1998wu,Aoki:1999ff,Schilling:Pisa99}.
The string breaking distance $\rb$ around which the static potential
should start flattening off, could nevertheless be estimated in the so
called quenched\footnote{
In this approximation the effects of internal quark loops are
neglected. In practical Monte Carlo simulations the computational
effort is considerably reduced.}
approximation of QCD to be \cite{Sommer:1996fr} 
\bes\label{rbquenched}
 \rb & \approx & 2.7\rnod \,,
\ees
where the scale $\rnod\approx0.5\,\fm$ was introduced in
\cite{Sommer:1993ce}. The static potential at short
distances and the mass of the static-light meson can be computed in
quenched QCD. The approximate
value in \eq{rbquenched} was obtained from the crossing point
of the linearly rising potential with twice the value of the meson
mass (which is expected to be the asymptotic value of the potential
after string breaking).

The investigation of the static potential in models other than QCD is
therefore relevant in order to understand its origin
and identify possible failures of the methods used to extract it.
First studies of string breaking were performed with a
hopping-parameter expansion in SU(2) gauge theory with Wilson fermions
\cite{Joos:1983qb}. In the Schwinger model, which is quantum
electrodynamics (QED) in two dimensions,
the exact solution for the static potential
can be given in the limit of zero fermion mass \cite{Becher:1983ft}: 
$V(r)=(e\sqrt{\pi}/2)\{1-\exp(-er/\sqrt{\pi})\}$, where
$e$ is the charge of the static sources. String breaking was
established by numerical simulation in the Schwinger model
\cite{Potvin:1985gw,Dilger:1992yn}.
Numerical evidence of the screening of the static potential was also
found in the U(1) Higgs model (scalar QED) in two dimensions
\cite{Jochen:PhD}.
The flattening of the static potential
at large distances is also expected in the confinement ``phase'' of the
SU(2) Higgs model.
Indeed, early simulations yielded some
qualitative evidence for string breaking
\cite{Evertz:1986vp,Bock:1990kq}.

String breaking can also be studied in Yang-Mills theories using
static sources in the adjoint representation of the gauge group. The
gauge field itself is responsible for the screening of the
sources and the formation of hadrons called ``gluelumps''. 
An important numerical investigation concerning this screening has
been carried out by
C. Michael in \cite{Michael:1992nc}, where it has been
noted that string breaking can be a {\em mixing}
phenomenon. The static potential is extracted from a matrix
correlation in which two types of states enter,
the adjoint ``string'' and the ``two-gluelump''. 
However, due to large errors, no clear evidence for
string breaking could be given.
The first numerical evidence
using the mixing method for string
breaking in non-Abelian gauge theories with dynamical matter fields,
was given in the four-dimensional \cite{Knechtli:1998gf} and
three-dimensional \cite{Philipsen:1998de} SU(2) Higgs model by the
computation of the potential between static quarks.
Most recently, the
extraction of the static adjoint potential in the three-dimensional 
SU(2) Yang-Mills theory \cite{Stephenson:1999kh,Philipsen:1999wf}
shows also evidence for string breaking.

Finally, we want to mention that string breaking has been seen in finite
temperature QCD \cite{DeTar:1998qa}, where the static potential can be
extracted from Polyakov loop correlators.

\vspace{1.5cm}

The status quo before our work, was that no clear evidence for string
breaking in non-Abelian gauge theories was established.
In our research, we investigate the potential between static quarks in the
four dimensional SU(2) Higgs model on the lattice. In \chapt{su2higgs},
we describe the model. The parameter
space of the theory is divided in two ``phases'', the confinement and
the Higgs ``phase''. In the confinement ``phase'', the properties are
similar to QCD: screening of external charges by the dynamical Higgs
field is expected. We describe the error analysis of the statistical 
measurements.

In \chapt{slmesons}, we concentrate on the determination of the mass
spectrum of the static-light mesons, which are expected to be the
asymptotic states after string breaking. We describe the variational
method that we use for extracting the energy spectrum from a matrix
correlation function constructed with a basis of states that can mix. 
We will use the same method for the determination of the static
potential.
The basis of states is enlarged by the use of smeared fields: we present 
a study of different smearing procedures for the Higgs field.

In \chapt{stringbreak}, we introduce the matrix correlation function
from which
we extract the static potential. We use two ``types'' of states:
``string states'' and ``two-meson states''. The variational method
determines the best linear combination approximating the true
eigenstates of the Hamiltonian.
We present the results for $\beta=4/g^2=2.4$, where $g$
is the gauge coupling. We are able to determine the ground state and first
excited state static potential with good accuracy. This allows us to study
the overlaps of the approximate eigenstates, determined from our basis
of states, with the true eigenstates. For comparison with the recent
studies of string breaking in QCD, we analyse what happens if we use
only the Wilson loops for determining the ground state static potential.

In \chapt{contlim}, we address the question of the ``continuum limit'':
in order to investigate lattice artifacts in our results, we would like to
reproduce the physical situation of \chapt{stringbreak}
on a coarser lattice at $\beta=2.2$. In the parameter space of the
model, this would define a line of constant physics (LCP), along which
two dimensionless physical quantities are kept constant under
variation of the lattice spacing. The static potential provides us with
a first quantity sensitive to the mass of the dynamical Higgs field.
We study the definition of a second quantity sensitive to the quartic
Higgs coupling. Although we are not able to match precisely
the parameters along the LCP, we find a parameter region in which the
discussion of the scaling properties of different quantities, in
particular the static potentials, is possible.

In \chapt{conclusions}, we summarise the results of our work
and give some prospectives for future investigations.

A number of more technical information is relegated to appendices.
In \App{notation}, we explain the notation conventions
that we use throughout the work. In \App{apptm}, we construct 
the transfer matrix operator for the SU(2) Higgs model and prove its
positivity, which is the condition for a real energy spectrum of the
theory. The connection between path integral expectation values and
vacuum expectation values of corresponding time ordered operators is
also shown. In \App{mcsu2higgs}, we describe our algorithms for the
Monte Carlo simulation of the SU(2) Higgs model.
In \App{bessel}, we explain the implementation of the
one-link integral method which allows the reduction of the statistical
variance of the correlation functions. Finally, \App{parallel} is
devoted to the description of the parallelised computer program that
we use for the Monte Carlo simulations.

\chapter{The SU(2) Higgs model \label{su2higgs}}

\section{Definition of the model}

The SU(2) Higgs model on a four-dimensional Euclidean lattice\footnote{
For a detailed description of the notation we refer to appendix
\ref{notation}.}
is defined by means of a
gauge field of dimensionless SU(2) matrices $U(x,\mu)$ and a complex
Higgs field $\Phi(x)$ in the fundamental representation of
the gauge group SU(2) and with canonical mass dimension one.
The full action is
\bes\label{action1}
  & & S = S_{\rmW} + \sum_xa^4\bigg\{
  \sum_{\mu}\left(\nabla_{\mu}\Phi(x)\right)^{\dagger}
  \left(\nabla_{\mu}\Phi(x)\right) \nonumber \\
  & & \qquad +\; m_0^2\Phid(x)\Phi(x) + \lambda_0\left[\Phid(x)\Phi(x)\right]^2
  \bigg\} \,,
\ees
where $S_{\rmW}$ is the Wilson action \eq{wilsonaction} 
for the SU(2) gauge field. Introducing
the dimensionless lattice fields $\Phi_{\rmL}=(a/\sqrt{\kappa})\Phi$
(we drop the subscript L in the following) together with the new couplings
$\lambda=\kappa^2\lambda_0$ and $\kappa=(1-2\lambda)/(8+a^2m_0^2)$ the
action can be written as
\bes\label{action2}
  & & S = S_{\rmW} + \sum_x\bigg\{ \Phid(x)\Phi(x) +
  \lambda\left[\Phid(x)\Phi(x)-1\right]^2 \nonumber \\
  & & -\kappa\sum_{\mu}\left[
  \Phid(x)U(x,\mu)\Phi(x+a\muh) + \Phid(x+a\muh)\Ud(x,\mu)\Phi(x)
  \right] \bigg\} \,.
\ees
The physics of the model is controlled by the three dimensionless
bare parameters $\beta\equiv 4/g^2,\;\kappa,\;\lambda$. We will use
the parametrisation \eq{action2} throughout the work.
We can rewrite
\eq{action2} using the $2\times 2$ matrix notation for the Higgs
field $\vp(x)=\rho(x)\alpha(x),\; \rho(x)\ge0,\; \alpha(x)\in\SUtwo$
defined in \sect{higgsnotation}:
\bes\label{action3} 
 & & S=S_{\rmW}+
 \sum_x\bigg\{\frac{1}{2}\tr(\vp^{\dagger}(x)\vp(x)) + \lambda
 \left[\frac{1}{2}\tr(\vp^{\dagger}(x)\vp(x))-1\right]^2 \nonumber \\
 & & \qquad 
 -\kappa\sum_{\mu}\tr(\vp^{\dagger}(x)U(x,\mu)\vp(x+a\hat{\mu}))
 \bigg\} \, .
\ees
The gauge and Higgs field are both represented as $2\times 2$ matrices
which are equal to a real constant times an SU(2) matrix. This is
particularly useful for programming purposes, see \sect{communication}.

The Euclidean expectation value of an observable\footnote{
We put square brackets to denote the
dependence on the whole field configuration and not only on a
particular field variable.}
$O[U,\Phi]$ is defined by
the Feynman path integral
\bes\label{expectvalue}
 \langle O \rangle & = & 
 \frac{1}{Z}\int\mass O[U,\Phi]\,\rme^{-S[U,\Phi]} \,,
\ees
where $Z$ is the partition function
\bes\label{partitionfunc}
 Z & = & \int\mass \rme^{-S[U,\Phi]} \,,
\ees
$\DU$ denotes the product measure
$\prod\limits_{x,\mu}\rmd U(x,\mu)$ ($\rmd U$ is the Haar measure on
SU(2)) and $\Dphi$ denotes 
$\prod\limits_x\rmd\phi_1(x)\cdot\cdot\cdot\rmd\phi_4(x)$.
The fields
$\phi_i(x)\;(i=1,2,3,4)$ are the four real components of $\Phi(x)$
defined in \eq{higgsnotation1}.

The expectation values in \eq{expectvalue} can be shown to correspond
to Euclidean vacuum expectation values of corresponding time-ordered
operators for large enough time extent of the lattice. This
correspondence can be established through the definition of a time
evolution operator called transfer matrix. How this is done for the
SU(2) Higgs model is the subject of \App{apptm}.

The lattice formulation of a quantum field theory provides a
mathematically well-defined, non-perturbative and 
completely finite regularisation of the theory. 
An analytical solution of \eq{expectvalue} is in the most
cases not possible, but it
can be computed with Monte Carlo algorithms.
The overall aim \cite{MontMuen} is to generate a representative
ensemble of field configurations
$\{[U_n,\Phi_n],\;n=1,2,...,N\}$ for the path integral
\eq{expectvalue} by employing a stochastic process. 
Representative means that the
probability distribution of the configurations in the ensemble is the
Boltzmann distribution $\exp(-S[U,\Phi])$.
We evaluate then \eq{expectvalue} as the ensemble average
\bes\label{ensembleaverage}
 \overline{O} & = & \frac{1}{N}\sum_{i=1}^N O[U_n,\Phi_n]\,.
\ees
The value $\overline{O}$ has a statistical error
$\Delta(O)$. Moreover, in most cases one is interested in
secondary quantities, which are functions of the primary averages
$\overline{O}$.

The methods that we use for the Monte Carlo simulation of the model 
are described in detail in \App{mcsu2higgs}.
The determination of the statistical errors will be the
subject of \sect{staterr}.

\section{Symmetries}

We dedicate a section to the discussion of symmetries of the action
and integration measure which reflect themselves in useful properties of the
expectation values defined by \eq{expectvalue}.

Let us start from the four-component $\phi^4$ theory without gauge
interactions.
The four components of the scalar field $\phi_i(x)\;(i=1,2,3,4)$ 
can be put in a $2\times 2$ matrix as described in
\sect{higgsnotation}. In this notation the action of the pure scalar
theory is
\bes\label{actionphi}
 & & S_{\phi}=
 \sum_x\bigg\{\frac{1}{2}\tr(\vp^{\dagger}(x)\vp(x)) + \lambda
 \left[\frac{1}{2}\tr(\vp^{\dagger}(x)\vp(x))-1\right]^2 \nonumber \\
 & & \qquad -\kappa\sum_{\mu}\tr(\vp^{\dagger}(x)\vp(x+a\hat{\mu}))
 \bigg\} \, .
\ees
The model is symmetric under the global O(4) transformation 
$\phi^{\prime}_i(x)=R_{ij}\phi_j(x)$, where $R\in \rmO(4)$.
In the matrix notation for the scalar field this transformation 
is equivalent to a global $\SUtwo_{\rmL}\otimes\SUtwo_{\rmR}$ 
symmetry defined as
\bes\label{su2timessu2}
 \vp(x) & \longrightarrow & A_{{\rm L}}^{\dagger}\vp(x)A_{{\rm R}}
 \quad , \quad A_{{\rm L,R}}\in\SUtwo \,.
\ees
The SU(2) Higgs model is obtained by gauging the subgroup
$\SUtwo_{{\rm L}}$, i.e. by ``promoting'' the global symmetry
associated with the group $\SUtwo_{{\rm L}}$ to a local
symmetry. According to the gauge principle \cite{Weyl:1929fm},
this automatically requires
the appearance of the gauge field $U(x,\mu)$ and of interactions.
We end up precisely with the action in \eq{action3}. 
Under gauge transformation defined by the field of SU(2) matrices
$\{\Lambda(x)\equiv A_{\rmL}(x)\}$ the fields transform as
\bes
 U^{\Lambda}(x,\mu) & = & \Lambda^{\dagger}(x)U(x,\mu)\Lambda(x+a\muh)
 \label{gtu} \\
 \vp^{\Lambda}(x) & = & \Lambda^{\dagger}(x)\vp(x) \, . \label{gtvp}
\ees
The SU(2) Higgs model has
a residual global SU(2) symmetry defined by the diagonal subgroup 
$A_{{\rm L}}=A_{{\rm R}}=A$:
\bes
 \vp(x) & \longrightarrow & A^{\dagger}\vp(x)A \, , \label{isovp} \\
 U(x,\mu) & \longrightarrow & A^{\dagger}U(x,\mu)A \, . \label{isou}
\ees
This symmetry is called the (weak) isospin. 

Let us now discuss the symmetry under gauge transformation. 
The action of the SU(2) Higgs model \eq{action3} is
invariant under the gauge transformations \eq{gtu} and \eq{gtvp} per
construction. The integration measure in \eq{expectvalue} as well: for the
gauge field the property $\rmd U=\rmd U^{\Lambda}$ is a consequence of 
the invariance of the Haar measure:
$\rmd U=\rmd (UV)=\rmd (VU)$ for $V\in SU(2)$. For the Higgs field
we note that
\bes\label{measurephi}
 \int_{\blackboardrrm^4}\rmd \phi_1(x)\cdot\cdot\cdot\rmd \phi_4(x)
 & \propto & \int_0^{\infty}\rho^3\rmd\rho\int_{\SUtwo}\rmd\alpha \, ,
\ees
if we write the Higgs field in the notation of \eq{higgsnotation2}
$\vp(x)=\rho(x)\alpha(x)$. From
$\vp^{\Lambda}(x)=\rho(x)(\Lambda^{\dagger}(x)\alpha(x))$ the
invariance of the measure follows immediately.
Defining $O^{\Lambda}[U,\Phi]=O[U^{\Lambda},\Phi^{\Lambda}]$ 
as the gauge transform of the observable $O$, we therefore
have the property
\bes\label{gaugeinvO}
 \langle O^{\Lambda} \rangle & = & \langle O \rangle \,.
\ees
In the continuum SU(2) Higgs model one speaks of Higgs and W-boson
fields which are not gauge invariant: a particular gauge is fixed and
perturbation theory can then be applied. In the non-perturbative
lattice scheme one does not need to fix the gauge and
gauge-invariant definitions of interpolating
fields for the Higgs and W-bosons have then to be used.
A gauge invariant composite Higgs field is defined as
\bes\label{higgsginv}
 H(x) & = &
 \Phi^{\dagger}(x)\Phi(x)\;=\;\frac{1}{2}\tr(\vp^{\dagger}(x)\vp(x)) \,.
\ees
It is an isospin 0 scalar field. A gauge invariant W-boson field is
defined as
\bes
 W_{r\mu}(x) & = & -i\tr(\tau_rV_{\mu}(x)) \quad (r=1,2,3) \,,
 \label{wbosonginv} \\
 V_{\mu}(x) & = & \alpha^{\dagger}(x)U(x,\mu)\alpha(x+a\muh) \,,
 \label{ginvlink}
\ees
where $V_{\mu}(x)$ is called the gauge invariant link variable.
The field in \eq{wbosonginv} is an isospin 1 (with isospin index $r$) 
spin 1 vector field. The isospin property can
be seen transforming the field under the isospin transformations 
\eq{isovp} and \eq{isou} and using the relation
\bes\label{adjointsu2}
 A\tau_rA^{\dagger} & = & R(A)_{sr}\tau_s\,, \quad
 A\in\SUtwo,\;R(A)\in{\rm SO(3)} \,,
\ees
which defines $R(A)$ as the adjoint (isospin=1) representation of the
isospin group SU(2).

Another symmetry of the action and the integration measure is under
complex conjugation of the field variables,
$U_{ab}(x)\rightarrow U_{ab}^*(x),\;
\Phi_a(x)\rightarrow\Phi_a^*(x)$. 
For the measure this symmetry is a consequence of the equivalence of the
representation 2 and $2^*$ of SU(2) \eq{equiv}.
For an observable of the form
$L=\Phi^{\dagger}(x)U(x,y)\Phi(y)$, where $U(x,y)$ is a link path
connecting $y$ with $x$, we can write
$L^*=(\Phi^*(x))^{\dagger}U^*(x,y)\Phi^*(y)$ from which it follows
\bes\label{ccinvO}
 \langle L^* \rangle & = & \langle L \rangle \,.
\ees
The reality of the action and the measure gives the result $\langle L
\rangle\in\blackboardrrm$.

\section{Phase diagram}

The bare parameters $\kappa,\;\lambda$ are restricted to the ranges
$\kappa\ge0,\;\lambda\ge0$. A negative value of $\lambda$ would lead
to an action which is not bounded from below and therefore the path
integral \eq{expectvalue} would not be well defined. With the help of
the transformation
\bes\label{kappasymm}
 \Phi(x) & \longrightarrow &
 \left\{ \bea{ll} -\Phi(x) &
   \mbox{if}\quad(-1)^{\sum_{\mu}x_{\mu}/a}=1 \\
   \Phi(x) & \mbox{if}\quad(-1)^{\sum_{\mu}x_{\mu}/a}=-1 \ea
 \right.
\ees
it is easy to see that the partition function \eq{partitionfunc} has the
following symmetry
\bes
 Z(\beta,\kappa,\lambda) & = & Z(\beta,-\kappa,\lambda) \,.
\ees
Replacing $\kappa$ with $-\kappa$ corresponds merely to a change in
the description of the model and leaves the physics of the model
invariant: with the help of the transformation \eq{kappasymm} we can
define a mapping of the observables of the model such that 
the expectation values \eq{expectvalue} computed with parameter $\kappa$
are reproduced by expectation values computed with parameter
$-\kappa$. Without loss of generality we can therefore restrict
ourselves to the parameter range $\kappa\ge0$.
%%%%%%%%%%%%%%%%%%%%%%%%%%%%%FIGURE%%%%%%%%%%%%%%%%%%%%%%%%%%%%%%%%%%%
\begin{figure}[tb]
\hspace{0cm}
\vspace{-1.0cm}
\centerline{\epsfig{file=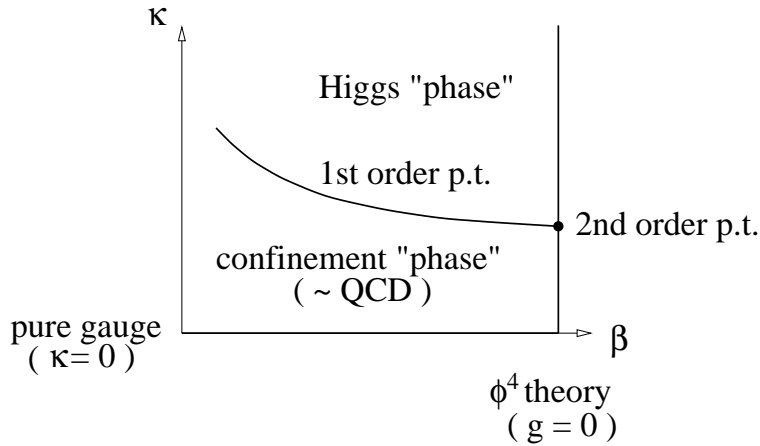,width=10cm}}
\vspace{0.5cm}
\caption{The phase diagram of the SU(2) Higgs model in the
  $(\beta,\kappa)$ plane is shown for a typical value of $\lambda$.
\label{phasediagram}}
\end{figure}
%%%%%%%%%%%%%%%%%%%%%%%%%%%%%%%%%%%%%%%%%%%%%%%%%%%%%%%%%%%%%%%%%%%%%%

\fig{phasediagram} shows the phase structure of the SU(2) Higgs
model \cite{MontMuen,Langguth:1986dr}. 
In the $(\beta,\kappa)$ plane there is a line
which is believed to be a first order phase transition line separating
the {\em confinement ``phase''} at small values of $\kappa$
from the {\em Higgs ``phase''} at larger $\kappa$. The situation in
\fig{phasediagram} is for a typical value of $\lambda$. 
The $\beta=\infty$ ($g=0$) boundary of the phase diagram
is the pure scalar $\phi^4$ model,
in which there is a second order phase transition line
$\kappa_c(\lambda)$ separating the phase where the O(4) symmetry 
is spontaneously broken ($\kappa>\kappa_c$) 
from the symmetric phase ($\kappa<\kappa_c$).
At $\kappa=0$ the Higgs field becomes infinitely heavy and
decouples from the gauge field: we are left with a pure SU(2) gauge
theory.
At small values of $\beta$ there is an analytic connection between the
two ``phases'': this is why we put ``phase'' in quotation
marks. As an analogy, we can think of the liquid and vapor regions in
the phase diagram of a fluid. Nevertheless, the physical properties
can be quite different in the two ``phases'', as can be
seen for instance by inspecting the static potential extracted 
from the Wilson loops at small distances.
In the confinement ``phase'', which is the continuation at finite
$\beta$ of the symmetric phase of the $\phi^4$ theory, the gauge field
has confinement properties like in QCD. The
static charges are bound by the color field string and the potential
rises, in good approximation, linearly as
in pure gauge theory \cite{Evertz:1986vp, Montvay:1986nk}.
In the Higgs ``phase'', which continues the spontaneously broken phase
of the $\phi^4$ theory, the Higgs mechanism is
at work: the gauge vector W-bosons become massive.
Far enough from the phase transition the interaction between 
static charges is mediated by the exchange of W-bosons 
and the potential has a Yukawa form
\cite{Montvay:1986nk,Langguth:1986dr}.

Because we are interested in the Higgs model as a test model for QCD,
we work in the confinement ``phase'' and at large values of the
gauge coupling $g$ ($\beta=4/g^2=2.2,\;2.4$). The situation
of the Standard Model Higgs sector is ``opposite'', in the sense that
it is in the Higgs ``phase'' and natural choices for the gauge coupling
are $\beta\simeq 8$ \cite{Langguth:1986dr}. The
simulation in the confinement phase at these small values of the gauge
coupling would be impossible because of the extremely large
correlation lengths. Close to the continuum, 
in the lowest order approximation of the massless 
perturbative renormalisation group equation, the lattice
spacing $a$ depends exponentially on $\beta$ \cite{Langguth:1986dr}
\bes\label{lattspacing}
 \ln\left(a\Lambda_{\rm L}\right) & \simeq &
 -\frac{12}{43}\pi^2\beta \,,
\ees
where $\Lambda_{\rm L}$ is the renormalisation group invariant
$\Lambda$-parameter in the lattice scheme.
The scale of a lattice gauge theory
simulation can be set by computing the physical length $\rnod$ 
\cite{Sommer:1993ce} from the force between static charges. In QCD
this length corresponds to $0.5\,\fm$. In \sect{resb24}, we will
present the results of the simulation of the SU(2) Higgs model for the
parameter set $\beta=2.4,\;\kappa=0.2759,\;\lambda=0.7$: the scale
$\rnod$ is approximately 5 lattice spacings.
If we evolve with \eq{lattspacing} the
lattice spacing from $\beta=2.4$ to $\beta=8$ we find
\bes\label{beta8}
 \frac{\rnod}{a} & \simeq & 10^7 \quad (\beta=8) \,.
\ees

\section{Monte Carlo simulation \label{simulation}}

The general principles of a simulation of a quantum field theory on a
space-time lattice are explained in reference \cite{MontMuen}. 
Here and in \App{mcsu2higgs}, we give a detailed description of 
the Monte Carlo updating algorithms that we use for the simulation of
the SU(2) Higgs model.

We use a hybrid over-relaxation algorithm (HOR)
\cite{Gupta:1988pf,Apostolakis:1991km,Hasenbusch:1992tx,Wolff:1992ze,
Creutz:1987xi,Brown:1987rr,Decker:1990hp,Gupta:1988yw,Booth:1992kk,
Wolff:1992ri} 
which is a mixture of heatbath and over-relaxation algorithms.
These algorithms are local in the sense that in each step
only one field variable $\psi$ 
(a gauge link $U(x,\mu)$ or a Higgs field variable $\Phi(x)$) 
is updated. A sequence of local steps, updating all field
variables, is called a sweep. The updating of the field variables is a
stochastic process: the change $\psi^{\prime}$ of the field variable
$\psi$ happens with a given transition probability
$p(\psi\to\psi^{\prime})$. In order that the field configurations
generated reach the equilibrium distribution $\exp(-S)/Z$, where
$S$ is the action and $Z$ the partition function \eq{partitionfunc},
it is sufficient that the updating algorithms satisfy the conditions of
local detailed balance and ergodicity. Local detailed balance means
that 
\bes\label{detbal}
 p(\psi\to\psi^{\prime})\,\rme^{-S(\psi)} & = &
 p(\psi^{\prime}\to\psi)\,\rme^{-S(\psi^{\prime})} \,,
\ees
where $S(\psi)$ is the part of the action depending on the field
variable $\psi$. Ergodicity means that each field configuration can be
reached by a finite number of updating sweeps.

In the heatbath algorithm, the new value $\psi^{\prime}$ for the field
variable $\psi$ is chosen independently of the original value
$\psi$ according to the transition probability
\bes\label{heatbath}
 p(\psi\to\psi^{\prime}) & = &
 \frac{\rme^{-S(\psi^{\prime})}}{\int\rmd\psi\,\rme^{-S(\psi)}} \,.
\ees
In analogy with thermodynamics, we can
imagine that the field variable is brought in contact with an
infinite ``heat bath'' in the equilibrium distribution \eq{heatbath}.

In the over-relaxation algorithm, the new value $\psi^{\prime}$ for the
field variable $\psi$ leaves the action $S(\psi)$ invariant
(this is called a microcanonical change)
\bes\label{overrel}
 S(\psi^{\prime}) & = & S(\psi) \,.
\ees
The change $\psi\to\psi^{\prime}$ is proposed with some arbitrary 
probability $p_{\rm C}(\psi\to\psi^{\prime})$ and is accepted with 
probability
\bes\label{metropolis}
 p_{\rm A}(\psi\to\psi^{\prime}) & \propto & 
 \min\left\{ 1,\frac{p_{\rm
     C}(\psi^{\prime}\to\psi)\rme^{-S(\psi^{\prime})}}{ 
     p_{\rm C}(\psi\to\psi^{\prime})\rme^{-S(\psi)}}\right\}
\ees
(the factors $\exp(-S)$ cancel if the change is exactly microcanonical).
The transition probability is $p=p_{\rm A}p_{\rm C}$.
The aim of the over-relaxation is to speed up the updating process by
choosing the new field variable $\psi^{\prime}$ as far as possible
from $\psi$ (a kind of reflection, see below). A consequence of
constant action is that the algorithm is non-ergodic. In the HOR
algorithms this difficulty is cured by combining over-relaxation with
heatbath updating sweeps.

It is often not possible to implement the heatbath and over-relaxation
algorithms exactly. What is then done is to propose a new value
$\psi^{\prime}$ of the field variable $\psi$ with an approximation of
the algorithm and accept the change with a probability that corrects
for the approximation done.

The updating of the SU(2) Higgs model we have chosen was inspired by
reference \cite{Bunk:1994xs}. It consists of cycles, that we call
iterations, composed each of one heatbath sweep for the gauge field,
followed by one heatbath sweep for the Higgs field and $N_{\rm OR}$
times an over-relaxation block composed by one over-relaxation sweep
for the gauge field and three over-relaxation sweeps for the Higgs
field. The integrated autocorrelation times (see \sect{staterr}) 
depend on the order in which the
field variables are updated during the sweep
\cite{deDivitiis:1995yz}. 
For the link variables we use the SF-updating of \cite{deDivitiis:1995yz}, 
in which the outermost loop runs over the direction $\mu$ and the
internal loops run over the lattice points in lexicographic order (we
refer to \App{parallel} for the details). The updating sweeps of the 
Higgs field variables process the lattice points in the same lexicographic
order. In the $HOR$ algorithms the parameter $N_{\rm OR}$ should be
chosen so as to minimise the autocorrelation times of the quantities
of interest. Our choice $N_{\rm OR}=1$ was motivated by a rough study 
of the integrated autocorrelation times of observables like
plaquette, Higgs length squared and gauge invariant links. They were
found to be minimal for $N_{\rm OR}=1,2$. From the study of these
``cheap'' (referred to the computer time needed for the measurements) 
observables we could draw useful conclusions for the measurements of
the observables in which we are interested (see \sect{binning}).

In \App{mcsu2higgs}, we give a detailed description of the different
parts of the HOR algorithm that we use for the simulation of the SU(2)
Higgs model. Attention is also paid to the generation of random
numbers needed for the implementation of the algorithms.

\section{Statistical error analysis \label{staterr}}

An essential part of the Monte Carlo simulations are the estimates of
the errors of the observables computed as in \eq{ensembleaverage},
called primary quantities, and of secondary quantities,
which are arbitrary functions of primary quantities. Besides the naive
statistical error, associated with the finite number of measurements $N$
and proportional to $1/\sqrt{N}$, there are other fundamental 
sources for errors related to the updating algorithms used \cite{Sokal:MCM}:
\begin{itemize}
\item[$\bullet$] Initialisation bias. The algorithm needs a number of
  thermalisation steps before it ``forgets'' the arbitrary 
  initial configuration and
  reaches the thermal equilibrium where the field configurations
  are distributed according to the Boltzmann factor $\exp(-S)$.
\item[$\bullet$] Autoccorelation in equilibrium. When thermal
  equilibrium is reached, the field configurations generated by the updating
  algorithm are correlated. This causes the statistical error of
  $\overline{O}$ in \eq{ensembleaverage} to be a factor 
  $2\tau_{\rm int}(O)$ larger than in an ensemble of independent
  configurations. The quantity $\tau_{\rm int}(O)$ is called the
  integrated autocorrelation time for the observable $O$.
\end{itemize}

The dependency on the initial (arbitrary) configuration can be avoided
by waiting a ``large enough'' number of updating steps before starting the
measurements. By measurements we mean the evaluation of the
observables on the field configurations generated by the
algorithm. What is ``large enough'' can be
estimated from the integrated autocorrelation times of the
observables. They can be very different for different observables. In
practice, the observed autocorrelation times have almost the same order
of magnitude and
a number of thermalisation steps equal $20$ to $100$ times the maximum
observed autocorrelation time
$\tau_{int,max}$ is a sensible choice.

The central role in the determination of the statistical errors is
played by the integrated autocorrelation times. How to estimate them is the
subject of this section.

\subsection{Primary quantities}

We consider a sequence of measurements 
$A_i\equiv A[U_i,\Phi_i],\;i=1,...,N$ of the
observable $A\equiv A[U,\Phi]$
performed on a large ensemble of field configurations
$\{[U_i,\Phi_i],\;i=1,...,N\}$
already in the equilibrium distribution.
We denote by
\bes\label{pav}
 \overline{A} & = & \frac{1}{N}\sum_{i=1}^N A_i
\ees
the ensemble average of $A$. The exact path integral expectation value
of $A$ is denoted by $\langle A\rangle$.
If the measurements are statistically independent the value of
$\overline{A}$ is normally distributed around the expectation value
$\langle A\rangle$ with variance
\bes\label{pvarnaive}
 \var(\ovA) & = & \overline{A^2}-\ovA^2 \;=\; 
 \overline{(A-\ovA)^2} \,.
\ees
The (naive) statistical error is then given by
\bes\label{perrnaive}
 (\Delta_{\rm naive}(\ovA))^2 & = & \frac{\var(A)}{N-1} \,.
\ees
In general, there are correlations in the sequence of generated field
configurations (and hence in the measurements), called
autocorrelations and \eq{perrnaive} underestimates the statistical error.
In order to obtain reliable statistical errors we follow
\cite{Sokal:MCM,Wolff:CP2}.

The (unnormalised) autoccorelation function is defined as 
\bes\label{autocorr}
 \Gamma_A(i-j) & = & \langle (A_i-\langle A\rangle)(A_j-\langle
 A\rangle) \rangle_{\rm MC} \;=\; \Gamma_A(j-i) \,,
\ees
where $\langle\cdot\cdot\cdot\rangle_{\rm MC}$ denotes the average over
infinitely many independent ensembles of configurations in thermal
equilibrium. The autocorrelation function $\Gamma_A$
depends only on the distance between the measurements $|t|=i-j$.
Typically, it decays exponentially
\bes\label{expdecayauto}
 \Gamma_A(t) & \sim & \exp(-|t|/\tau) \qquad \mbox{for large $t$} \,.
\ees
The integrated autocorrelation time is defined as
\bes\label{intautotime}
 \tau_{\rm int}(A) & = & \frac{1}{2}\sum_{t=-\infty}^{\infty}
 \frac{\Gamma_A(t)}{\Gamma_A(0)} \,,
\ees
where $\Gamma_A(0)=\langle (A-\langle A\rangle)^2\rangle=\var(A)$ is the
variance\footnote{
We note that $\langle A_i\rangle_{\rm MC}=\langle A\rangle$.}
of the observable $A$. Here,
``time'' refers to the ``Monte Carlo time'' of the simulation and
labels the measurements.

The goal is to estimate the effects of the autocorrelations based on a
finite (but large) sequence of measurements $A_i,\;i=1,...,N$. The
ensemble average $\ovA$ in \eq{pav} has statistical error, corrected for
autocorrelations, given by
\bes\label{perr}
 (\Delta(\ovA))^2 & = & \frac{1}{N^2}\sum_{i,j=1}^N \Gamma_A(i-j) \nonumber
 \\
 & = & \frac{1}{N}\sum_{t=-(N-1)}^{N-1} \left(1-\frac{|t|}{N}\right)
 \Gamma_A(t) \nonumber \\
 & \approx & \frac{1}{N}(2\tau_{\rm int}(A))\Gamma_A(0) \qquad
 \mbox{for}\quad N\gg\tau \,.
\ees
Comparing with \eq{perrnaive}, we see that the statistical error is a
factor $\sqrt{2\tau_{\rm int}(A)}$ larger than for independent
measurements. Stated differently, the number of ``effectively
independent measurements'' in a run of length $N$ is roughly
$N/(2\tau_{\rm int}(A))$.
The ``natural'' estimator of $\Gamma_A(t)$ is
\bes\label{eautocorr}
 \Gamma_A(t) & \approx & \frac{1}{N-|t|}\sum_{i=1}^{N-|t|}
 (A_i-\ovA)(A_{i+|t|}-\ovA) \,.
\ees
In order to get a good estimator of $\tau_{\rm int}(A)$, one sums
the terms in \eq{intautotime} (with $\Gamma_A$ computed according to
\eq{eautocorr}) up to $|t|\le M$, where $M$ is a suitably chosen
cut-off \cite{Madras:1988ei}. This cut-off is necessary since the
``signal'' for $\Gamma_A(t)/\Gamma_A(0)$ gets lost in the ``noise'' for
$|t|\gg\tau$.

In the following subsection, we describe an alternative method for
estimating the error, the binning method. Knowing $\Delta(\ovA)$ one
can use \eq{perr} together with \eq{perrnaive} to estimate the
integrated autocorrelation time:
\bes\label{eintautotime}
 \tau_{\rm int}(A) & = & \frac{1}{2}\left(
 \frac{\Delta(\ovA)}{\Delta_{\rm naive}(\ovA)} \right)^2 \,.
\ees

\subsection{Binning \label{binning}}

An easy method to analyse the data of a Monte Carlo simulation is
the binning method. The measurements $A_i,\;i=1,...,N$
are first averaged into blocks of length $B$ called bins
\bes\label{bin}
 A_{b,B} & = & \frac{1}{B}\sum_{i=1+(b-1)B}^{bB} A_i\,, \quad
 b=1,...,N_B=[N/B]\,.
\ees
If $N$ is divisible by $B$, the average over the blocked measurements
is the same as the average $\overline{A}$ in \eq{pav}.
The variance computed from the blocked measurements is
\bes\label{pvarbin}
 \var(\ovA,B) & = & \frac{1}{N_B}\sum_{b=1}^{N_B}
 \left(A_{b,B}-\frac{1}{N_B}\sum_{b^{\prime}=1}^{N_B} A_{b^{\prime},B}
 \right)^2 \,.
\ees
The blocked measurements still suffer from
autocorrelations. The error $\Delta(\ovA,B)$ of the average $\overline{A}$,
estimated through the variance \eq{pvarbin}, is
\bes\label{perrbin}
 (\Delta(\ovA,B))^2 & = & \frac{\var(\ovA,B)}{N_B-1} \,.
\ees
It increases with the bin length $B$:
if the integrated autoccorelation time $\tau_{\rm int}(A)$ is
small with respect to $B$, the systematic effect due to autocorrelations is
proportional to $\tau_{\rm int}(A)/B$ \cite{Jansen:1998mx}. The
relative statistical uncertainty of the error estimate \eq{perrbin} is
approximately given by $(2N_B)^{-1/2}$ \cite{Jansen:1998mx}.
Increasing the value of $B$ the
error \eq{perrbin} flattens and oscillates around
its correct value $\Delta(\ovA)$,
if the number of measurements is large enough to see this.
The integrated autocorrelation time can then be estimated as in
\eq{eintautotime}, with $\Delta_{\rm naive}(\ovA)\equiv
\Delta(\ovA,B=1)$. The error of this estimate is dominated
by the uncertainty of $\Delta(\ovA)$ and is given by
\bes\label{erreintautotime}
 \frac{\Delta(\tau_{\rm int}(A))}{\tau_{\rm int}(A)} & = & 
 2\,\frac{\Delta(\Delta(\ovA))}{\Delta(\ovA)} \,.
\ees
%%%%%%%%%%%%%%%%%%%%%%%%%%%%%FIGURE%%%%%%%%%%%%%%%%%%%%%%%%%%%%%%%%%%%
\begin{figure}[tb]
\hspace{0cm}
\vspace{-1.0cm}
\centerline{\epsfig{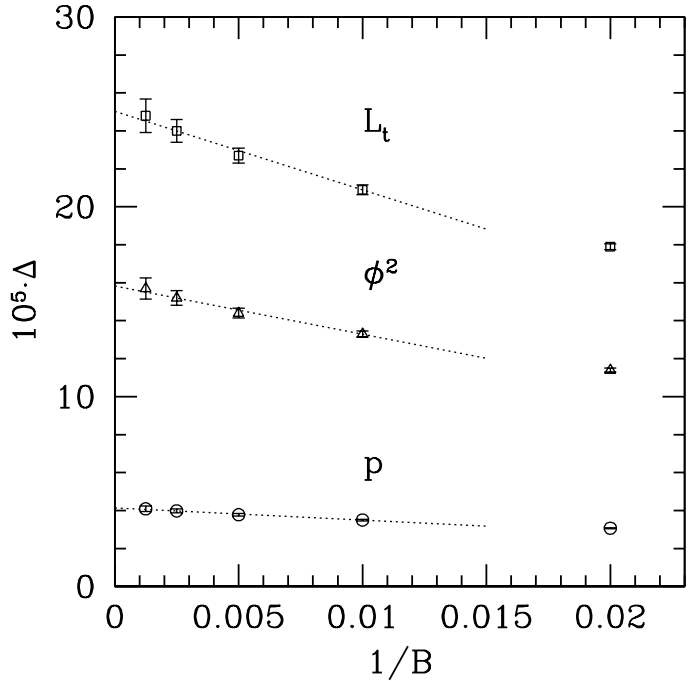}}
\vspace{0.5cm}
\caption{Here, we show the dependence of the error estimates
  \eq{perrbin} on the inverse
  bin length $1/B$. The systematic effects due to autocorrelations,
  which are proportional to $\tau_{\rm int}/B$, are clearly
  visible. The observables are defined in \eq{plaq},
  \eq{higgslensq} and \eq{tsphilink} and were measured on a $8^4$
  lattice with parameters $\beta=2.2$, $\kappa=0.274$ and
  $\lambda=0.5$. The statistics is 320,000 measurements.
  \label{f_autocorr_plahig}}
\end{figure}
%%%%%%%%%%%%%%%%%%%%%%%%%%%%%%%%%%%%%%%%%%%%%%%%%%%%%%%%%%%%%%%%%%%%%%

In order to illustrate the binning method, in
\fig{f_autocorr_plahig} we show
the error estimate \eq{perrbin} as function of the inverse bin
length $1/B$ for different observables.
The measurements are performed on a $8^4$ lattice, for
the parameter set $\beta=2.2$, $\kappa=0.274$ and $\lambda=0.5$,
after each iteration updating (the Monte Carlo time unit
is therefore 1 iteration).
The observables are the plaquette 
\bes\label{plaq}
 p & = & \frac{1}{6\,\Omega}\sum_x\sum_{\mu<\nu} \{1-\frac{1}{2}
 \tr P_{\mu\nu}(x)\} \,,
\ees 
the Higgs field length squared
\bes\label{higgslensq}
 \phi^2 & = &  \frac{1}{\Omega}\sum_x \Phid(x)\Phi(x) \,,
\ees
and the time-like\footnote{
Time and space are the same on a $L^4$ lattice.}
gauge invariant link
\bes\label{tsphilink}
 L_t & = & \frac{1}{\Omega}\sum_x
 \frac{1}{2}\tr[\vp^{\dagger}(x)U(x,0)\vp(x+a\hat{0})] \,.
\ees
We make use of translation invariance and isotropy on the
lattice to average the observables: this helps to reduce the
statistical errors.
Reliable error estimates for all these observables can be read off at
$1/B=1/400$. The number of blocked measurements is then $N_B=800$
giving a relative statistical uncertainty of the error estimates of 2.5\%.
The systematic uncertainty can be estimated from the dotted lines in
\fig{f_autocorr_plahig}, precisely from the difference between the
errors at $1/B=1/400$ and $1/B=0$, the latter being extrapolated.
For all observables represented
the relative systematic uncertainty of their error estimates is 4\%.
From \eq{eintautotime} we obtain
the integrated autocorrelation times
\bes\label{tauestimatesb22}
 & \tau_{\rm int}(p) \; = \; 12.1 \,,\quad
 \tau_{\rm int}(\phi^2) \; = \; 10.6 \,,\quad 
 \tau_{\rm int}(L_t) \; = \; 14.2 \,,
\ees
in units of iterations. From \eq{erreintautotime} we can
estimate the relative uncertainty of the integrated
autocorrelation times to be 8\%, coming from the systematic
uncertainty of the errors of the observables.

The same analysis, repeated for a $32^4$ lattice and parameter set 
$\beta=2.4$, $\kappa=0.2759$, $\lambda=0.7$, on 24,000 measurements,
gives the integrated autocorrelation times (in units of
iterations)
\bes\label{tauestimatesb24}
 & \tau_{\rm int}(p) \; = \; 4.9 \,,\quad
 \tau_{\rm int}(\phi^2) \; = \; 6.0 \,,\quad 
 \tau_{\rm int}(L_t) \; = \; 9.9 \,.
\ees
The relative uncertainty is again dominated by the systematic effects
and is of 12\% for $\tau_{\rm int}(p)$, 22\% for 
$\tau_{\rm int}(\phi^2)$ and 14\% for $\tau_{\rm int}(L_t)$. 
The autocorrelation times have all the same order of magnitude: we
conclude that measurements effectuated only after 30 iterations
should almost be statistically independent. This is in fact
confirmed by the measurements of the matrix correlation for the static
potential and the meson mass, see \sect{resb24}.

\subsection{Secondary quantities: jackknife binning \label{jackknife}}

Secondary quantities are defined as 
\bes\label{sob}
 y & = & f(A^{(1)},A^{(2)},\cdot\cdot\cdot) \,,
\ees
where $f$ is an arbitrary function of the
primary quantities $A^{(1)},\;A^{(2)},\;\cdot\cdot\cdot$.
The function $f$ can be complicated, such as the extraction of
eigenvalues of a matrix correlation, see \sect{variation}.
The best estimate of a secondary quantity is
\bes\label{secondary}
 \overline{y} & = & f(\overline{A^{(1)}},\overline{A^{(2)}},
 \cdot\cdot\cdot) \,.
\ees
To estimate the statistical error
of $\overline{y}$ one can in principle use the binning method
described in \sect{binning}: the quantities
$y_{b,B}=f(A^{(1)}_{b,B},A^{(2)}_{b,B},\cdot\cdot\cdot)$ are inserted in 
\eq{pvarbin} and \eq{perrbin} at the place of $A_{b,B}$.
The problem in practice, is often that the bins are too small
(because of the time costs of the measurements)
and they fluctuate too much around $\overline{y}$.
This problem can be overcome with the method of jackknife binning.

For the primary quantities,
we consider the bins $A_{b,B},\;b=1,...,N_B$ and build the jackknife
averages
\bes\label{jackcompl}
 A_{b,\bar{B}} & = & \frac{1}{N_B-1}\sum_{b^{\prime}\neq b}
 A_{b^{\prime},B}\,, \quad b=1,...,N_B \,,
\ees
obtained by omitting a single bin in all possible ways. The index
$\bar{B}$ means that $A_{b,\bar{B}}$ is the complement of the bin $A_{b,B}$.
Evaluating the secondary
quantity $y$ with the jackknife averages \eq{jackcompl}
we obtain the jackknife estimators
\bes\label{jackest}
 y_{b,\bar{B}} & = & f(A^{(1)}_{b,\bar{B}},A^{(2)}_{b,\bar{B}},
 \cdot\cdot\cdot) \,,
\ees
with an average
\bes\label{jackav}
 \overline{y_{\bar{B}}} & = & \frac{1}{N_B}\sum_b y_{b,\bar{B}} \,.
\ees
The error estimate for $\overline{y}$ can be obtained from
\cite{MontMuen,Wolff:CP2}
\bes\label{serr}
 (\Delta(\overline{y},B))^2 & = & (N_B-1)\left(
 \frac{1}{N_B}\sum_b y_{b,\bar{B}}^2\, -\, \overline{y_{\bar{B}}}^2 \right)
 \,.
\ees
For a primary quantity $y\equiv A$, \eq{serr} reproduces \eq{perrbin}.
The error estimate \eq{serr} can be studied under variation of the
bin length $B$ as in \sect{binning}. Increasing $B$, the
error estimate flattens and oscillates around the correct error.
The integrated autocorrelation time for the secondary quantity $y$
can then be estimated as in \eq{eintautotime},
the naive error being the error \eq{serr} for $B=1$.

The jackknife error analysis is our standard method for estimating
statistical errors.

\chapter{Static-light mesons \label{slmesons}}

As already described in the introduction,
we expect the static potential $V_0$ to be described in terms
of a pair of weakly interacting static-light mesons
at large separations $r$ of the static charges.
A static-light meson
is a bound state of a static charge and the dynamical 
Higgs field. The interaction between two such mesons is expected to be
of Yukawa-type, mediated by the exchange of light color singlet bound
states of Higgs and gauge fields. We denote by $\mu$ the mass of one
static-light meson: the static potential is expected to reach the
asymptotic (in $r$) value
\bes\label{potasympt}
 \lim_{r\to \infty}V_0(r) = 2 \mu \,.
\ees
In the Hamiltonian formalism explained in \App{apptm}, the static-light
mesons live in the sector of the Hilbert space with one static
charge in the fundamental representation of the gauge group. We denote
by $\Oop^{\rmM}_i(\vec{x})$ a set of operators labelled by $i$ that,
when applied to the vacuum state $|0\rangle$, create meson-type states
\bes\label{mesonstate}
 |i\rangle & = & \Oop^{\rmM}_i(\vec{x})|0\rangle \quad (i=1,2,3,...)
\ees
localised around the position $\vec{x}$ of the static charge.
These operators carry a color\footnote{
Color is the quantum number associated with the gauge group.}
index $a=1,2$ and
transform, under gauge transformation defined in \eq{gtrsfuop} and
\eq{gtrsfphiop}, as
\bes\label{mesonop}
 \Rop^{\dagger}(\Lambda) [\Oop^{\rmM}_i(\vec{x})]_a \Rop(\Lambda)
 & = & \Lambda^{\dagger}_{aa^{\prime}}(\vec{x})
 [\Oop^{\rmM}_i(\vec{x})]_{a^{\prime}} \, ,
\ees
where $\Rop(\Lambda)$ denote the operator representation of the
gauge transformation $\{\Lambda(\vec{x})\in\SUtwo\}$. The transfer matrix in
the sector with one static charge has a spectral representation
\bes\label{tmmes}
 \trans_{\rmq} & = & \sum_{n_{\rmq}} |n_{\rmq}\rangle
 \rme^{-aE_n^{(\rmq)}} \langle n_{\rmq}| \,,
\ees
where $|n_{\rmq},a\rangle$ is a basis of eigenstates of the Hamilton
operator $\ham$ in this sector with energies $E_n^{(q)}$ independent
of the color $a$ (the sum over the color multiplicity of the states is
implicit in \eq{tmmes}). The mass
$\mu$ of a static-light meson is defined as the difference between
the ground state energy $E_0^{(\rmq)}$ and the vacuum energy $E_0^{(0)}$
\bes\label{mumass}
 \mu & = & E_0^{(q)} - E_0^{(0)} \, .
\ees
It can be extracted from the correlation
\bes\label{mucorrop}
  C^{\rmM}_{ij}(t) & = & \frac{1}{Z}\Tr\left(\sum_b
   [\Oop^{\rmM}_j(\vec{x})^{\dagger}]_b \trans_{\rmq}^{t/a}
   [\Oop^{\rmM}_i(\vec{x})]_b \trans_0^{(T-t)/a} \right) \,,
\ees
where $\trans_0=\sum_{n_0} |n_0\rangle\exp(-aE_n^{(0)})\langle n_0|$
is the transfer matrix in the zero charge
sector (see \sect{chargelatt}),
$Z=\Tr(\trans_0^{T/a})$ is the partition function and $T$
the physical time extension of the lattice.
In the limits
\bes\label{tasympt}
 (T-t)(E_1^{(0)}-E_0^{(0)})\gg1 & \mbox{and} &
 t(E_1^{(\rmq)}-E_0^{(\rmq)})\gg1 \,,
\ees
the correlation in \eq{mucorrop} has the asymptotic behavior
\bes\label{mucorrop2}
  C^{\rmM}_{ij}(t) & \sim & \alpha_j^*\alpha_i\rme^{-t\mu} \,,
\ees
where
$\alpha_i\equiv[\alpha_i]_{ab}=
\langle0_{\rmq},a|[\Oop^{\rmM}_i(\vec{x})]_b|0\rangle$ and
the trace over the color indices of the states and of the meson operators
is implicit in \eq{mucorrop2}.
In analogy with the reconstruction theorem proved in
\sect{reco}, one can show that \eq{mucorrop} can be rewritten in the path
integral formalism as the expectation value
\bes\label{mucorr}
  C^{\rm M}_{ij}(t) & = & \langle [O^{\rmM}_j(x+t\hat 0)^*]_a\,
  U(x,x+t\hat 0)^{\dagger}_{ab}\,[O^{\rmM}_i(x)]_b \rangle \,.
\ees
The static charge is represented by a straight time-like Wilson line
$U(x,x+t\hat 0)^{\dagger}$ connecting $x$ with $x+t\hat{0}$.
The meson state $|i\rangle$ is
represented by the composite field $O^{\rmM}_i(x)$ involving Higgs and
gauge fields at equal time $x_0$. To any of such fields we can uniquely
associate a field operator in the Hilbert space by replacing the
fundamental fields with the multiplicative field operators defined in
\eq{opu} and \eq{opphi}.
The operator associated with $O^{\rmM}_i(x)$ is precisely
$\Oop^{\rmM}_i(\vec{x})$.
The only restriction in the choice of the fields $O^{\rmM}_i(x)$
is imposed by the transformation property under gauge transformation
defined in \eq{gaugetrsfu4} and \eq{gaugetrsfphi4}: the field
$O^{\rmM}_i(x)$ must be in the fundamental representation of the color
gauge group.
In addition to the local field $\Phi(x)$ we can choose for $O^{\rmM}_i(x)$
linear combinations which take into account contributions from the
neighboring Higgs fields (smeared fields) and also more
general composite fields, with the intent to reproduce the wave function
of the meson.
The physical picture is that of a cloud of dynamical Higgs
and gauge fields surrounding and bound to the static charge.

Actually, \eq{mucorr} defines a matrix correlation function. In
\sect{variation}, we describe a variational method for extracting from
$C_{ij}^{\rmM}(t)$ not only the ground state meson mass $\mu$, but also
the energy spectrum of the excited states.
The idea behind the method is that it is possible to find,
from the basis of states $|i\rangle$ defined in \eq{mesonstate},
linear combinations approximating the eigenstates of
the Hamiltonian in the sector with one static charge. The success of
the variational method is therefore based on the ``quality'' of the
basis of states. This fact makes the study of smearing operators
important and it is the subject of \sect{mesontype}.

Static-light mesons in QCD are a good approximation for
B-mesons: the mass $m_b$ of the $b$ quark is large compared to
$\Lambda_{\rm QCD}\sim\,0.2\,\GeV$ and in this sense the $b$ quark can be
considered a heavy quark. The corrections to the static limit
$m_b\to\infty$ are of order $\Lambda_{\rm QCD}/m_b$ and can be
computed in the framework of the heavy quark effective field theory,
see for example references \cite{Neubert:1994ib,Neubert:1994mb}.

\section{Variational method \label{variation}}

From the matrix correlation function $C_{ij}(t)$ in \eq{mucorrop}
(we drop the label M), constructed
with the basis of states $|i\rangle$ defined in \eq{mesonstate},
it is possible to extract the energy spectrum
in the charge sector of the Hilbert space with one static charge,
where the static-light mesons ``live''.
We denote the eigenstates of the Hamiltonian in this charge sector by
$|\alpha\rangle$. These states have at least a two-fold degeneracy:
they carry a color index $a=1,2$ and their energy is independent
of the color since the Hamiltonian is gauge invariant.
Due to gauge invariance of the correlation matrix in
\eq{mucorrop}, the color multiplicity is simply factored out.
In the following therefore,
we drop the color indices of the states and operators.
Moreover, we restrict our considerations to states with spin 0.
This restriction is implemented in the way the states are
constructed, for example the smearing procedures that we employ treat
each spacial direction in the same way.
The eigenvalues of the Hamiltonian are discrete because we are on the
lattice, so that, in summary, the index $\alpha=0,1,2,...$ labels
the energy levels $E_{\alpha}^{(\rmq)}$ which we assume are not
degenerate. Only the energy differences 
\bes\label{eigen}
 W_{\alpha}\;=\;E_{\alpha}^{(\rmq)}-E_0^{(0)}\,, \quad
 W_{\alpha}\;<\;W_{\alpha+1} \quad
 (\alpha=0,1,2,...) \,,
\ees
have a physical meaning.
For the eigenstates we choose the normalisation
\bes\label{normeig}
 \langle\alpha|\alpha^{\prime}\rangle & = & \delta_{\alpha\alpha^{\prime}}
 \,.
\ees
Taking the limit $T\to\infty$ in \eq{mucorrop}, we can write
the matrix correlation function $C_{ij}(t)$ as
\bes\label{mucorrtd}
 C_{ij}(t) & = & \sum_{\alpha} 
 \langle 0|\Oop^{\rmM}_j(\vec{x})^{\dagger}|\alpha\rangle 
 \langle \alpha|\Oop^{\rmM}_i(\vec{x})|0\rangle\,\rme^{-tW_{\alpha}} \,.
\ees
In practice, the limit $T\to\infty$ is reached when
$T(E_1^{(0)}-E_0^{(0)})\gg1$, which means that $T$ must be larger than
the inverse mass gap in the zero charge sector. This is always the
case for the situations that we consider, as we discuss in
\sect{kappamatch}.

For matrices of the type in \eq{mucorrtd} a general lemma for the
extraction of the energies $W_{\alpha}$ has been proved in
\cite{phaseshifts:LW}. In this reference, a variational
method is proposed, which is superior to a straightforward application
of the lemma. It consists in solving the generalised
eigenvalue problem:
\bes\label{genev}
  C_{ij}(t)v_{\alpha,j}(t,t_0) & = & 
  \lambda_{\alpha}(t,t_0)C_{ij}(t_0)v_{\alpha,j}(t,t_0) \, , \quad 
  \lambda_{\alpha} > \lambda_{\alpha+1} \, ,
\ees
where $t_0$ is fixed and small (in practice we use $t_0=0$).
The generalised eigenvalues
$\lambda_{\alpha}(t,t_0)$ are computed as the
eigenvalues of the symmetric matrix $\bar{C}=C(t_0)^{-1/2}C(t)C(t_0)^{-1/2}$
and the vectors
\bes\label{genevec}
 \bar{v}_{\alpha,i}\;=\;[C(t_0)^{1/2}]_{ij}v_{\alpha,j}(t,t_0) 
 & \mbox{with} &
 \bar{v}_{\alpha,i}\bar{v}_{\alpha^{\prime},i}\;=\;
 \delta_{\alpha\alpha^{\prime}}
\ees
are the orthonormal eigenvectors of $\bar{C}$.
The positivity of the transfer matrix
ensures that $C(t)$ is positive definite for all $t$.
In \cite{phaseshifts:LW} it is proven that the energies
$W_{\alpha}$ are given by the expressions
\bes\label{spectrum}
  a W_{\alpha} & = & \ln(\lambda_{\alpha}(t-a,t_0) 
  /\lambda_{\alpha}(t,t_0)) +
  \rmO\left(\rme^{-t\Delta W_{\alpha}}\right) \, ,
\ees
where $\Delta W_{\alpha}=
\min\limits_{\beta\neq\alpha}|W_{\alpha}-W_{\beta}|$. It is
expected that, for a good basis of states,
the coefficients of the higher exponential corrections
in \eq{spectrum} are suppressed so that the
energies can be read off at moderately large values of $t$ from the
right-hand side of \eq{spectrum}.

From \eq{mucorr} and \eq{ccinvO} it follows that the matrix 
$C_{ij}(t)$ is real. Taking the complex conjugate of \eq{mucorrtd}, one
immediately sees that $C_{ij}(t)$ is symmetric. In a Monte Carlo
simulation these properties are satisfied only in the limit of infinite
statistics. We make use of the reality property and measure in the
simulation only the real part of the matrix elements. When we analyse
the data we symmetrise the matrix by hand. The eigenvalues of
$\bar{C}=C(t_0)^{-1/2}C(t)C(t_0)^{-1/2}$ are numerically obtained with the
Jacobi method for symmetric matrices \cite{NumRec}.

The variational method \eq{genev} and \eq{spectrum} is our standard
method for extracting the energy spectrum. What we have stated here
about this method is valid for
{\em any} charge sector of the Hilbert space. One has to start from a
basis $|i\rangle$ of states belonging to that charge sector, see
\sect{chargesect}. The matrix
correlation $C_{ij}(t)$ corresponds to matrix elements
$\langle j|\trans^n|i\rangle,\;n\equiv t/a$
of powers of the transfer matrix operator $\trans$
projected into the charge sector. How this works in detail,
is shown in \sect{reco}
for the sector with a static charge and a static anti-charge
in the fundamental representation of the gauge group. The energy
spectrum in this sector, the static potentials,
is the main subject of our work.

\section{One-link integral \label{onelink}}

Before describing our choice for the meson-type fields, we would
like to discuss a feature of the measurement of the matrix correlation
function $C_{ij}(t)$ \eq{mucorr} which is independent of that choice.
As we see from \eq{mucorrop2}, the values of the matrix elements
$C_{ij}(t)$ fall down exponentially for large $t$.
In order to measure these values in a Monte Carlo simulation with
statistical significance, also the variance of the matrix elements
should decrease exponentially\footnote{
The alternative is an exponential {\em increase} 
of the number of measurements.}
with $t$.
To achieve this, a method called ``one-link integral'' or
``multi-hit'' has been proposed in \cite{Parisi:1983hm}, which has
proven successful.

The general principle is to replace the observable $O$, for which one wants to 
decrease the statistical error by another one $O_{\rmI}$, with the same
expectation value but much smaller variance. Such an observable
$O_{\rmI}$ is called improved estimator. In the case of the matrix
correlation function $C_{ij}(t)$, we observe that it depends linearly
on the time-like links. When measuring $C_{ij}(t)$, we can substitute
the time-like links
by their expectation values in the fixed configuration of
the other field variables. These expectation values are called
one-link integrals.\footnote{
In general, the substitution in an observable of links with
their one-link integrals can be made under the following restrictions:
the observable must depend linearly on the links in question and no pair of
substituted links can belong to the same plaquette.}
For a given time-like link $U(x,0)$
we write the action \eq{action3} like
\bes\label{actionu0link}
 S & = & -\frac{\beta}{2}\tr\{U(x,0)W^{\dagger}(x,0)\} + \nonumber
 \\ & & \mbox{terms independent of $U(x,0)$} \, , \\
 W(x,0) & = & V(x,0) +
 \frac{2\kappa}{\beta}\vp(x)\vp^{\dagger}(x+a\hat 0) \, ,
\ees
where $V(x,0)$ is the sum of the products of links over the six
``staples'' around the link $U(x,0)$
\bes\label{staples0}
 V(x,0) & = & \sum_{k=1}^3
 \{U(x,k)U(x+a\kh,0)U^{\dagger}(x+a\hat{0},k) + \nonumber \\
 & & \quad U^{\dagger}(x-a\kh,k)U(x-a\kh,0)U(x-a\kh+a\hat{0},k)\} \, .
\ees
We denote the part of the action depending on $U(x,0)$ in
\eq{actionu0link} by $S(U(x,0))$.
The expectation value of $U(x,0)$, with all the other field variables
kept fixed, is given by
\bes\label{onelinkintegral}
 \overline{U}(x,0) & = & \frac{\int\rmd
   U(x,0)\,U(x,0)\,\rme^{-S(U(x,0))}}{\int\rmd
   U(x,0)\,\rme^{-S(U(x,0))}} \nonumber \\
 & = & \frac{I_2(\rho)}{I_1(\rho)}
 \frac{W(x,0)}{\sqrt{\det(W(x,0))}} \, ,
\ees
where $\rho=\beta\sqrt{\det(W(x,0))}$ and $I_n$ are the modified
Bessel functions of the specified integer order. The derivation of
\eq{onelinkintegral} and the numerical
evaluation of the ratio of Bessel functions is
discussed in \App{bessel}.

An exponential decrease of the variance of an observable
with the number of links that
are substituted by their one-link integrals, is reported for example in
\cite{Bali:1995de}. The observables considered there are Wilson loops
of time extent $t$ and the number of integrated links is
$2t/a$. The variance of Wilson loops computed with the
one-link integrals decay exponentially with $t$.

\section{Meson-type operators \label{mesontype}}

We studied different bases of meson-type fields $O_i^{\rmM}(x)$
by measuring in Monte Carlo simulations the matrix correlation function
$C_{ij}(t)$ defined in \eq{mucorr} and computing from it the energy
spectrum of the static-light mesons using the variational method described
in \sect{variation}.
All composite fields $O_i^{\rmM}(x)$, constructed with field variables
taken at equal time $x_0$ and transforming under gauge transformation
defined by \eq{gaugetrsfu4} and \eq{gaugetrsfphi4} as
\bes\label{gtrsfmesonfield}
 [O_i^{\rmM,\Lambda}(x)]_a & = &
 \Lambda_{aa^{\prime}}^{\dagger}(x)[O_i^{\rmM}(x)]_{a^{\prime}} \,,
\ees
can be considered.
Our aim was to find the best field basis for describing
the ground state of the static-light mesons.
For these studies we simulated the SU(2) Higgs model on a $20^4$ lattice
with parameters $\beta=2.2$, $\kappa=0.274$ and $\lambda=0.5$. This
parameter point is in the confinement ``phase'' of the model.
At the end of the section we show the results for the mass of the
ground and first excited meson state for a simulation at $\beta=2.4$.
The measurement of the matrix correlation is improved by the use of the
one-link integral method described in \sect{onelink}.
%%%%%%%%%%%%%%%%%%%%%%%%%%%%%FIGURE%%%%%%%%%%%%%%%%%%%%%%%%%%%%%%%%%%%
\begin{figure}[tb]
\hspace{0cm}
\vspace{-1.0cm}
\centerline{\epsfig{file=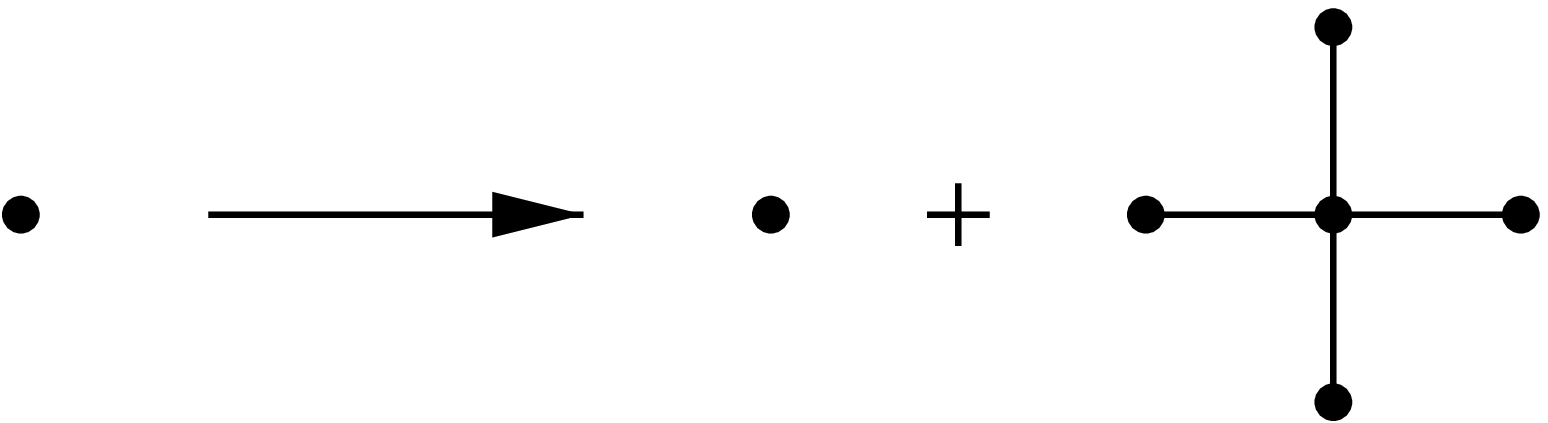,width=7.5cm}}
\vspace{0.5cm}
\caption{Here, we show the smearing procedure $S_1$ for the Higgs field
  (circles), defined in \eq{smearophiggs1}.
  The lines represent the link connections.
\label{smearingh1}}
\end{figure}
%%%%%%%%%%%%%%%%%%%%%%%%%%%%%%%%%%%%%%%%%%%%%%%%%%%%%%%%%%%%%%%%%%%%%%
%%%%%%%%%%%%%%%%%%%%%%%%%%%%%FIGURE%%%%%%%%%%%%%%%%%%%%%%%%%%%%%%%%%%%
\begin{figure}[tb]
\hspace{0cm}
\vspace{-1.0cm}
\centerline{\epsfig{file=plots/meson_comp.epsi,width=10cm}}
\vspace{-0.0cm}
\caption{Here, we compare the extraction of the mass $\mu$ of a
  static-light meson using different smearing operators defined in
  \eq{smearophiggs1} and \eq{smearophiggs2}.
\label{meson_comp}}
\end{figure}
%%%%%%%%%%%%%%%%%%%%%%%%%%%%%%%%%%%%%%%%%%%%%%%%%%%%%%%%%%%%%%%%%%%%%%

We first studied a basis containing the fundamental Higgs field
$\Phi(x)$ and smeared Higgs fields obtained by iterating the
application of a smearing operator $S_1$ to the Higgs field.
The smearing operator $S_1$ is defined as
\bes\label{smearophiggs1}
  S_1\,\Phi(x) & = &  
  \Phi(x)+\sum_{|x-y|=a \atop x_0=y_0}U(x,y)\Phi(y) \, , 
\ees
where $U(x,y)$ is the link connecting $y$ with $x$,
and is schematically represented in \fig{smearingh1}.
The Higgs field $\Phi(x)$ is substituted by the sum of itself and
of the Higgs fields sitting on the nearest neighbor
sites (in the same timeslice) parallel-transported to $x$. Iterating the
smearing operator $S_1$ we obtain smeared Higgs fields
\bes\label{smhiggsf1}
 \Phi^{(m)}_1(x) & = & S_1^m\,\Phi(x) \, ,
\ees
where $m=0,1,2,...$ denotes the number of smearing iterations and 
is called the smearing level ($m=0$ corresponds to the fundamental
Higgs field). We measured a matrix correlation function with a
basis of smeared Higgs fields corresponding to smearing
levels 0,1 and 2 of $S_1$. The result for the ground state extracted
according to \eq{spectrum} is shown in \fig{meson_comp}. 
We were not able to reach a plateau for the ratio 
$\ln(\lambda_{\alpha}(t-a)/\lambda_{\alpha}(t))$ within the
range of $t$ considered (up to 8 in lattice unit).
%%%%%%%%%%%%%%%%%%%%%%%%%%%%%FIGURE%%%%%%%%%%%%%%%%%%%%%%%%%%%%%%%%%%%
\begin{figure}[tb]
\hspace{0cm}
\vspace{-1.0cm}
\centerline{\epsfig{file=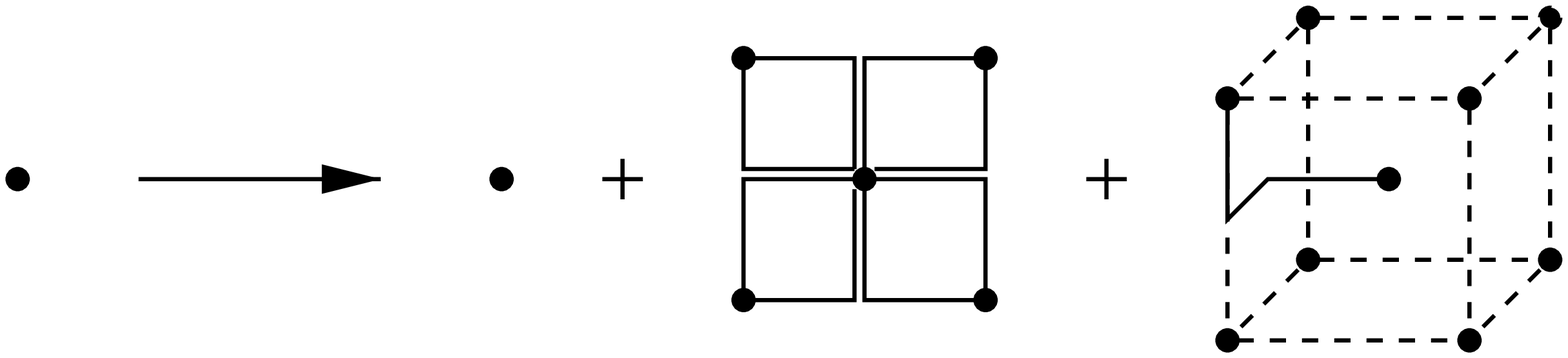,width=12cm}}
\vspace{0.5cm}
\caption{Here, we show the smearing procedure $S_2$ for the Higgs field
  (circles), defined in \eq{smearophiggs2}. 
  The full lines represent the link connections.
\label{smearingh2}}
\end{figure}
%%%%%%%%%%%%%%%%%%%%%%%%%%%%%%%%%%%%%%%%%%%%%%%%%%%%%%%%%%%%%%%%%%%%%%
%%%%%%%%%%%%%%%%%%%%%%%%%%%%%FIGURE%%%%%%%%%%%%%%%%%%%%%%%%%%%%%%%%%%%
\begin{figure}[tb]
\hspace{0cm}
\vspace{-1.0cm}
\centerline{\epsfig{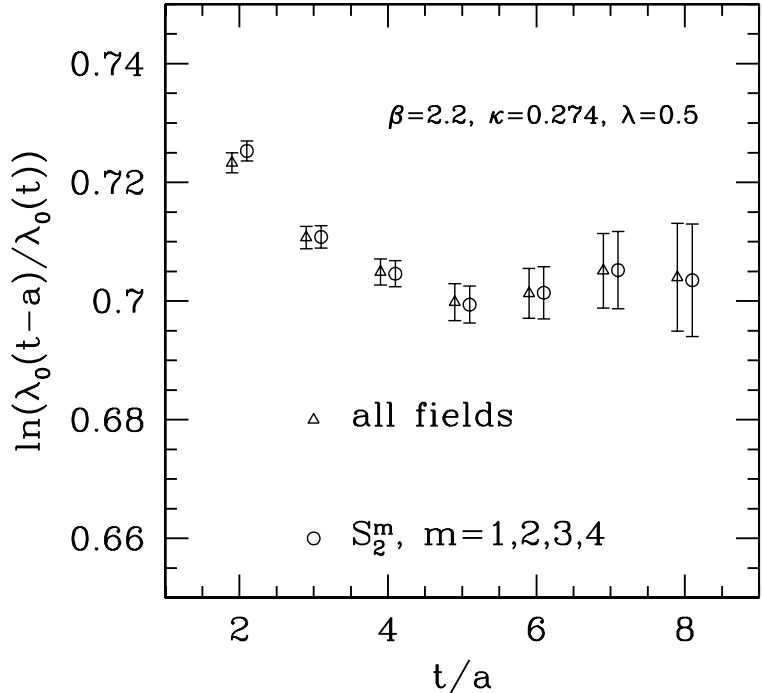}}
\vspace{-0.0cm}
\caption{Here, we compare the extraction of the mass $\mu$ of a
  static-light meson using all the fields \eq{meop1}--\eq{meop2}
  and only the smeared Higgs fields corresponding to smearing levels
  1,2,3,4 of $S_2$ \eq{smearophiggs2}.
\label{meson_all}}
\end{figure}
%%%%%%%%%%%%%%%%%%%%%%%%%%%%%%%%%%%%%%%%%%%%%%%%%%%%%%%%%%%%%%%%%%%%%%

We then investigated a larger basis of meson-type fields, defining in
particular a smearing operator $S_2$ as
\bes\label{smearophiggs2}
  S_2\,\Phi(x) & = &
  \mathcal{P}\{\mathcal{P}\Phi(x) + 
  \mathcal{P}\sum_{|x-y|=\sqrt{2}a \atop x_0=y_0}
  \overline{U}(x,y)\Phi(y) + \nonumber \\ 
  & & \quad\, \mathcal{P}\sum_{|x-y|=\sqrt{3}a \atop x_0=y_0}
  \overline{U}(x,y)\Phi(y)\} \, , 
\ees
where $\mathcal{P}\Phi = \Phi/\sqrt{\Phi^{\dagger}\Phi}$
and  $\overline{U}(x,y)$ represents
the average over the shortest link connections between $y$ and $x$.
This smearing procedure is schematically represented in \fig{smearingh2}.
Contributions from Higgs fields sitting on the corners of the squares and
the cube of side length $a$ around $x$ (lying in the same timeslice as
$x$) are taken into account. Through
iteration of $S_2$ we obtain the smeared Higgs fields
\bes\label{smhiggsf2}
 \Phi^{(m)}_2(x) & = & S_2^m\,\Phi(x) \, ,
\ees
where $m=0,1,2,...$ is the smearing level.
We considered the following
basis of meson-type fields $O^{\rmM}_i(x),\; i=1,2,...,11$:
\bes
 O^{\rmM}_1(x) & = & \mathcal{P}\Phi(x) \, , \label{meop1} \\
 O^{\rmM}_2(x) & = & \mathcal{P}\sum_{|x-y|=a \atop x_0=y_0} 
  U(x,y)\Phi(y) \, , \\
 O^{\rmM}_3(x) & = & \mathcal{P}\sum_{|x-y|=\sqrt{2}a \atop x_0=y_0}
  \overline{U}(x,y)\Phi(y) \, , \\
 O^{\rmM}_4(x) & = & \mathcal{P}\sum_{|x-y|=\sqrt{3}a \atop x_0=y_0}
  \overline{U}(x,y)\Phi(y) \, , \\
 O^{\rmM}_i(x) & = & \Phi^{(i-4)}_2(x)\,, \quad i=5,6,7,8 \, , \\
 O^{\rmM}_9(x) & = & \Phi(x)\times\frac{1}{6}\sum_{k=1}^3\{
  \Phi^{\dagger}(x-a\kh)U(x-a\kh,k)\Phi(x)+ \nonumber \\
 & & \qquad\qquad\qquad \Phi^{\dagger}(x)U(x,k)\Phi(x+a\kh) \}
  \, , \\
 O^{\rmM}_{10}(x) & = & \Phi(x)\times\frac{1}{12}\sum_{1\le k<l\le3}
  \{P_{kl}(x) + P_{kl}(x-a\kh) + \nonumber \\
 & & \qquad\qquad\qquad  P_{kl}(x-a\kh-a\hat{l}) + 
  P_{kl}(x-a\hat{l})\} \, , \\
 O^{\rmM}_{11}(x) & = & \Phi(x)\times(\Phi^{\dagger}(x)\Phi(x)) \, . 
  \label{meop2} 
\ees
The fields $O^{\rmM}_i(x),\;i=1,...,8$ have been already described above.
The field $O^{\rmM}_9(x)$ is constructed from $\Phi(x)$
by multiplying it with a ``cloud'' of gauge invariant links. The field
$O^{\rmM}_{10}(x)$ is $\Phi(x)$ multiplied with the sum of the plaquettes
around $x$. Finally, $O^{\rmM}_{11}(x)$ is $\Phi(x)$
multiplied with its length squared.
In \fig{meson_all}, the result for the extraction of the mass of a
static-light meson using the fields $O^{\rmM}_i(x),\;i=1,...,11$ is shown
(triangles). Note the enlarged scale on the y-axis as compared to
\fig{meson_comp}.
We obtain a nice plateau already at moderately large values of
$t$. The situation remains practically unchanged (also the statistical
errors) if we remove from the basis all fields except the smeared fields
obtained by iterations of the smearing operator $S_2$. This means that
this smearing procedure contains all relevant features for describing
the ground state which could be obtained by using the larger
basis.

When the generalised eigenvalue problem \eq{genev} is solved, the
optimal linear combination of the basis fields $O_i^{\rmM}(x)$
describing the ground state can be expressed in terms of the
components of the vector $v_0$ as $\sum_iv_{0,i}O_i^{\rmM}(x)$.
Therefore, we call $v_0$ the ground state wave function.
Using all the fields \eq{meop1}--\eq{meop2} for constructing 
the matrix correlation function, we observe that
$v_{0,1}$, $v_{0,3}$ and $v_{0,4}$ have
approximately the same value. This is why we defined $S_2$ in
\eq{smearophiggs2} with all coefficients in the sum
equal to 1.

Another interesting fact we can learn from the ground state
wave function $v_0$, is that the field $O^{\rmM}_2$, with nearest neighbor
contributions, has the lowest coefficient $v_{0,2}$. This explains our
original difficulties in extracting the meson ground state. In
\fig{meson_comp}, a direct comparison of the smearing operators $S_1$
and $S_2$, shows clearly that
the contributions from the excited states are much more suppressed when
we use $S_2$.
%%%%%%%%%%%%%%%%%%%%%%%%%%%%%FIGURE%%%%%%%%%%%%%%%%%%%%%%%%%%%%%%%%%%%
\begin{figure}[tb]
\hspace{0cm}
\vspace{-1.0cm}
\centerline{\epsfig{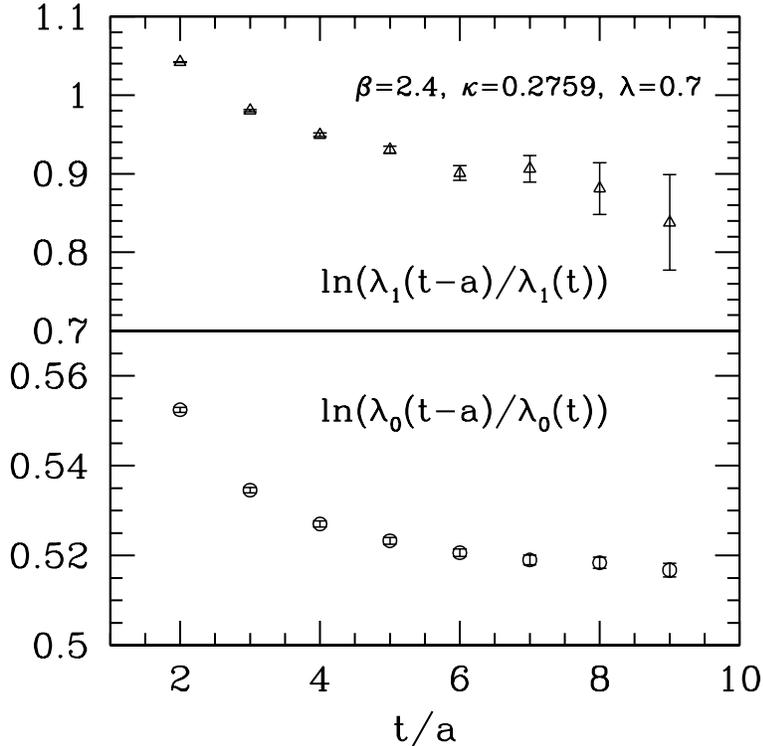}}
\vspace{-0.0cm}
\caption{Here, we show the extraction of the mass of the ground 
  and first excited meson state at $\beta=2.4$. 
  The basis of meson-type fields was obtained using the smearing
  procedure $S_2$ with smearing levels $m=1,3,5,7,10,15$.
  The simulation was performed on a $32^4$ lattice and the statistics
  is of 800 measurements.
\label{meson}}
\end{figure}
%%%%%%%%%%%%%%%%%%%%%%%%%%%%%%%%%%%%%%%%%%%%%%%%%%%%%%%%%%%%%%%%%%%%%%
%%%%%%%%%%%%%%%%%%%%%%%%%%%%%FIGURE%%%%%%%%%%%%%%%%%%%%%%%%%%%%%%%%%%%
\begin{figure}[tb]
\hspace{0cm}
\vspace{-1.0cm}
\centerline{\epsfig{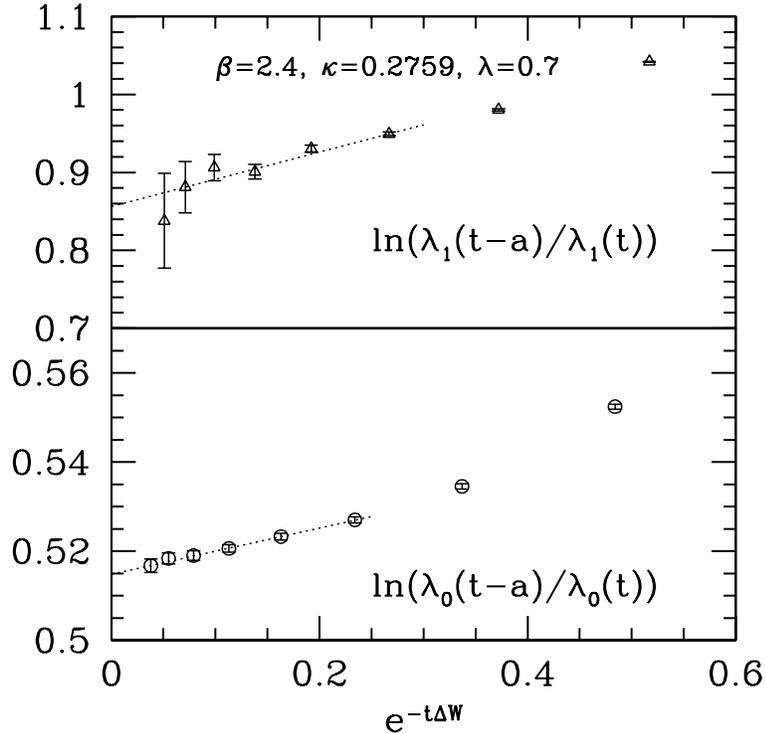}}
\vspace{-0.0cm}
\caption{Here, we show the same data as in \fig{meson} but
  plotted against the correction term $\exp(-t\Delta W)$ in
  \eq{spectrum}. We use $\Delta W=\mu^*-\mu$ for the ground
  state and $\Delta W=\mu^{**}-\mu^*$ for the first excited state, see
  (\ref{muspectrumb24}). \label{f_meson_exp}}
\end{figure}
%%%%%%%%%%%%%%%%%%%%%%%%%%%%%%%%%%%%%%%%%%%%%%%%%%%%%%%%%%%%%%%%%%%%%%

In \fig{meson}, we show the results for the static-light meson spectrum
that we obtained for the parameters
$\beta=2.4,\;\kappa=0.2759,\;\lambda=0.7$ (in the confinement
``phase'') on a $32^4$ lattice. Details
about this simulation will be given in \chapt{stringbreak}. For the
measurement of the matrix correlation function we used a basis with
the six fields
\bes\label{basisme24}
 \Phi_2^{(m)}(x)\,, \quad m=1,3,5,7,10,15 \,,
\ees
obtained by iterating the smearing procedure $S_2$. As we will see in
\chapt{stringbreak}, the lattice spacing at $\beta=2.4$ is reduced by
almost a factor two with respect to the lattice spacing at
$\beta=2.2$. Therefore at $\beta=2.4$ smeared fields with
high smearing levels $m$
are expected to play a more important role than at $\beta=2.2$. This
expectation is confirmed by the simulation.
In order to determine with confidence the static-light meson masses,
we plot in \fig{f_meson_exp} the logarithmic ratios on the right-hand
side of \eq{spectrum} as
functions of the correction terms $\exp(-t\Delta W)$. This enables us to
choose the best time $t$ for reading off the masses from the logarithmic
ratios and to estimate the systematic errors associated with this
choice. For the mass of the ground state,
we must take the largest value $t/a=9$.
For the mass of the first excited state, we
can take $t/a=8$. In both cases, the systematic errors\footnote{
The systematic errors for the masses are estimated from the
difference between the mass read off at the chosen value of $t$ and
the crossing point of the dotted lines in \fig{f_meson_exp}
with the y-axis ($t=\infty$).}
are of the same magnitude as the statistical errors. However, these
errors are small.  The results for the meson spectrum are
\bes\label{muspectrumb24}
 a\mu\;=\;0.517(2)\,, & a\mu^*\;=\;0.88(3)\,, & a\mu^{**}\;=\;1.21(9) \,.
\ees
We note that the convergence of the right-hand
side of \eq{spectrum} is not so ``critical'' in the case of the static
potentials considered in \chapt{stringbreak}.

\chapter{String breaking \label{stringbreak}}

We now introduce a method, which -- as we will demonstrate in
the following sections -- allows to compute 
the static potential, $V_0(r)$, at all relevant distances in the 
theory with matter fields. 
Before explaining the details, we would like to mention
the basic point, which has first been noted by C. Michael
\cite{Michael:1992nc}. Mathematically, the method is based
on the existence of the transfer matrix \cite{Luscher:TM} and 
the fact that it 
can be employed also when external static sources are present 
(see e.g. \cite{Borgs_Seiler}). We have already used this fact in
\chapt{slmesons} for the computation of the static-light meson spectrum.

As we show in detail in \sect{reco},
in the path integral a static source at position $\vec{x}$, together
with an anti-source at position $\vec{x}_r\,=\,\vec{x}+r\hat{k}$,
are represented by straight time-like Wilson lines fixed at these
space-positions. These Wilson lines have to be present in any (matrix) 
correlation
function from which one wants to compute the potential energy of
these charges. The space-like parts of the correlation functions,
which are again Wilson lines  when one considers standard Wilson loops, 
do not determine which intermediate states appear in the spectral 
representation of the correlation functions.
They do, however, influence the weight with which different states
contribute. For these space-like parts,
we therefore use both Wilson lines which will
have large overlap with string-type states and Higgs fields
with a dominant overlap  with meson-type states. Combining them in a 
matrix correlation
function, the correct linear combination which gives the ground state
in the presence of charges can be found systematically
by the variational method described in \sect{variation}. 

\begin{figure}[tb]
\hspace{0cm}
\vspace{-1.0cm}
\centerline{\epsfig{file=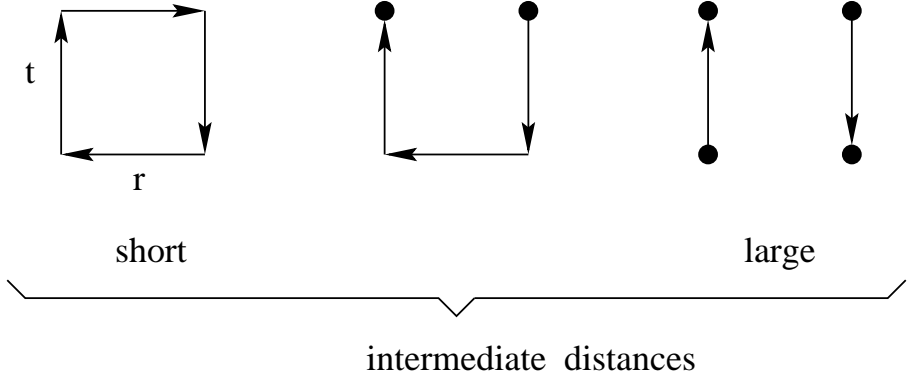,width=12cm}}
\vspace{0.5cm}
\caption{Here, the correlation functions used to determine the static
potential are shown, from left to right: $C_{\rm WW}$, $C_{\rm WM}$ and 
$C_{\rm MM}$. The lines represent the Wilson lines, 
the filled circles the Higgs field.
\label{f_corr}}
\end{figure}

Let us now give precise definitions of the correlation functions,
which are illustrated in \fig{f_corr}.
%%%%%%%%%%%%%%%%%%%%%%%%%%%%%FIGURE%%%%%%%%%%%%%%%%%%%%%%%%%%%%%%%%%%%
%%%%%%%%%%%%%%%%%%%%%%%%%%%%%%%%%%%%%%%%%%%%%%%%%%%%%%%%%%%%%%%%%%%%%%
For small values of $r$ or in the pure
gauge theory, 
the static potential can be efficiently computed by means of Wilson
loops $C_{\rm WW}(r,t)$ defined as
\bes\label{wilsonloop}
  \langle \tr[U(x,x_r)\, 
  U(x_r,x_r+t\hat{0})\,U^{\dagger}(x+t\hat{0},x_r+t\hat{0})\,
  U^{\dagger}(x,x+t\hat{0})] \rangle \,,
\ees
where $x_r\,=\,x+r\hat{k}$ and $U(x,y)$ denotes the product of 
gauge links along the straight line connecting $y$ with $x$. 
For distances significantly larger
than the string breaking distance $\rb$, 
where the relevant states correspond to weakly interacting 
mesons of mass $\mu$, we expect that the potential is close to the  
value $\lim_{r\to \infty}V_0(r) = 2 \mu$ and 
can be extracted from the correlation function $C_{\rm MM}(r,t)$ defined
as
\bes\label{potbigr}
  \langle \Phi^{\dagger}(x+t\hat 0)
  U^{\dagger}(x,x+t\hat 0)\Phi(x) \,\,\, 
  \Phi^{\dagger}(x_r)U(x_r,x_r+t\hat 0)\Phi(x_r+t\hat 0)
  \rangle \, .
\ees
In order to investigate all (and in particular the intermediate)
distances, we introduce a (real, see \eq{ccinvO})
symmetric matrix correlation function $C_{ij}(r,t)$, $i,j \in \{\rm W,M\}$
with $C_{\rm WM}(r,t)$ given by
\bes\label{potallr}
 \langle \Phi^{\dagger}(x+t\hat 0)\,
 U^{\dagger}(x,x+t\hat 0)\,U(x,x_r) 
 \, U(x_r,x_r+t\hat 0)\,\Phi(x_r+t\hat 0) \rangle \, .
\ees

In \App{apptm}, we construct an Hamiltonian formalism for the SU(2)
Higgs model with which we can derive the results described above.
We summarise here the main points of the derivation.
The state vectors forming the Hilbert space of the theory
are represented by wave functionals of
the fundamental field variables $\Phi$ and $U$. 
The Hilbert space is classified in charged sectors 
according to the
gauge transformation property of the state vectors: this
transformation is related by Gauss' law to the
presence of external static charges.
The physical (gauge invariant) states live in the vacuum sector with
no static charges.
The static potential $V_0(r)$ is defined as the energy of the ground
state (normalised to the vacuum energy) in the sector with 
a static quark and a static anti-quark\footnote{
A static quark (anti-quark) is a charge in the (complex conjugate of
the) fundamental representation of the gauge
group. In the case of SU(2) there is no distinction between quark and
anti-quark.}
separated by a distance $r$.
In \sect{transfmat}, we construct a time evolution operator, the
transfer matrix in the temporal gauge. We prove that it is
strictly positive: this allows the definition of the Hamiltonian and
ensures the reality of the energy spectrum.
The energy levels in a charged sector can be extracted by
evaluating powers of the appropriate transfer matrix
operator\footnote{
The transfer matrix operator in the temporal gauge 
is restricted to a specific charged sector by multiplying it with the
projection operator into the sector.}
between states belonging to this sector.
These matrix elements can be shown to
correspond to expectation values in the path integral formalism. The
reconstruction of these expectation values from the operator
expressions is proved in \sect{reco} for the sector with a
pair of static charges. For charged states generated
by the operators given in \eq{corropex} applied to the vacuum,
the results are precisely the expectation values
\eq{wilsonloop}, \eq{potbigr} and \eq{potallr}.

\section{Matrix correlation \label{matrixcorr}}

The states generated by the operators in \eq{corropex} do not have a space
extension. In a physical picture we expect
the string-type states to be a flux tube
\cite{Bali:1995de,Luscher:1981iy,Sommer:1987uz,Wosiek:1987kx, 
      Caselle:1996fh,Pennanen:1997qm}
of gauge fields binding the static charges.
To mimic this situation, we introduce smeared gauge fields.
For the meson-type states, we expect that at large separation of the 
static charges we can describe the system
in terms of two weakly interacting mesons.
Therefore, we use the one-meson wave functions (determined as
described in \sect{mesontype}) to construct two-meson states.
The one-meson states have
a space extension due to the smearing of the Higgs
field. For a high number of smearing iterations, there is
effectively an ``interaction'' between the mesons in the two-meson
state due to the overlap of the smeared Higgs fields.

The states entering in the correlation functions for determining the
static potential are restricted by the transformation property under
gauge transformation and must depend on field variables in the same
timeslice.
For the string-type states we use smeared Wilson lines. They consist
of the product of the smeared space-like links along the straight line
connecting the static charges.
We define the smearing operator $S$ following reference \cite{smear:ape}  
\bes\label{smearopgauge}
  S\,U(x,k) & = & \mathcal{P}\{U(x,k) + 
  \epsilon\sum_{j\neq k=1}^3[
  U(x,j)U(x+a\hat{j},k)U^{\dagger}(x+a\hat{k},j) + 
  \nonumber \\
  & & U^{\dagger}(x-a\hat{j},j)U(x-a\hat{j},k)
  U(x+a\hat{k}-a\hat{j},j)]\} \, ,
\ees
where $\mathcal{P}$ denotes the projection into SU(2).
The four space-like ``staples'' around the link $U(x,k)$ are added to
it with a weight $\epsilon$ which is set to the numerical
value $\epsilon\,=\,1/4$ and the sum is projected back into SU(2).
The smeared space-like links corresponding to a number $m$ of smearing
iterations are given by
\bes\label{smearlink}
  U^{(m)}(x,k) & = & S^m\,U(x,k) \, .
\ees

For the meson-type states we use the following construction.
We determine the spectrum of the static-light mesons, using the
variational method of \sect{variation}, from the
matrix correlation function in \eq{mucorr} constructed with the
field basis $O^{\rmM}_i(x)=\Phi_2^{(n_i)}(x)\;(i=1,2,...,N)$.
The smeared Higgs fields $\Phi_2^{(n_i)}(x)$ are defined in \eq{smhiggsf2}
and the numbers $n_i\;(i=1,2,...,N)$ denote the
smearing levels.
The eigenvectors $v_{\alpha}\in\blackboardrrm^N\;(\alpha=0,1,2,...)$,
obtained by solving
the generalised eigenvalue problem \eq{genev} for large $t$,
are the wave functions
describing approximately (because of the finite basis of fields and 
the finite time $t$) the true eigenstates of the Hamiltonian.
We define the fields
\bes\label{mesoneig}
 \Psi_{\alpha}(x) & = & 
 \sum_{i=1}^N v_{\alpha,i}\Phi_2^{(n_i)}(x)  \quad
 (\alpha=0,1,2,...) \,,
\ees
corresponding to the approximate meson eigenstates. The fields we
choose as basis for the two-meson states are defined as
\bes\label{twomeson}
 \left[\Psi_{\alpha}(x)\right]_a\cdot
 \left[\Psi_{\beta}^*(x_r)\right]_b \, , \quad
 \alpha,\beta\,=\,0,1,2 \,,
\ees
where $x$ and $x_r\,=\,x+r\hat{k}$ are the positions
of the static charges and $a,b=1,2$ are the color indices. 
The values $\alpha=0,1,2$ refer to the ground,
first and second excited one-meson state. The field basis in \eq{twomeson}
contains combinations with $\alpha\neq\beta$ 
which are not symmetric under interchange of the positions $x$ and
$x_r$ of the static charges.
Because we expect the ground two-meson state to be symmetric,
we project into the symmetric 
linear combinations of the fields in
\eq{twomeson} when we analyse the data of the simulations. The ``mixed''
states (for example of one meson in the ground state and one meson in
the first excited state) can be important when looking at the 
asymptotic behavior (in $r$) of excited static potentials
\cite{prniedermayer}.

The matrix correlation function,
from which the spectrum of the Hamiltonian in presence of a pair of
static charges can be determined, is constructed
with the field basis
\bes\label{potbasis}
 [O_i(x,x_r)]_{ab} =  
 \left\{\bea{cl} U^{(m_i)}_{ab}(x,x_r) &
 i=1,2,...,N_{\rmU} \\
 \left[\Psi_{\alpha_i}(x)\right]_a\,\left[\Psi_{\beta_i}^*(x_r)\right]_b &
 i=N_{\rmU}+1,...,N_{\rmU}+9 \ea \right.
\ees
where $U^{(m_i)}(x,x_r)$ is the product of smeared gauge links
(with smearing level $m_i$) along the straight line connecting $x_r$
with $x$ and the pairs of indices
$(\alpha_i=0,1,2;\beta_i=0,1,2)$ label the 9 combinations of
meson-type states. Constructing correlations like \eq{wilsonloop},
\eq{potbigr} and \eq{potallr}, but inserting for the space-like parts
the fields $O_i$ at time $x_0$ and $O_j$ at time $x_0+t\hat{0}$, we
obtain a matrix correlation $C_{ij}(t,r)$.
Its spectral representation is given in \eq{corrAB}.
We denote the energy levels, called static potentials, by
$V_{\alpha}(r),\;\alpha=0,1,2,...$.
In the notation of \App{apptm}, $V_{\alpha}(r)\equiv 
E_{\alpha}^{(\qqb)}(r)-E_0^{(0)}$. The corresponding eigenstates of
the Hamiltonian are denoted by
$|\alpha\rangle\equiv|\alpha,a,b\rangle$ with color indices $a,b$.
Taking the limit of infinite time extension of the lattice
$T\to\infty$ in \eq{corrAB}, we obtain
\bes\label{potcorr}
 C_{ij}(t,r) & = & \sum_{\alpha} 
 \langle 0|\Oop^{\dagger}_j(r)|\alpha\rangle 
 \langle \alpha|\Oop_i(r)|0\rangle\,\rme^{-tV_{\alpha}(r)} \,.
\ees
The operators $\Oop_i(r)$ correspond to the fields \eq{potbasis}.
The trace over the color indices of the states and operators 
is implicit in \eq{potcorr}.
For fixed separation $r$, we extract from $C(t,r)$ the potentials
$V_{\alpha}(r)$ using the variational method described in \sect{variation}.

\section{Results at $\beta=2.4$ \label{resb24}}

Inspired by the investigations in reference \cite{Evertz:1986vp}, we
decided to simulate the SU(2) Higgs model in the confinement ``phase''
near the phase transition line. At fixed $\beta$,
the mass $\mu$ of a static-light
meson decreases with increasing $\kappa$. However, the slope (string
tension) of the approximately linear piece of the static potential 
for small distances remains constant near the phase transition
\cite{Evertz:1986vp}. Thus, string breaking is expected to occur at
smaller separations of the static charges for larger values of
$\kappa$.

The first results that we obtained
\cite{Knechtli:1998gf,Knechtli:1998bd}
were from a simulation at $\beta=2.2,\;\kappa=0.274,\;\lambda=0.5$ on
a $20^4$ lattice. We observed string breaking at a distance 
$r_b/a\,\approx\,5$. We decided then to study the
system with a better lattice resolution at $\beta=2.4$.

The results that we describe in the following
are obtained on a $32^4$ lattice for the parameter set
\bes\label{parameterb24}
 \beta\,=\,2.4\,, \quad \kappa\,=\,0.2759\,, \quad \lambda\,=\,0.7 \, .
\ees
%%%%%%%%%%%%%%%%%%%%%%%%%%%%TABLE%%%%%%%%%%%%%%%%%%%%%%%%%%%%%%%%%%%%%%%
\begin{table}
 \centerline{
 \begin{tabular}{|l|l|} \hline
  fields & smearing levels $m$ \\ \hline\hline
  $S^m\,U$ (see \eq{smearopgauge}) & 7,10,15 \\ \hline
  $S_2^m\,\Phi$ (see \eq{smearophiggs2}) & 1,3,5,7,10,15 \\ \hline
 \end{tabular}}
\caption{Here, we list the smearing levels for the gauge and 
  Higgs fields used in the simulation with parameters
  $\beta=2.4,\;\kappa=0.2759,\;\lambda=0.7$. \label{t_smear}}
\end{table}
%%%%%%%%%%%%%%%%%%%%%%%%%%%%%%%%%%%%%%%%%%%%%%%%%%%%%%%%%%%%%%%%%%%%%%%%
The field basis is constructed according to
\eq{potbasis} from smeared gauge ($N_{\rmU}=3$)
and Higgs fields, whose smearing
parameters are summarised in table \tab{t_smear}. 
The parameters for the simulation were fixed after some trial runs.

The simulation was performed on a parallel computer CRAY T3E.
The $32^4$ lattice is partitioned in the $xy$-plane on $8\times8$
processors. We started the simulation from
thermalised field configurations.
The matrix correlation function $C_{ij}(t,r)\;(i,j=1,2,...,12)$ is
measured in the $zt$-plane
starting from each point of the lattice
up to $r_{\rm max}=15a$ and $t_{\rm max}=9a$.
The time-like links are replaced by their one-link integrals
(see \sect{onelink}). At $r=1$
this replacement is only possible for one time-like Wilson line.
The matrix correlation function is measured every 30
iterations of updating (we recall that one iteration updating 
is composed by one heatbath sweep for both gauge and Higgs field, one
over-relaxation sweep for the gauge field and three over-relaxation
sweeps for the Higgs field): this choice was motivated by the values 
of the integrated autocorrelation times given in (\ref{tauestimatesb24}).
The CPU time cost of 30 iterations is 80
seconds per processor, for one measurement of the matrix correlation
function 260 seconds per processor.
We collected a statistic of 800 measurements.
When we analyse the data, we project the matrix correlation function
$C_{ij}(t,r)$ into the symmetric linear combinations of the two-meson
fields, thereby reducing its dimension from 12 to 9.

Autocorrelations in the measurements of the matrix
correlation functions for the static potentials and the static-light meson
spectrum are practically absent: the statistical errors, 
computed by a jackknife analysis (see \sect{jackknife}), remain constant
when we group the measurements in bins of length 1,2 or 4.
%%%%%%%%%%%%%%%%%%%%%%%%%%%%%FIGURE%%%%%%%%%%%%%%%%%%%%%%%%%%%%%%%%%%%
\begin{figure}[tb]
\hspace{0cm}
\vspace{-1.0cm}
\centerline{\epsfig{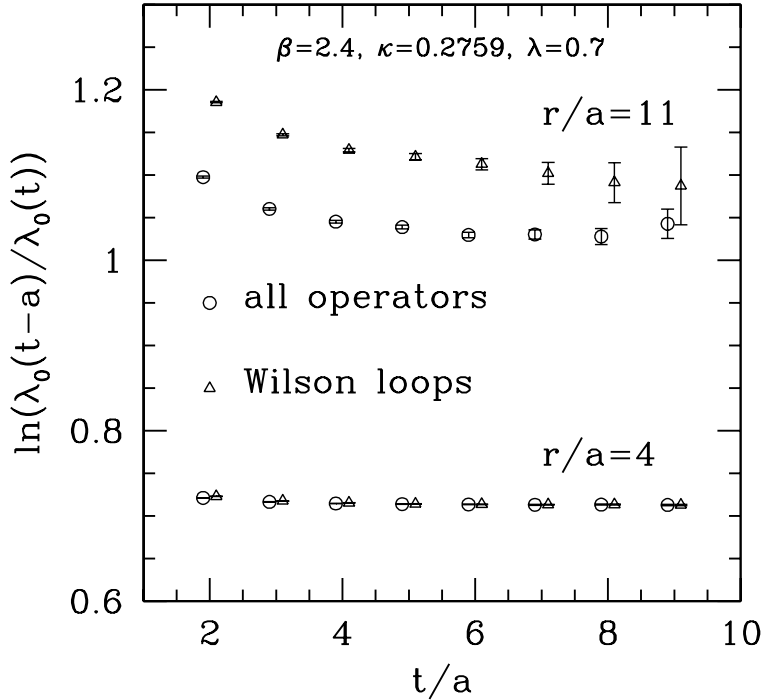}}
\vspace{-0.0cm}
\caption{Here, we compare the static potential computed from
  \eq{potentials}, using the full matrix correlation function (circles)
  and only the sub-block with the (smeared) Wilson loops (triangles).
  Two representative values of $r$ are shown. \label{f_potential_t}}
\end{figure}
%%%%%%%%%%%%%%%%%%%%%%%%%%%%%%%%%%%%%%%%%%%%%%%%%%%%%%%%%%%%%%%%%%%%%%

\subsection{Static potential $\mathbf{V_0}$}

The static potentials $V_{\alpha}(r)\;(\alpha=0,1,2,...)$ are 
extracted from the matrix correlation function $C_{ij}(t,r)$ using the
variational method described in \sect{variation}. We rewrite
\eq{spectrum} as
\bes\label{potentials}
   a V_{\alpha}(r) & = & \ln(\lambda_{\alpha}(t-a,t_0) 
  /\lambda_{\alpha}(t,t_0)) +
  \rmO\left(\rme^{-t\Delta V_{\alpha}(r)}\right) \, ,
\ees
where $\Delta V_{\alpha}(r)=
\min\limits_{\beta\neq\alpha}|V_{\alpha}(r)-V_{\beta}(r)|$ and the
eigenvalues $\lambda_{\alpha}(t,t_0)$ are obtained by solving the
generalised eigenvalue problem \eq{genev} with the matrix correlation
function at fixed $r$. We choose $t_0=0$. 

At all separations of the static charges we compute the static
potential $V_0(r)$ using the full $9\times9$ matrix correlation function.
As an example, the convergence of \eq{potentials} for
$r=4a$ and $r=11a$ is shown in
\fig{f_potential_t} (circles). At all distances $r$ we can read off with
confidence and very good statistical precision (per mille level)
values for the potential at $t=7a$ which agree fully with $t=6a$. 
We compare these results with what we obtain by considering only the
$3\times3$ sub-block of the matrix correlation function corresponding to the 
(smeared) Wilson loops. The
resulting potential estimates (triangles in \fig{f_potential_t}) are
very good at short distances but have large correction terms at long
distances. Without a very careful analysis one might extract a
potential which is too high at large distances, when one uses the
Wilson loops alone.

\subsection{Scale $\mathbf{\rnod}$}

If we want to compute a dimensionful quantity in a lattice gauge theory
simulation, we get a dimensionless number expressing this
quantity in units of the lattice spacing $a$ as a result.
Therefore, we need to fix
one dimensionful quantity to its physical value in order to get the
value of the overall scale $a$ of the simulation. 

In a pure SU($N$) lattice gauge theory, there is 
only one bare parameter, the gauge
coupling constant $g$ (or equivalently $\beta=2N/g^2$). In the
vicinity of the continuum limit,
the relation with the scale $a$ is given by
the perturbative renormalisation group:
\bes
 & & a\,=\,\frac{1}{\Lambda_{\rm L}}\, 
 \rme^{-1/(2b_0g^2)}\, (b_0g^2)^{-b_1/(2b_0^2)}\,
 \{1+\rmO(g^2)+\rmO(a^2)\} \, , \label{rgscale} \\
 & & b_0\,=\,\frac{11N}{48\pi^2}\,, \quad b_1=\frac{34N^2}{3(16\pi^2)^2}
 \, . \label{betagaugecoeff}
\ees
The coefficients $b_0,\,b_1$ are the universal one- and two-loop 
coefficients of the beta function.
As a result of \eq{rgscale}, we
see that the continuum limit is reached when $g\to0$.
The solution of the renormalisation group equation introduces
an integration constant $\Lambda_{\rm L}$ 
(called the lattice $\Lambda$-parameter) 
with the dimension of a mass.
The development of a dimensionful scale in a theory, which
at the classical level does not contain any scale, is called
dimensional transmutation.

The \eq{rgscale} is not useful to set the scale $a$ of a lattice gauge
theory simulation because of the $\rmO(g^2)$ corrections on the
right-hand side. An efficient and
precise way of doing it is described in reference
\cite{Sommer:1993ce} and is based on the force between static quarks.
A system of two static quarks is
approximately realised in nature in the $\bar{c}c$ and $\bar{b}b$
bound states. The spectra of states of the $J/\psi$ and
$\Upsilon$ systems are found to be well described by means of a single 
effective non-relativistic potential \cite{Eichten:1981mw}. There are
a number of successful potential models (references are given in
\cite{Eichten:1981mw}).
In lattice QCD we can compute the static potential $V_0(r)$ and from it
the static force $F(r)\,=\,\rmd V_0(r)/\rmd r$.
The distance $\rnod$ is defined through
\bes\label{scaler0}
 r^2\,F(r)|_{r=\rnod} & = & 1.65 
\ees
and in the phenomenological potentials corresponds to the value
\bes\label{valuer0}
 \rnod & \simeq & 0.5\,\fm \, .
\ees
The scale $a$ of a lattice QCD simulation can be set by computing
the static force and solving \eq{scaler0} to obtain the value of $\rnod$
in lattice units. Although the phenomenological interpretation of this
scale is valid only for QCD, the static force can be computed in any
lattice gauge theory and \eq{scaler0} has a solution provided the
distance $\rb$, at which the gauge string breaks, is larger than
$\rnod$. Due to the clean definition of $\rnod$
and the good statistical precision
with which it can be computed, results in lattice gauge theories are
often quoted in this unit.

In order to solve \eq{scaler0} using lattice measurements of the
static potential, 
we have first to define the static force on the lattice. We follow
\cite{Sommer:1993ce} and define
\bes\label{latforce}
  a\,F(r_I) & = & V_0(r)-V_0(r-a) \, ,
\ees
where the argument $r_I$ is chosen such that in perturbation theory we
have
\bes\label{latforcepert}
  F(r_I) & = & \frac{3}{4}\frac{g^2}{4\pi r_I^2} + O(g^4) \,.
\ees
To lowest order perturbation theory, the lattice artifacts
are exactly eliminated: they remain (probably quantitatively reduced) 
only in the higher $O(g^4)$ terms. The force defined as in
\eq{latforce} is called a tree-level improved observable. 

To solve \eq{scaler0} we need to interpolate the force, which is known
only for discrete values $r/a$.
The general form for our interpolations is
\bes\label{forceinterp}
  r^2\,F(r) & = & f_0\,r^{-2} + f_1 + f_2\,r + f_3\,r^2 \, ,
\ees
which corresponds to the potential
\bes\label{potinterp}
  r\,V(r) & = & -\frac{f_0}{3}r^{-2} - f_1 + f_2\,r\log(r) + f_3\,r^2
  \, .
\ees
The term with coefficient $f_1$ is the Coulomb term and the
coefficient $f_3$ corresponds to the ``string tension'' (linear term in
the potential).
To check for systematic errors we used three interpolations: 
(A) two-point interpolation with $f_0=f_2=0$, (B) three-point
interpolation with $f_2=0$ and (C) three-point interpolation with
$f_0=0$. With the coefficients of the
interpolations determined,
we evaluate the expression for $\rnod/a$ obtained by solving
\eq{scaler0}. If this value lies in the 
interval defined by the interpolation points it will be the solution
of the equation. We want to point out that $r^2\,F(r)$ in the theory
with matter fields is not monotonic (because of string breaking we have
$\lim_{r\to\infty}F(r)=0$): we expect that there are two solutions for
$\rnod$ and the smaller one is to be selected. One more comment about
\eq{latforce}: for $V_0(r)$ we used the values of the ratio 
$\lambda_0(t-a)/\lambda_0(t)$ in \eq{potentials} and repeated the 
computation of the force and $\rnod$ for three values $t/a=6,7,8$. 
The results for $t/a=8$
agree fully with $t/a=7$ and are quoted in the following.

For the parameter set in (\ref{parameterb24}) we obtain from
all three interpolations (A), (B) and (C) the result
\bes\label{r0b24}
 \rnod/a & = & 5.29(6) \, .
\ees
Comparing this number with the values of $\rnod/a$ computed in
quenched QCD \cite{Guagnelli:1998ud}, we see that
$\beta=2.4$ in the SU(2) Higgs model corresponds to $\beta\approx6$ in
QCD.
%%%%%%%%%%%%%%%%%%%%%%%%%%%%%FIGURE%%%%%%%%%%%%%%%%%%%%%%%%%%%%%%%%%%%
\begin{figure}[tb]
\hspace{0cm}
\vspace{-1.0cm}
\centerline{\epsfig{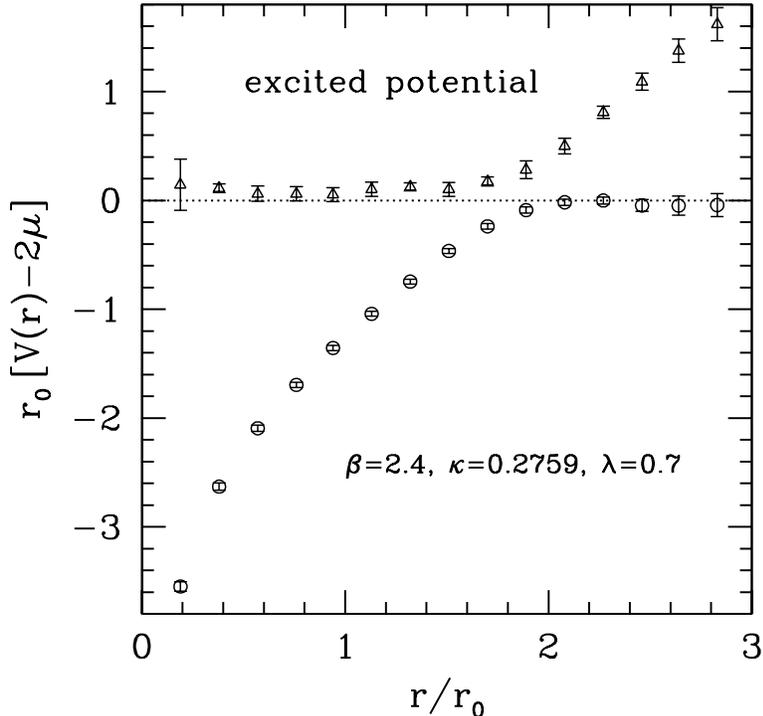}}
\vspace{-0.0cm}
\caption{Here, the renormalised ground state and first excited state static
  potentials in units of $\rnod$ are shown as functions
  of the separation of the
  static charges. String breaking is clearly visible at
  $\rb\approx1.9\rnod$ together with the crossing of the energy
  levels. \label{f_invariant}}
\end{figure}
%%%%%%%%%%%%%%%%%%%%%%%%%%%%%%%%%%%%%%%%%%%%%%%%%%%%%%%%%%%%%%%%%%%%%%

\subsection{Renormalised static potentials $\mathbf{\rnod\,[V(r)-2\mu]}$}

The renormalisation of Wilson
loops $W[C]$ ($C$ is the contour of the loop) in the continuum
pure gauge theory is considered 
in references \cite{Dotsenko:1980wb,Brandt:1981kf}. 
For a smooth contour $C$, it is shown that self-energy graphs diverge
linearly in the cut-off $\Lambda$ with a coefficient proportional to the
length $L(C)$ of the contour. The divergence originate from the
space-time integration region in which the vertices of the graph 
are close together.
These divergencies are present in all
orders of perturbation theory and can be exponentiated
\bes\label{wilsonloopcont}
 W[C] & = & \rme^{-c\Lambda L(C)}\,\times\,W_{\rm ren}[C] \, ,
\ees
where $c$ is a number $\sim1$ and $W_{\rm ren}[C]$ is a finite function
of the renormalised gauge coupling.

In the lattice regularisation, the contributions of
self-energies of Wilson
lines diverge in the continuum like $\frac{1}{a}$. We are interested
in extracting the static potentials $V_{\alpha}(r)$ in \eq{potcorr}.
Therefore, we only have to worry about the divergent contributions
arising from the time-like Wilson lines representing the static
charges. From the considerations in the continuum, we expect that they
exponentiate with a coefficient proportional to $2t$. 
The same divergencies affect the correlation \eq{mucorr} 
for the static-light meson and exponentiate with a coefficient
proportional to $t$. Therefore, we expect that the quantity
\bes\label{renpot}
  a\,[V(r)-2\mu] \, ,
\ees
where $\mu$ is the (unrenormalised) mass of a static-light meson,
is free of divergent self-energy contributions
and allows the definition of renormalised static
potentials. 

In \fig{f_invariant}, we represent the dimensionless
potentials $\rnod\,[V(r)-2\mu]$ for the ground state and the first
excited state. For the static potentials $aV(r)$ we take the values of
the ratios $\ln(\lambda_{\alpha}(t-a)/\lambda_{\alpha}(t))\;
(\alpha=0,1)$ in \eq{potentials} at large $t$.
The computation of the mass $\mu$ is
discussed at the end of \sect{mesontype}.
The ground state potential shows an approximate
linear rise at small distances: around distance
\bes\label{rb}
 \rb & \approx & 1.9\,\rnod
\ees
the potential flattens. The string breaks!
As expected, for large distances the potential approaches
the asymptotic value $2\mu$. The first excited potential comes
very close to the ground state potential around $\rb$ and rises linearly at
larger distances. The scenario of string breaking as a level crossing
phenomenon \cite{Drummond:1998ar} is confirmed beautifully.

For later purposes, we define a dimensionless renormalised quantity
$F_1$ as
\bes\label{F1}
 F_1 & = & \rnod\,[2\mu-V_0(\rnod)] \, .
\ees
The value $V_0(\rnod)$ was computed using the interpolation
\eq{potinterp} with three parameters $f_1$, $f_2$ and $f_3$
($f_0=0$). We find the value $F_1=1.26(2)$.

\subsection{Overlaps}

Overlaps of variationally determined (see \sect{variation})
wave functions $v_0$ 
are a certain measure for the efficiency of a basis of fields
used to construct the matrix correlation functions.
To give a precise definition of the overlap, we 
define the projected correlation function
\bes\label{projcorr}
  \Omega(t) \; = \; v_{0,i}C_{ij}(t)v_{0,j} \; = \; 
  \sum_{\alpha} \omega_{\alpha}
  \rme^{-tV_{\alpha}(r)} \, , 
\ees
with normalisation\footnote{
The property $\Omega(t_0)=\bar{v}_{0,i}\bar{v}_{0,i}=1$
follows from \eq{genevec}. We use $t_0=0$.}
$\Omega(0)=1$ and
$\alpha$ labels the states in the sector of the Hilbert
space with two static charges.
The positive coefficients $\omega_{\alpha}$ can be 
derived from \eq{potcorr}
\bes\label{overlaps}
  \omega_{\alpha} \; = \; 
  |\langle \alpha|\Oop_v(r)|0\rangle|^2 & \mbox{with} &
  \Oop_v(r) \; = \; v_{0,i}\Oop_i(r) \, ,
\ees
and may be interpreted as the overlap of the true
eigenstates of the Hamiltonian $|\alpha\rangle$ 
with the approximate ground state
characterized by $v_0$. The ``overlap'' is an abbreviation
commonly used to denote the ground state overlap, $\omega_0$.
%%%%%%%%%%%%%%%%%%%%%%%%%%%%%FIGURE%%%%%%%%%%%%%%%%%%%%%%%%%%%%%%%%%%%
\begin{figure}[tb]
\hspace{0cm}
\vspace{-1.0cm}
\centerline{\epsfig{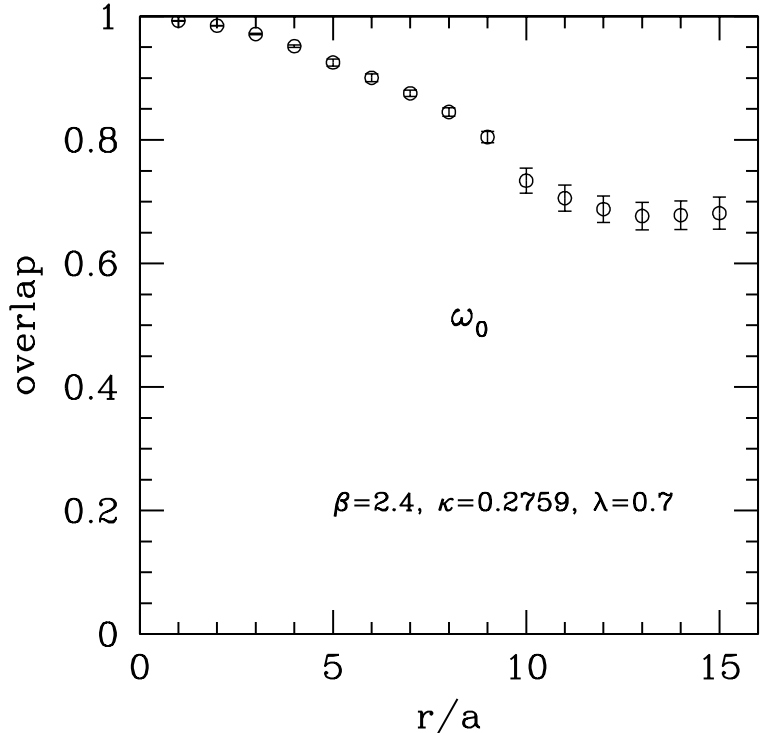}}
\vspace{-0.0cm}
\caption{Here, we show the overlap $\omega_0$ determined from the
  projected correlation function \eq{overlap}. The full matrix
  correlation is used. \label{f_overlap_qq}}
\end{figure}
%%%%%%%%%%%%%%%%%%%%%%%%%%%%%%%%%%%%%%%%%%%%%%%%%%%%%%%%%%%%%%%%%%%%%%

We compute $v_0$ by solving the generalised eigenvalue problem with
$C(t=7a)$.
We determine $\omega_0$ straightforwardly 
from the correlation function $\Omega(t)$ by noting that
\bes\label{overlap}
 \ln\omega_0 & \approx &
 \frac{t+a}{a}\ln\Omega(t)-\frac{t}{a}\ln\Omega(t+a) 
 \quad \mbox{(t large)} \, .
\ees
We extract safe values for $\omega_0$
at $t=7a$, which agree fully with $t=6a$
and are shown in \fig{f_overlap_qq}.
Our basis of fields \eq{potbasis} is big (and good) enough such that  
$\omega_0$ exceeds about 60\% for all distances.
%%%%%%%%%%%%%%%%%%%%%%%%%%%%%FIGURE%%%%%%%%%%%%%%%%%%%%%%%%%%%%%%%%%%%
\begin{figure}[tb]
\hspace{0cm}
\vspace{-1.0cm}
\centerline{\epsfig{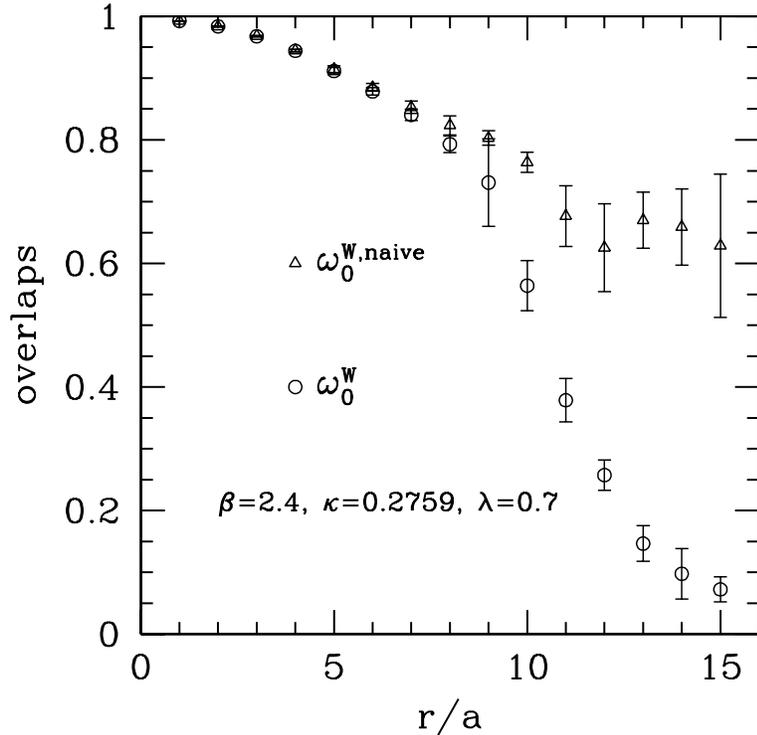}}
\vspace{-0.0cm}
\caption{Here, we show the overlap
  $\omega_0^{\rm W}$ determined from the Wilson
  loops sub-block of the matrix correlation function.
  A ``naive'' way of
  extracting it (triangles), using \eq{overlap},
  gives an erroneous large overlap
  at long distances. A safe estimate (circles) is obtained from \eq{overlapwl}.
  \label{f_overlap_wl}}
\end{figure}
%%%%%%%%%%%%%%%%%%%%%%%%%%%%%%%%%%%%%%%%%%%%%%%%%%%%%%%%%%%%%%%%%%%%%%

It is interesting to consider also
the overlap for the Wilson loops alone, i.e.
we again restrict the matrix correlation function to the
$3\times3$ sub-block associated with (smeared) Wilson loops. 
Let us denote the corresponding projected correlation function by
$\Omega_{\rm W}(t)$ and the overlap by $\omega_0^{\rm W}$. 
The computation of $\omega_0^{\rm W}$ is more difficult and tricky
because it turns out to be very small at large $r$. 
In \fig{f_overlap_wl}, we present the results for two estimates of
$\omega_0^{\rm W}$.
The triangles correspond to the estimate from \eq{overlap}, 
with $\Omega(t)$ replaced by $\Omega_{\rm W}(t)$.
The circles correspond to the more reliable estimate using the
information from the full matrix correlation:
the expression
\bes\label{overlapwl}
  \omega_0^{\rm W} & \simttoinfty & \omega_0
  {\Omega_{\rm W}(t) \over \Omega(t)}
\ees
converges reasonably fast and $\omega_0^{\rm W}$ can be estimated
from the r.h.s. for large $t$.
Using \eq{overlapwl}, we see that
(smeared) Wilson loops alone have an overlap which drops at 
intermediate distances and
they are clearly inadequate to extract the ground state at large $r$.
On the contrary, using \eq{overlap} we get an
overlap above 50\% at large distances: what is estimated here, is
actually the coefficient $\omega_1^{\rm W}$, i.e. the overlap of the
(smeared) Wilson loops with the first excited state (this statement
is supported by direct calculation, see \sect{s_mixing}). Because
$\omega_1^{\rm W}$ turns out to be so large,
one should consider $\Omega_{\rm W}$
at much larger values of $t$ in order to extract the overlap
$\omega_0^{\rm W}$ using \eq{overlap}.
This might explain the problems encountered in QCD
for observing string breaking from the analysis of a correlation function
with Wilson loops only.

\subsection{Level crossing \label{s_mixing}}

Finally, we want to get an insight into the interplay between ``string
states'' and ``two-meson states'' in the string breaking
phenomenon. The results shown in \fig{f_invariant} support the idea of
crossing between the energy levels associated with these
states. We try to quantify this statement.

We consider the diagonal sub-blocks of the matrix correlation
function \eq{potcorr}
corresponding to string-type states (fields $i=1,2,3$ in
\eq{potbasis}) and to meson-type states (fields $i=4,5,...,12$ in
\eq{potbasis}) separately. We
solve the generalised eigenvalue problem \eq{genev} separately
with these restricted matrix correlation functions for fixed $r$
and determine approximate ground state wave functions $v_0^{\rmW}$ for the
string-type states and $v_0^{\rmM}$ for the meson-type states. With
the help of these wave functions we construct a projected matrix
correlation function
\bes\label{projcorrwm}
 \Omega_{kl}(t)\,=\,v_{0,i}^kC_{ij}(t)v_{0,j}^l\,=\,
 \sum_{\alpha} \langle\psi_l|\alpha\rangle
 \langle\alpha|\psi_k\rangle \rme^{-tV_{\alpha}(r)} \quad
 (k,l=\rmW,\rmM)\,,
\ees
where
\bes\label{smstates}
 |\psi_k\rangle\;=\;\left(v_{0,i}^k\Oop_i(r)\right)|0\rangle \,.
\ees
In \eq{smstates}, the string-type ($k=\rmW$) and meson-type ($k=\rmM$) 
states are defined in terms of the operators that create them
when applied to the vacuum $|0\rangle$.
The definitions in \eq{smstates} follow directly from
\eq{projcorrwm} and \eq{potcorr}. 
%%%%%%%%%%%%%%%%%%%%%%%%%%%%%FIGURE%%%%%%%%%%%%%%%%%%%%%%%%%%%%%%%%%%%
\begin{figure}[tb]
\hspace{0cm}
\vspace{-1.0cm}
\centerline{\epsfig{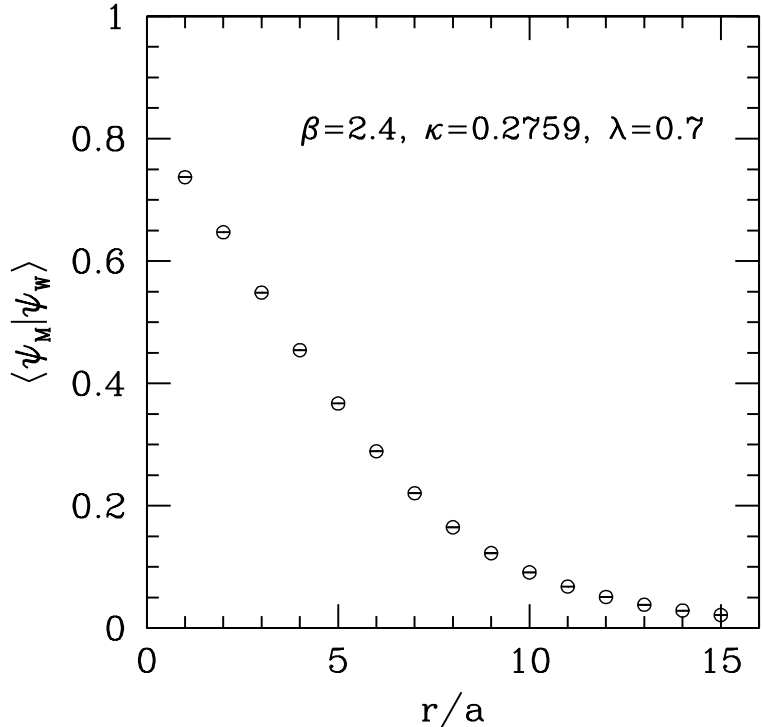}}
\vspace{-0.0cm}
\caption{Here, the scalar product of the
 string-type and meson-type states
 defined in \eq{smstates} is shown as function of the separation $r$
 of the static charges. \label{f_scprod}}
\end{figure}
%%%%%%%%%%%%%%%%%%%%%%%%%%%%%%%%%%%%%%%%%%%%%%%%%%%%%%%%%%%%%%%%%%%%%%

Taking $t=0$ in \eq{projcorrwm}, we get
\bes\label{scalarproduct}
 \Omega_{kl}(0) & = & \langle\psi_l|\psi_k\rangle \,.
\ees
The projected matrix correlation in \eq{projcorrwm} at time $t=0$ 
is equivalent to the scalar product of the states defined in
\eq{smstates}. From \eq{genevec}, one convinces himself that the
normalisation of the states
$\langle\psi_k|\psi_k\rangle=1\;(k=\rmW,\rmM)$ is a direct consequence
of the solution of the generalised eigenvalue problem \eq{genev} with
$t_0=0$. The scalar product $\langle\psi_{\rmM}|\psi_{\rmW}\rangle$ is
represented in \fig{f_scprod} as a function of the separation $r$ of
the static charges. We see that string-type and meson-type states are
orthogonal only for large values of $r$.
%%%%%%%%%%%%%%%%%%%%%%%%%%%%%FIGURE%%%%%%%%%%%%%%%%%%%%%%%%%%%%%%%%%%%
\begin{figure}[tb]
\hspace{0cm}
\vspace{-1.0cm}
\centerline{\epsfig{file=plots/overlaps0.epsi,width=10cm}}
\vspace{-0.0cm}
\caption{Here, the overlaps of the string-type (circles) and meson-type
 (triangles) states, defined in \eq{smstates},
 with the ground state of the Hamiltonian are shown
 as functions of the separation $r$ of
 the static charges. \label{f_overlaps0}}
\end{figure}
%%%%%%%%%%%%%%%%%%%%%%%%%%%%%%%%%%%%%%%%%%%%%%%%%%%%%%%%%%%%%%%%%%%%%%
%%%%%%%%%%%%%%%%%%%%%%%%%%%%%FIGURE%%%%%%%%%%%%%%%%%%%%%%%%%%%%%%%%%%%
\begin{figure}[tb]
\hspace{0cm}
\vspace{-1.0cm}
\centerline{\epsfig{file=plots/overlaps1.epsi,width=10cm}}
\vspace{-0.0cm}
\caption{Here, the overlaps of the string-type (circles) and meson-type
 (triangles) states, defined in \eq{smstates},
 with the first excited eigenstate of the Hamiltonian are shown
 as functions of the separation $r$ of
 the static charges. \label{f_overlaps1}}
\end{figure}
%%%%%%%%%%%%%%%%%%%%%%%%%%%%%%%%%%%%%%%%%%%%%%%%%%%%%%%%%%%%%%%%%%%%%%

The coefficients
\bes\label{smoverlaps}
 \omega_k(\alpha) & \equiv & \langle\alpha|\psi_k\rangle \quad
 (k=\rmW,\rmM) \,,
\ees
in the expansion \eq{projcorrwm},
express the overlap of the string-type and meson-type states with the
true eigenstates of the Hamiltonian.
The matrix $\Omega_{kl}(t)$ is real, which means that the coefficients
$\omega_{\rmW}(\alpha)$ and $\omega_{\rmM}(\alpha)$ have the same
complex phase that can be absorbed into a redefinition of the state
$|\alpha\rangle$. Therefore, we assume that the coefficients
$\omega_k(\alpha)$ are real. Moreover, we can choose the
sign conventions $\omega_{\rmW}(0)>0$ and
$\omega_{\rmW}(1)>0$.
We truncate the sum in \eq{projcorrwm}
after $\alpha=1$ and consider the diagonal matrix elements
$\Omega_{kk}(t)$ for two fixed times $t=t_1$ and $t=t_2$: inserting the
known values for $V_0(r)$ and $V_1(r)$, we get a linear system of equations
for $\omega_k^2(0)$ and $\omega_k^2(1)$. The solutions read
\bes
 \omega_k^2(0) & = & \frac{\Omega_{kk}(t_1)\rme^{t_1V_1(r)} -
   \Omega_{kk}(t_2)\rme^{t_2V_1(r)}} {\rme^{t_1\Delta V(r)} -
   \rme^{t_2\Delta V(r)}} \quad (k=\rmW,\rmM)\,, \label{ovwm0} \\
 \omega_k^2(1) & = & \rme^{t_1V_1(r)}
   [\Omega_{kk}(t_1)-\omega_k^2(0)\rme^{-t_1V_0(r)}] \quad
   (k=\rmW,\rmM)\,, \label{ovwm1}
\ees
where $\Delta V(r)\equiv V_1(r)-V_0(r)$.
The sign of the coefficients
$\omega_{\rmM}(0)$ and $\omega_{\rmM}(1)$ is not fixed yet. 
From the solutions \eq{ovwm0} and \eq{ovwm1}, we
can compute the off-diagonal matrix elements $\Omega_{\rmW\rmM}(t_1)$
and $\Omega_{\rmW\rmM}(t_2)$ using \eq{projcorrwm} for the four
different sign combinations. Comparing with the
values that we get from the simulation, we can establish the right
sign combination. We find that for all $r$,
$\omega_{\rmM}(0)>0$ and $\omega_{\rmM}(1)<0$ (in our sign convention).
The overlaps \eq{ovwm0} and \eq{ovwm1} of the
string-type (circles) and meson-type (triangles) states with the
ground state of the Hamiltonian are shown in \fig{f_overlaps0} and
with the first excited eigenstate of the Hamiltonian in
\fig{f_overlaps1}. The results correspond to the choice
$t_1/a=5$ and $t_2/a=7$.
Other choices of $t_1$ and $t_2$ give
results which are compatible within the statistical errors.
String-type states have a large overlap at short distances
with the ground state and at large distances
with the first excited state.
Meson-type states have a large overlap at short
distances with the first excited state and at large distances
with the ground state. In addition, we observe that the overlap
of the meson-type states with the ground state is also large at very short
distances. The explanation for this fact
is found by looking at \fig{f_scprod}, which clearly shows that
string-type and meson-type states have an
overlap with each other at short distances.
In the string breaking region around $r/a=9-10$, the
overlaps of the string-type and meson-type states have similar
magnitude, both when the ground state or the first excited state is
considered. This fact is reflected in the crossing
of the energy levels \fig{f_invariant}.
Here, we would like to point out that the overlaps represented in
\fig{f_overlaps0} and \fig{f_overlaps1} are {\em not} quantities which
have a continuum limit. They are specific to the $\beta$-value and the
other parameters (e.g. of the smearing) that we consider.

\chapter{Scaling \label{contlim}}

\begin{figure}[tb]
\hspace{0cm}
\vspace{-1.0cm}
\centerline{\epsfig{file=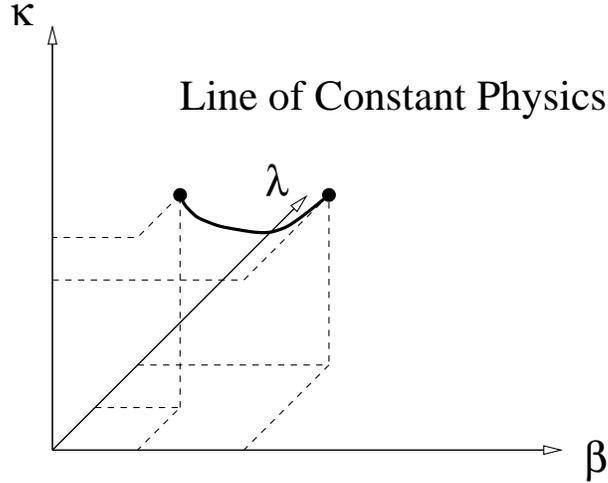,width=8cm}}
\vspace{0.5cm}
\caption{Here, a line of constant physics (LCP) in the SU(2) Higgs
  model is shown.
  The lattice spacing is changed along the LCP keeping two
  dimensionless physical quantities $F_1$ and $F_2$ at fixed values.
\label{f_lcp}}
\end{figure}

The lattice spacing $a$ is mainly determined by the choice of the
parameter $\beta=4/g^2$, where $g$ is the gauge coupling.
Dimensionless physical quantities\footnote{
We mean ratios of physical quantities with the same mass dimension.} 
$F$ such as $F_1=\rnod\,[2\mu-V_0(\rnod)]$ 
contain a dependence on the value of the lattice spacing  
which vanishes in the continuum limit. For a scalar theory we can write
\cite{Symanzik:1983dc,Symanzik:1983gh}
\bes\label{scaling}
 F(a) & = & F(0)+\rmO\left((a/\rnod)^2\right) \, ,
\ees
where $\rmO((a/\rnod)^2)$ summarises terms that contain at least two
powers\footnote{
The conjecture that $\rmO(a/\rnod)$ corrections in \eq{scaling}
are absent for any observable in a scalar theory, is well accepted.}
of $a/\rnod$ and may be modified by logarithmic corrections.
When the corrections in \eq{scaling} (called lattice artifacts) 
become so small that $F$ is almost
independent of $a$, we call this fact scaling.
In order to investigate the presence of lattice artifacts in a
dimensionless physical quantity, it is not sufficient to change the 
value of $\beta$
in a model with three bare parameters such as the SU(2) Higgs model. The
physics of the model is influenced by the choice of all three
parameters, for each observable differently. For an estimate of
the correction terms in \eq{scaling} we must vary the lattice
spacing $a$ (by changing $\beta$) and tune the bare parameters
$\kappa$ and $\lambda$ to keep two dimensionless physical quantities 
$F_1$ and $F_2$ constant. 
This procedure corresponds to the renormalisation of $\kappa$ and
$\lambda$ and defines in the parameter space a so called Line of
Constant Physics (LCP) characterised by
\bes\label{lcp}
  F_i(\beta,\kappa,\lambda) & = & \mbox{constant}, \quad i=1,2 \,.
\ees
The situation is schematically represented in \fig{f_lcp}.
Other dimensionless physical quantities $F_3,\,F_4,\,\cdot\cdot\cdot$
will, in principle, show correction terms as in \eq{scaling}.
This will give a measure for the scaling behavior of the theory in the
investigated range of the lattice spacing.
%%%%%%%%%%%%%%%%%%%%%%%%%%%%%FIGURE%%%%%%%%%%%%%%%%%%%%%%%%%%%%%%%%%%%
%%%%%%%%%%%%%%%%%%%%%%%%%%%%%%%%%%%%%%%%%%%%%%%%%%%%%%%%%%%%%%%%%%%%%%

In \chapt{slmesons} and \chapt{stringbreak} we presented results
for the static-light meson spectrum and the static potentials obtained
for the parameter set
$\beta=2.4,\;\kappa=0.2759,\;\lambda=0.7$ on a $32^4$ lattice.
We want to reproduce the physical situation on a coarser lattice
corresponding to the choice $\beta=2.2$.
A sensible choice for the dimensionless physical quantities to keep
constant, would be to
find $F_1$ strongly dependent on $\kappa$ and $F_2$ strongly dependent on
$\lambda$. We have
already found the right quantity $F_1=\rnod\,[2\mu-V_0(\rnod)]$ \eq{F1}: it
mainly depends on the value of the mass of the dynamical Higgs field 
which is in turn determined by the choice of $\kappa$.
We now have to face the problem of finding $F_2$.

Physics shows a dependence on the physical size of
the system and on the boundary conditions. We regularise the SU(2)
Higgs model on a periodic lattice,
which corresponds to a torus in the continuum and impose periodic
boundary conditions on the fields.
The physical lattice size $L/\rnod$ is part of the
definition of a dimensionless physical quantity $F$. The variation of
the lattice spacing in \eq{scaling} must be accompanied by a change in
the number of lattice points to keep the physical size of the torus
constant. The dependence of a physical quantity on the size of the
torus is called finite size effect.

It is well accepted -- supported
by the weak gauge coupling expansion \cite{Montvay:1987dw}
and by early numerical
simulations \cite{Langguth:1987vf,Hasenfratz:1988uc} --
that the SU(2) Higgs model is a
trivial theory \cite{MontMuen},
which means that the continuum {\em limit} is a free field
theory. Nevertheless, the model can exhibit in a large range of values of the
lattice spacing scaling properties of a non-trivial almost continuum
theory. The interpretation of such a behavior is that the SU(2) Higgs
model in this range describes an effective low-energy field
theory and lattice results are perfectly relevant for a continuum
Higgs model.

\section{Matching of $\kappa$ \label{kappamatch}}

Our choice of the first physical quantity to use for the matching of
the bare parameters along the LCPs is $F_1=\rnod\,[2\mu-V_0(\rnod)]$ \eq{F1}.
The first matching condition reads
\bes\label{matching1}
  F_1 \; = \; \rnod\,[2\mu-V_0(\rnod)] \; = \; F_1^*\equiv1.26 \,,
\ees
where the numerical value is found at $\beta=2.4$ with a statistical
error $\Delta F_1^*=0.02$. Since $F_1$ is
mainly sensitive to the value of the meson mass $\mu$, which in turn
is determined by the mass of the dynamical Higgs field, the quantity
$F_1$ is surely sensitive to the parameter $\kappa$.
For different values of $\lambda$, we match the
parameter $\kappa$ at $\beta=2.2$ by computing $F_1$ on a $20^4$ lattice
for two values $\kappa=\kappa_1$ (giving $F_1^{(1)}$ with statistical
error $\Delta F_1^{(1)}$) and $\kappa=\kappa_2$
(giving $F_1^{(2)}$ with statistical error $\Delta F_1^{(2)}$).
From a linear interpolation
\bes\label{lininterpF1}
 F_1(\kappa)|_{\lambda} & = & a_0(\lambda) + \kappa\cdot
 a_1(\lambda) \,,
\ees
we determine $\kappa^*$ such that $F_1(\kappa^*)=F_1^*$:
\bes\label{kappastar}
 \kappa^* & = & \frac{\kappa_1(F_1^*-F_1^{(2)}) +
 \kappa_2(F_1^{(1)}-F_1^*)}{F_1^{(1)} - F_1^{(2)}} \, .
\ees
Propagating the independent errors $\Delta F_1^{(1)}$, $\Delta F_1^{(2)}$ and
$\Delta F_1^*$, we get the error for the value
$\kappa^*$ obtained from \eq{kappastar}.
%%%%%%%%%%%%%%%%%%%%%%%%%%%%%FIGURE%%%%%%%%%%%%%%%%%%%%%%%%%%%%%%%%%%%
\begin{figure}[tb]
\hspace{0cm}
\vspace{-1.0cm}
\centerline{\epsfig{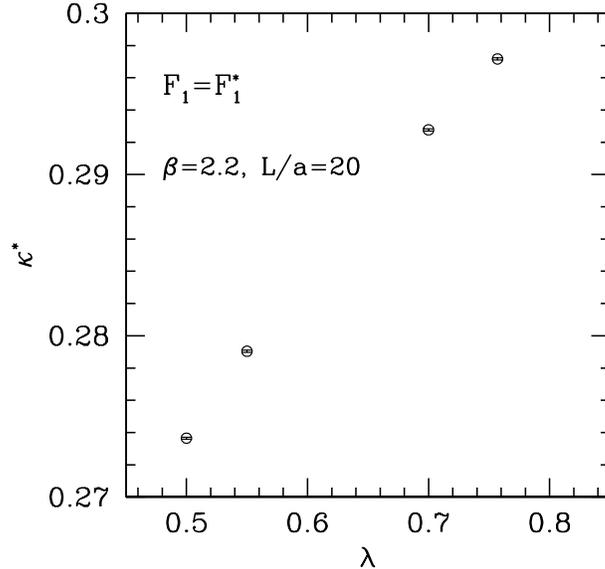}}
\vspace{-0.0cm}
\caption{Here, the matching of $\kappa$ at $\beta=2.2$ on a $20^4$ lattice 
  using the condition \eq{matching1} is shown for different values of
  $\lambda$. The matched value $\kappa^*$ is obtained from
  \eq{kappastar}. \label{f_lcpkl}}
\end{figure}
%%%%%%%%%%%%%%%%%%%%%%%%%%%%%%%%%%%%%%%%%%%%%%%%%%%%%%%%%%%%%%%%%%%%%%
The results for $\lambda$=0.5, 0.55, 0.7, 0.757 are shown in
\fig{f_lcpkl}.
As a by-product of these simulations, we obtain
the typical value of $\rnod/a$ at $\beta=2.2$, for $F_1$ values 
near $F_1^*$:
\bes\label{r0b22}
 \rnod/a & \approx & 2.8 \,.
\ees
Using \eq{r0b24}, we can derive
the change in the lattice spacing between $\beta=2.2$ and $\beta=2.4$:
\bes\label{aratio}
 {a(\beta=2.2) \over a(\beta=2.4)} & \approx & 1.9 \, .
\ees
The choice of a $20^4$ lattice at $\beta=2.2$ and a $32^4$ lattice at
$\beta=2.4$ corresponds only approximately to constant physical volume:
we have $L/\rnod\approx7$ at $\beta=2.2$ and $L/\rnod\approx6$
at $\beta=2.4$. But these lattices can be considered relatively large
for the extraction of the meson spectrum and the static potential: the
relevant quantity, as we discussed in \sect{variation}, is the product
$Lm_{\rmH}$ which must be much larger than 1. The Higgs mass
$m_{\rmH}$, that we define as the mass gap in the zero charge sector
of the Hilbert space, can be estimated in our simulations to be
$\rnod m_{\rmH}\approx1-1.5$, see \fig{f_c5mhl}. Therefore, we do not expect
relevant finite size effects.

\section{Matching of $\lambda$ \label{lambdamatch}}

In this section, we propose dimensionless physical quantities and study
their $\lambda$ dependence. What we are looking for, are
gauge invariant quantities with a well-defined continuum limit and
sensitive to a variation of the parameter $\lambda$, once
the parameter $\kappa$ is matched as described in \sect{kappamatch}.

In the region of small bare couplings, one can use bare lattice
perturbation theory to determine the LCPs. By
using an expansion for small gauge coupling one can derive, at the
1-loop level, the following equations \cite{Montvay:1987dw}
\bes\label{pertlcp}
 \frac{\rmd\lambda_0(\tau)}{\rmd\tau} & = & \frac{1}{16\pi^2}\left[
 96\lambda_0^2+\frac{9}{32}g^4-9\lambda_0g^2+\cdot\cdot\cdot \right] \,,
 \nonumber \\
 \frac{\rmd g^2(\tau)}{\rmd\tau} & = & \frac{1}{16\pi^2}\left[
 -\frac{43}{3}g^4+\cdot\cdot\cdot \right] \,,
\ees
where $\tau\equiv\ln(am_{\rmR})^{-1}$ is the logarithm of the inverse
of a renormalised mass\footnote{
The terms on the right-hand sides of \eq{pertlcp} are universal, any
renormalised mass can be taken in the definition of
$\tau$. The non-universal corrections are proportional to $a/\rnod$.}
in lattice units, $g^2=4/\beta$ and
$\lambda_0=\lambda/(4\kappa^2)$. The dots stand for terms of higher
order $\rmO(\lambda_0^3,\lambda_0^2g^2,\lambda_0g^4,g^6)$.

The change $\tau_2-\tau_1=\ln(a_1/a_2)$ in \eq{pertlcp} can be
determined using \eq{aratio}. If we then use \eq{pertlcp},
with $\beta_1=2.2$ and $\lambda_1=0.5$ as initial conditions,
to compute the change in $\lambda$ at $\beta_2=2.4$, we obtain\footnote{
The value of $\kappa$ is taken to keep $F_1=F_1^*$.}
the value $\lambda_2=0.8$,
which is not far from the value $\lambda=0.7$ that we used for the
simulations at $\beta=2.4$. The use of \eq{pertlcp} is {\em not}
justified in this case, because the bare couplings are not
small enough. Nevertheless, we take this crude estimate to start our
investigations and we consider at $\beta=2.2$ the range
$0.5\le\lambda<0.8$.

\subsection{Cumulants \label{s_cumulants}}

We consider the gauge invariant Higgs field $S(x)=\Phid(x)\Phi(x)$ and
the construction of a renormalised coupling from its connected
$p$-point (or Green) functions.
There are some subtleties due to the fact that
$S(x)$ is a composite field.

A composite field (or field operator) is a product of fields (field
operators) at the same space-time point. The renormalisation of
composite fields is a complicated issue and we refer to textbooks,
e.g. \cite{Collins,Muta}, for a detailed discussion. In general, a
renormalised composite field $A_{\rmR}(x)$ is expressed in terms
of unrenormalised fields $B(x)$ by
\bes\label{rencomp}
 A_{\rmR}(x) & = & \sum_{B} Z_{AB} B(x) \,,
\ees
where $Z_{AB}$ are the renormalisation constants which depends on the
bare couplings and on the cut-off, e.g. the lattice spacing $a$ in the
lattice regularisation. The fields $B$,
with which $A$ can mix under renormalisation, have canonical dimension
equal to or lower than that of $A$ and the same quantum numbers of $A$.
For example,
the gauge invariant Higgs field $S(x)=\Phid(x)\Phi(x)$
gets renormalised like
\bes\label{renhiggs}
 S_{\rmR}(x) & = & Z_1 + Z_S S(x) \,.
\ees
The canonical dimension of the field $S(x)=\Phid(x)\Phi(x)$
is two.\footnote{
On the lattice, the field $\Phi(x)$ is made dimensionless by absorbing
a factor $a/\sqrt{\kappa}$. With mass dimension of a field, we always
mean the canonical dimension in the continuum.}
Besides $1$ and $S(x)$ itself,
there is no other gauge invariant
scalar field of dimension two or less, with which $S(x)$
can mix under renormalisation.

The connected $p$-point (or Green) functions of $S(x)$
are defined by the path integral expectation values (see later)
\bes\label{connphiggs}
 G_c(x_1,...,x_p) & = & \langle S(x_1) \cdot\cdot\cdot S(x_p)
 \rangle_c \,.
\ees
The renormalised connected $p$-point functions of $S(x)$,
which we denote by $G_{c,\rmR}(x_1,...,x_p)$, are
related to \eq{connphiggs} and \eq{renhiggs} by
\bes\label{renconnphiggs}
 G_{c,\rmR}(x_1,...,x_p) \;=\; \langle S_{\rmR}(x_1) \cdot\cdot\cdot 
 S_{\rmR}(x_p) \rangle_c \;=\; Z_S^p\,G_c(x_1,...,x_p) \,,
\ees
The additive renormalisation $Z_1$ of the field $S(x)$ \eq{renhiggs}
cancels in the definition of the connected Green functions.
The multiplicative renormalisation factor $Z_S$
in \eq{renhiggs} is chosen so that
the renormalised connected $p$-point functions in \eq{renconnphiggs}
have a well-defined
continuum limit, provided all points $x_1,...,x_p$ are kept at
non-zero distance from one another \cite{Luscher:1998pe}.
The question is, what happens in the continuum limit, if some of
the points $x_1,...,x_p$ coincide?
This question can be addressed in the continuum with
the help of the operator-product expansion of Wilson
\cite{Wilson:1969ey}, according to which
\bes\label{opes2}
 \langle S_{\rmR}(x)S_{\rmR}(y)\rangle_c
 & _{\mbox{$\stackrel{\displaystyle\sim}
 {\scriptstyle y\to x}$}} & \frac{1}{|x-y|^4} \,.
\ees
The power of the divergence on the right-hand side of \eq{opes2}
corresponds to the naive dimensional counting.
One can easily convince himself that this is true for free fields: for
interacting fields, logarithmic correction factors may appear.
We can generalise \eq{opes2} to the form
\bes\label{opesk}
 \langle S_{\rmR}(x_1)\cdot\cdot\cdot S_{\rmR}(x_k)\rangle_c & 
 _{\mbox{$\stackrel{\displaystyle\sim}
 {\scriptstyle x_i\to y}$}} & \frac{1}{|x-y|^{2k}}
 \quad (x\to y)\,.
\ees
We consider now the zero-momentum field
\bes\label{contzerohiggs}
 s_{\rmR} & = & \frac{1}{V}\int\rmd^4x\,S_{\rmR}(x) \,,
\ees
where $V$ is the physical volume of the torus. 
The connected $p$-point functions $\langle
s_{\rmR}^p\rangle_c\;(p=2,3,4,...)$ can be written as
\bes\label{ppoints}
 \langle s_{\rmR}^p\rangle_c & = &
 \frac{1}{V^p} \int\rmd^4x_1\cdot\cdot\cdot\rmd^4x_p\,
 \langle S_{\rmR}(x_1)\cdot\cdot\cdot S_{\rmR}(x_p)\rangle_c
 \nonumber \\
 & = & \frac{1}{V^{p-1}} 
 \int\rmd^4y_1\cdot\cdot\cdot\rmd^4y_{p-1}\,
 \langle S_{\rmR}(y_1)\cdot\cdot\cdot S_{\rmR}(y_{p-1})
 S_{\rmR}(0)\rangle_c \,,
\ees
where in the second line we used the translation invariance property
of the connected Green functions.
In naive dimensional counting,
when $y_i\to0\;(i=1,...,p-1)$, from the $(p-1)$ integrations we get
$4(p-1)$ powers of $y$ whereas the integrand diverges 
in this limit like $|y|^{-2p}\,\;(y\to0)$, as can be seen from \eq{opesk}.
Therefore, the $p$-point functions \eq{ppoints} remain finite if 
\bes\label{pfinite}
 2p\;<\;4(p-1) & \Leftrightarrow & p\;>\;2 \,.
\ees
Only the 2-point function $\langle s_{\rmR}^2\rangle_c$ is naively
logarithmic divergent. This problem can be cured by defining
a modified connected 2-point function
\bes\label{2pointsreg}
 \langle\bar{s}^{(2)}_{\rmR}\rangle_c & = & 
 \frac{1}{V}\int\rmd^4y\,
 \langle S_{\rmR}(y)S_{\rmR}(0)\rangle_c\,
 \sin\left(\frac{y_0}{T}\pi\right)
\ees
where $T$ is the time extension of the torus: throughout
our work we use $T\equiv L\equiv V^{1/4}$.
The sine function avoids contributions coming from $y=0$.
In order to be sure that {\em all} divergencies associated with coincident
arguments in \eq{ppoints} are regularised for $p>2$,
it remains to consider the case of two coincident arguments,
e.g. $y_1\to y_2$. In contrast with $p=2$, there are still two
integrations, over $y_1$ and $y_2$, that regulate the divergent
behavior of the integrand, given in this limit by \eq{opes2}.
We are now able to give a definition of renormalised couplings in
the continuum
\bes\label{contcumulants}
 c_p & = & \frac{\langle s^p \rangle_c}{\left[\langle \bar{s}^{(2)}
   \rangle_c\right]^{p/2}} \quad (p=3,4,5,...) \,.
\ees
The multiplicative renormalisation of the field $S(x)$
cancels in the ratio and we can
then use the bare fields in \eq{contcumulants}.
We call the couplings defined by \eq{contcumulants} cumulants. 
We derive in the following expressions for the connected $p$-point 
functions of the zero-momentum field $s$.

In the path integral formalism
the $p$-point functions $G(x_1,...,x_p)=\langle S(x_1)\cdot\cdot\cdot
S(x_p) \rangle$ are
constructed \cite{ChengLi} from the generating functional
\bes\label{genfunc}
 Z[J] & = & \langle \exp\{\int\rmd^4x\, J(x)S(x)\} \rangle
\ees
by functional differentiation with respect to the sources $J(x)$:
\bes\label{ppoint}
 G(x_1,...,x_p) & = & \left. \frac{\delta^pZ[J]}{\delta
   J(x_1)\cdot\cdot\cdot\delta J(x_p)} \right|_{J=0} \, .
\ees
Combining \eq{genfunc} and \eq{ppoint} we can write
\bes
 Z[J] & = & \sum_{p=0}^\infty \frac{1}{n!}\int
 \rmd^4x_1\,...\,\rmd^4x_p\, J(x_1)\cdot\cdot\cdot J(x_p)\,
 G(x_1,...,x_p) \, .
\ees
The connected $p$-point functions $G_c(x_1,...,x_p)= 
\langle S(x_1)\cdot\cdot\cdot S(x_p)\rangle_c$
are constructed from the generating functional
\bes\label{genfuncc}
 W[J] & = & \ln\,Z[J]
\ees
in the same manner
\bes\label{ppointc}
 G_c(x_1,...,x_p) & = & \left. \frac{\delta^pW[J]}{\delta
   J(x_1)\cdot\cdot\cdot\delta J(x_p)} \right|_{J=0} \, .
\ees
Combining \eq{genfuncc} and \eq{ppointc} we can write
\bes
 W[J] & = & \sum_{p=0}^\infty \frac{1}{n!}\int
 \rmd^4x_1\,...\,\rmd^4x_p\, J(x_1)\cdot\cdot\cdot J(x_p)\,
 G_c(x_1,...,x_p) \, .
\ees
Expanding the right-hand side of \eq{genfuncc} in ``powers'' of $J$ one
can derive general relations between $G_c$ and $G$. From these
relations, we obtain that
the connected $p$-point functions of $s$ can be expressed for
$p=3,4,5$ by
\bes
  \langle s^3\rangle_c & = & \langle s^3\rangle - 3\langle
  s\rangle\langle s^2\rangle + 2\langle s \rangle^3 \, , \label{cum3}\\
  \langle s^4\rangle_c & = & \langle s^4\rangle - 4\langle
  s^3\rangle\langle s\rangle - 3\langle s^2\rangle^2 + 12\langle
  s^2\rangle\langle s\rangle^2 - 6\langle s \rangle^4 \, , \label{cum4}\\
  \langle s^5\rangle_c & = & \langle s^5\rangle - 5\langle
  s^4\rangle\langle s\rangle - 10 \langle s^3\rangle\langle s^2\rangle
  + 20\langle s^3\rangle\langle s\rangle^2 \nonumber \\ 
  & & + 30\langle s^2\rangle^2\langle s\rangle
  - 60 \langle s^2\rangle\langle s\rangle^3 + 24 \langle s\rangle^5 \,
  . \label{cum5}
\ees

Finally, we give a discretised version of the
definition of the cumulants that we use on the lattice. The
zero-momentum gauge invariant Higgs field is defined as
\bes\label{zerohiggs}
 s & = & \frac{1}{\Omega}\sum_x S(x) \,,
\ees
where $\Omega$ is the number of lattice points. The cumulants are then
defined as in \eq{contcumulants}
\bes\label{cumulants}
 c_p & = & \frac{\langle s^p \rangle_c}{\left[\langle \bar{s}^{(2)}
   \rangle_c\right]^{p/2}} \quad (p=3,4,5,...) \,,
\ees
where the modified connected 2-point function
$\langle\bar{s}^{(2)}\rangle_c$ is given by
\bes\label{2pointslat}
 \langle\bar{s}^{(2)}\rangle_c & = & 
 \frac{1}{\Omega^2}\sum_{x,y}
 \langle S(x)S(y)\rangle_c\,
 \sin\left(\frac{x_0-y_0}{T}\pi\right) \,,
\ees
with $\langle S(x)S(y)\rangle_c=\langle S(x)S(y)\rangle - \langle
S(x)\rangle\langle S(y)\rangle$. The connected
$p$-point functions $\langle s^p\rangle_c\;(p=3,4,5)$ are computed
according to \eq{cum3}, \eq{cum4} and \eq{cum5} from the expectation
values $\langle s^q\rangle\;(q=1,...,5)$.

\subsubsection{Dependence on the volume}

%%%%%%%%%%%%%%%%%%%%%%%%%%%%%FIGURE%%%%%%%%%%%%%%%%%%%%%%%%%%%%%%%%%%%
\begin{figure}[tb]
\hspace{0cm}
\vspace{-1.0cm}
\centerline{\epsfig{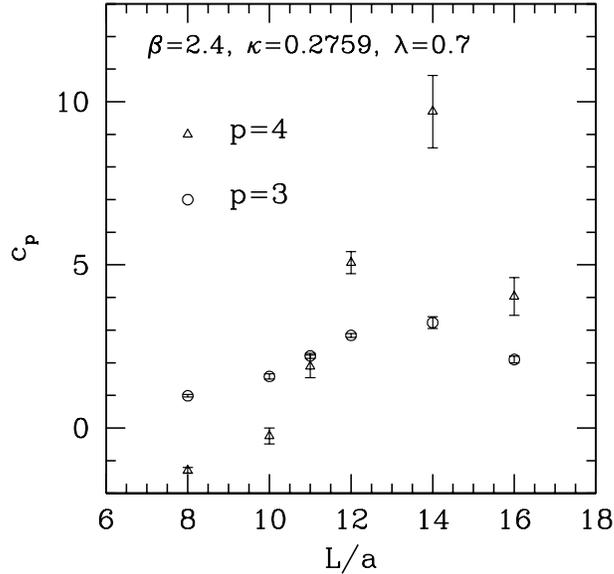}}
\vspace{-0.0cm}
\caption{Here, we show the volume dependence of the cumulants $c_3$ and $c_4$ 
 defined in \eq{cumulants}.
 \label{f_cumvol}}
\end{figure}
%%%%%%%%%%%%%%%%%%%%%%%%%%%%%%%%%%%%%%%%%%%%%%%%%%%%%%%%%%%%%%%%%%%%%%
The cumulants $c_p$ for large values of $p$ are very small:
in the Monte Carlo simulations we were able to obtain a significant signal 
up to $p=5$.
In \fig{f_cumvol}, we consider the volume dependence of $c_3$ and $c_4$
for the parameter set $\beta=2.4,\;\kappa=0.2759,\;\lambda=0.7$. We
observe that the cumulants are strongly varying functions of the lattice size
$L$: they are finite size couplings.
The use of the cumulants for the matching of the parameters along the
LCPs requires therefore a precise matching of the
physical volume.
We decided to compute the cumulants at $\beta=2.2$ on a $6^4$ lattice:
the matching of the physical volume at $\beta=2.4$ using \eq{aratio}
gives a lattice size $L/a\approx11.4$.

\subsubsection{Dependence on the parameter $\mathbf{\lambda}$}

%%%%%%%%%%%%%%%%%%%%%%%%%%%%%FIGURE%%%%%%%%%%%%%%%%%%%%%%%%%%%%%%%%%%%
\begin{figure}[tb]
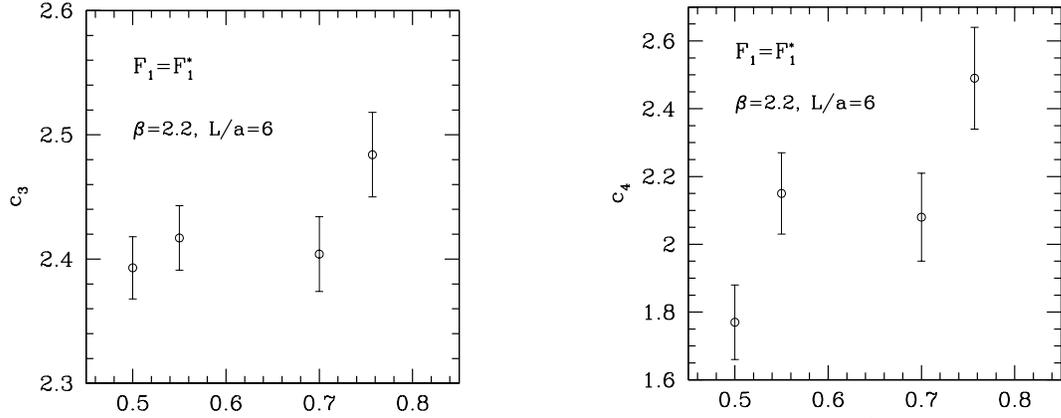

\hspace{0cm}
\vspace{-1.0cm}
\parbox{6.2cm}{
\centerline{\epsfig{file=plots/lcpc3l.epsi,width=6.0cm}}}
\hfill
\parbox{6.2cm}{
\centerline{\epsfig{file=plots/lcpc4l.epsi,width=6.0cm}}}
\vspace{0.5cm}
\caption{Here, we show the $\lambda$-dependence of the cumulants $c_3$ and
  $c_4$. The parameter $\kappa$ is matched such that $F_1=F_1^*$
  \eq{matching1}. \label{f_c3c4l}}
\end{figure}
%%%%%%%%%%%%%%%%%%%%%%%%%%%%%%%%%%%%%%%%%%%%%%%%%%%%%%%%%%%%%%%%%%%%%%
%%%%%%%%%%%%%%%%%%%%%%%%%%%%%FIGURE%%%%%%%%%%%%%%%%%%%%%%%%%%%%%%%%%%%
\begin{figure}[tb]
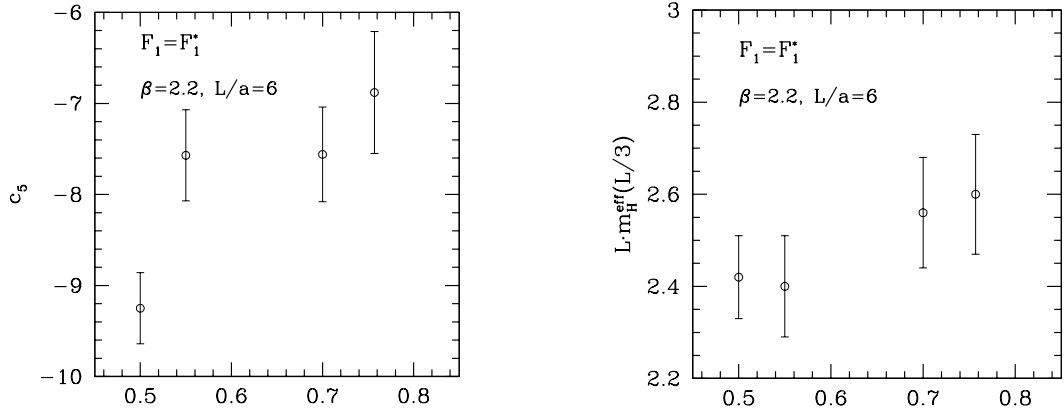

\hspace{0cm}
\vspace{-1.0cm}
\parbox{6.2cm}{
\centerline{\epsfig{file=plots/lcpc5l.epsi,width=6.0cm}}}
\hfill
\parbox{6.2cm}{
\centerline{\epsfig{file=plots/lcpmheffl.epsi,width=6.0cm}}}
\vspace{0.5cm}
\caption{Here, we show the $\lambda$-dependence of the cumulant $c_5$ and
  the effective Higgs mass $Lm_{\rmH}^{\rm eff}(L/3)$.
  The parameter $\kappa$ is matched such that $F_1=F_1^*$ 
  \eq{matching1}. \label{f_c5mhl}}
\end{figure}
%%%%%%%%%%%%%%%%%%%%%%%%%%%%%%%%%%%%%%%%%%%%%%%%%%%%%%%%%%%%%%%%%%%%%%

In \fig{f_c3c4l} and \fig{f_c5mhl}, we show the $\lambda$-dependence 
of the cumulants computed at $\beta=2.2$ on a $6^4$ lattice,
with the parameter $\kappa$ matched such that $F_1=F_1^*$ \eq{matching1}.
In the interval of $\lambda$-values considered and within our
statistical precision, the cumulants show a weak
$\lambda$-dependence. Especially $c_4$ seems to be sensitive to
$\lambda$. However,
for the practical purpose of matching the $\lambda$ parameter, this
situation is not suitable. As can be seen in \fig{f_cumvol}, the
cumulant $c_4$ computed at $\beta=2.4$ has a strong $L$-dependence
in the region around the matched lattice size $L/a\approx11.4$. The
uncertainty coming from an interpolation to this lattice size and the
statistical errors of $c_4$ at $\beta=2.2$ are the limiting factors
for the matching of $\lambda$.

\subsection{Higgs and W-boson correlations \label{hw_correlations}}

We describe in the following the construction of renormalised
quantities with a well-defined continuum limit from correlations
of gauge invariant Higgs and W-boson fields with zero spacial
momentum.

On each timeslice of the lattice, we construct a 
gauge invariant (composite) Higgs field
with zero spacial momentum $H(x_0)$ as follows:
\bes\label{hfieldzero}
 H(x_0) & = & \left(\frac{a}{L}\right)^3\sum_{\vec{x}} 
 \Phid(x_0,\vec{x})\Phi(x_0,\vec{x}) \, ,
\ees
where the sum is over the points $\vec{x}$ in the timeslice with time
coordinate $x_0$. We define
the correlation function $f_{\rmH}(t)$ as
\bes\label{fhcorr}
 f_{\rmH}(t) & = & \frac{a}{T}\sum_{x_0} \langle
 H(x_0+t)H(x_0)\rangle_c \, ,
\ees
where $T$ denotes the time extent of the lattice: throughout
our work we use $T\equiv L$. The connected expectation value of
the product of two observables $O_1$ and $O_2$ is
defined as usual by 
\bes\label{connexpv}
 \langle O_1\,O_2\rangle_c & = & \langle O_1\,O_2\rangle - \langle
 O_1\rangle\,\langle O_2\rangle \, .
\ees
The correlation function $f_{\rmH}(t)$ is multiplicatively renormalised
with renormalisation factor $Z_S^2$ defined in \eq{renhiggs}. Because
the Higgs fields entering the connected two-point function in
\eq{fhcorr} are taken at physical time separation $t$, the problems
that we encountered in \sect{s_cumulants}, due to the coincidence of
the arguments of fields in Green functions, are avoided here.
From the periodic boundary conditions in time direction, we derive
the property $f_{\rmH}(T-t)=f_{\rmH}(t)$. 
The correlation function $f_{\rmH}(t)$ is
symmetric with respect to the point $t=T/2$.

A gauge invariant W-boson field with zero spacial momentum 
is defined as
\bes\label{wfieldzero}
 W_{rk}(x_0) & = & \left(\frac{a}{L}\right)^3\sum_{\vec{x}}
 -i\tr(\tau_rV_k(x_0,\vec{x})) \, ,
\ees
where $V_k(x)\;(k=1,2,3)$ is the gauge invariant link defined in \eq{ginvlink}
and $\tau_r\;(r=1,2,3)$ are the isospin Pauli matrices. We restrict
the Lorentz index of the W-boson field to the spacial components $k$.
We define the correlation function $f_{\rmW}(t)$ as
\bes\label{fwcorr}
 f_{\rmW}(t) & = & \frac{a}{T}\sum_{x_0}\sum_{r,k} \langle
 W_{rk}(x_0+t)W_{rk}(x_0)\rangle_c \, .
\ees
The real field $W_{rk}(x)=-i\tr(\tau_rV_k(x))$ corresponds in the
naive continuum limit
to the field $\,\Re\tr(\tau_r\vp^{\dagger}(x)D_k\vp(x))$, where
$D_k=\pd_k+igA_{kr}(x)\frac{\tau_r}{2}$
is the gauge covariant derivative for the gauge field $A_{kr}(x)$ 
and $\vp(x)$ is the Higgs field in $2\times2$ matrix notation. 
The canonical dimension of the field
$W_{rk}(x)$ is therefore three and under renormalisation it can mix
with fields of equal or lower canonical dimension
which are gauge invariant and
have an isospin index $r$ and a Lorentz index $k$.
Because we are in a scalar theory we need
a derivative $\pd_k$ to obtain a Lorentz index.
The only possibility is the field
$\pd_k\tr(\tau_r\vp^{\dagger}(x)\vp(x))$, but it can be rewritten
like $\,2\Re\tr(\tau_r\vp^{\dagger}(x)D_k\vp(x))$.
The absence of mixing with other fields implies that
the correlation function $f_{\rmW}(t)$ is multiplicatively renormalised.
Again, we have per construction the symmetry property 
$f_{\rmW}(T-t)=f_{\rmW}(t)$.

From the correlation functions $f_i(t),\;i=\rmH,\rmW$ we construct the
quantities
\bes\label{hwlogratio}
 \mu_i(t/T) & = & \ln\left(\frac{f_i(t)}{f_i(T/2)}\right) \,, \quad
 i=\rmH,\rmW \, .
\ees
The multiplicative renormalisation of the correlation functions
cancels in the ratio and keeping the physical size $L\equiv T$
constant the $\mu_i$'s have a well-defined continuum limit. 

Using the transfer matrix formalism it is easy to show that the
correlation $f_{\rmH}(t)$ behaves like
\bes\label{tmfhcorr}
 f_{\rmH}(t) & \sim & {\rm
   const}\times\cosh\left((\frac{T}{2}-t)\,m_{\rmH}\right) \, ,
\ees
where $m_{\rmH}\equiv E_1^{(0)}-E_0^{(0)}$. The energies
$E_{\alpha}^{(0)}\;(\alpha=0,1,2,...)$ are the spectrum in the
zero charge sector of the Hilbert space. We consider $m_{\rmH}$ to be the
``Higgs mass'' in the confinement phase of the model. The relation in
\eq{tmfhcorr} is valid in the limits
\bes\label{tmfhcorrlim}
 t\,\Delta m_{\rmH} \gg 1 & \mbox{and} & (T-t)\Delta m_{\rmH} \gg 1 \,,
\ees
where $\Delta m_{\rmH}\equiv E_2^{(0)}-E_1^{(0)}$.
We can define an effective Higgs mass $m_{\rmH}^{\rm eff}(t)$ through
the relation
\bes\label{mheff}
 \cosh\left((\frac{T}{2}-t)m_{\rmH}^{\rm eff}(t)\right) & = &
 \frac{f_{\rmH}(t)}{f_{\rmH}(T/2)} \, .
\ees
The effective mass $m_{\rmH}^{\rm eff}(t)$ converges for large enough 
$t$ to the Higgs mass $m_{\rmH}$. We could think to use
$Lm_{\rmH}^{\rm eff}(t)$, determined at a fixed physical time $t$, as
the dimensionless physical quantity $F_2$ for the matching of $\lambda$.
In \fig{f_c5mhl}, we show the $\lambda$-dependence of
$L\,m_{\rmH}^{\rm eff}(L/3)$ computed at $\beta=2.2$ on a $6^4$
lattice with the parameter
$\kappa$ matched such that $F_1=F_1^*$ \eq{matching1}.
The effective mass is essentially a flat function
of $\lambda$ within the statistical precision and it would not be
suitable for the matching. The same conclusion is valid if we
consider an effective mass extracted from the correlation function
$f_{\rmW}(t)$. We note that these observations are in agreement
with the exploratory results of reference \cite{Montvay:1986nk}.

\section{Scaling}

The results of \sect{lambdamatch}, concerning the
search for a dimensionless physical quantity $F_2$ which is sensitive
to a variation of the parameter $\lambda$ once the parameter $\kappa$
is matched using the condition \eq{matching1}, can be summarised
in the following statements:
\begin{itemize}
\item The cumulant $c_4$, defined in \eq{cumulants},
  shows a $\lambda$-dependence, as can be seen from \fig{f_c3c4l}.
  However, it is a strongly varying function of the lattice size, see
  \fig{f_cumvol}. The use of $c_4$ for matching the parameter
  $\lambda$ requires a precise matching of the physical lattice size,
  which can be done only by interpolation and is therefore difficult.
\item The Higgs and W-boson effective masses are essentially flat
  functions of $\lambda$, see \fig{f_c5mhl}.
\end{itemize}
In the following we use the assumption,
supported by these results,
that the physics in the confinement ``phase'' of the SU(2) Higgs model is
weakly dependent on the parameter $\lambda$ once a dimensionless
physical quantity such as $F_1=\rnod\,[2\mu-V_0(\rnod)]$ is kept fixed. 
We compare the results for several dimensionless physical quantities
computed at $\beta=2.2$ and at $\beta=2.4$ for the parameter sets
\bes
 \beta=2.2, \qquad 0.5\le\lambda<0.8, \qquad F_1(\kappa)=F_1^*\equiv1.26
\ees
and
\bes
 \beta=2.4, \qquad \lambda=0.7, \qquad \kappa=0.2759 \,.
\ees
This comparison, by a change in the lattice spacing of
almost a factor two \eq{aratio}, gives a measure of the scaling
properties of the SU(2) Higgs model in the confinement ``phase''.
%%%%%%%%%%%%%%%%%%%%%%%%%%%%%FIGURE%%%%%%%%%%%%%%%%%%%%%%%%%%%%%%%%%%%
\begin{figure}[t]
\hspace{0cm}
\vspace{-1.0cm}
\centerline{\epsfig{file=plots/lcpmh.epsi,width=10cm}}
\vspace{-0.0cm}
\caption{Here, we show the scaling of the logarithmic ratio 
  $\mu_{\rmH}$ defined in
  \eq{hwlogratio}. The values at $\beta=2.2$ are computed with lattice
  size $L/\rnod\approx 2.14$.
 \label{f_lcpmh}}
\end{figure}
%%%%%%%%%%%%%%%%%%%%%%%%%%%%%%%%%%%%%%%%%%%%%%%%%%%%%%%%%%%%%%%%%%%%%%
%%%%%%%%%%%%%%%%%%%%%%%%%%%%%FIGURE%%%%%%%%%%%%%%%%%%%%%%%%%%%%%%%%%%%
\begin{figure}[h]
\hspace{0cm}
\vspace{-1.0cm}
\centerline{\epsfig{file=plots/lcpmw.epsi,width=10cm}}
\vspace{-0.0cm}
\caption{Here, we show the scaling of the logarithmic ratio
  $\mu_{\rmW}$ defined in
  \eq{hwlogratio}. The values at $\beta=2.2$ are computed with lattice
  size $L/\rnod\approx 2.14$.
 \label{f_lcpmw}}
\end{figure}
%%%%%%%%%%%%%%%%%%%%%%%%%%%%%%%%%%%%%%%%%%%%%%%%%%%%%%%%%%%%%%%%%%%%%%

\subsection{Cumulants}

Because of the mentioned strong dependence of the cumulants on the
physical lattice size and the difficulty to match it precisely between
$\beta=2.2$ and $\beta=2.4$, we can only give some indications in form
of the following table.
\[\bea{lcl}
 \beta=2.2 & \; & \beta=2.4 \\ & & \\
 c_3(L/\rnod=2.12)\;=\;2.39(3) \quad (\lambda=0.5) & \; &
 c_3(L/\rnod=2.08)\;=\;2.21(8) \\
 c_3(L/\rnod=2.16)\;=\;2.40(3) \quad (\lambda=0.7) & \; &
 c_3(L/\rnod=2.27)\;=\;2.84(6) \\ & & \\
 c_4(L/\rnod=2.12)\;=\;1.77(11) \quad (\lambda=0.5) & \; &
 c_4(L/\rnod=2.08)\;=\;1.9(3) \\
 c_4(L/\rnod=2.16)\;=\;2.08(13) \quad (\lambda=0.7) & \; &
 c_4(L/\rnod=2.27)\;=\;5.1(3) \\ & & \\
 c_5(L/\rnod=2.12)\;=\;-9.3(4) \quad (\lambda=0.5) & \; &
 c_5(L/\rnod=2.08)\;=\;-6.4(9) \\
 c_5(L/\rnod=2.16)\;=\;-7.6(5) \quad (\lambda=0.7) & \; &
\ea\]
The numbers are compatible but no precise conclusion can be made.

\subsection{Higgs and W-boson correlations}

In \fig{f_lcpmh} and \fig{f_lcpmw} we show the results for the
logarithmic ratios $\mu_{\rmH}$ and $\mu_{\rmW}$
defined in \eq{hwlogratio}. Because we cannot match
exactly the physical size of the lattice, which at $\beta=2.2$ has
the value $L/\rnod\approx2.14$, we
show for $\beta=2.4$ the results for two different lattice sizes
$L/\rnod=1.89$ and $L/\rnod=2.27$. Moreover, for $\beta=2.2$ we choose
two values $\lambda=0.5$ and $\lambda=0.7$. The results show rough
compatibility with scaling, even more pronounced for $\mu_{\rmW}$.
%%%%%%%%%%%%%%%%%%%%%%%%%%%%%FIGURE%%%%%%%%%%%%%%%%%%%%%%%%%%%%%%%%%%%
\begin{figure}[tb]
\hspace{0cm}
\vspace{-1.0cm}
\centerline{\epsfig{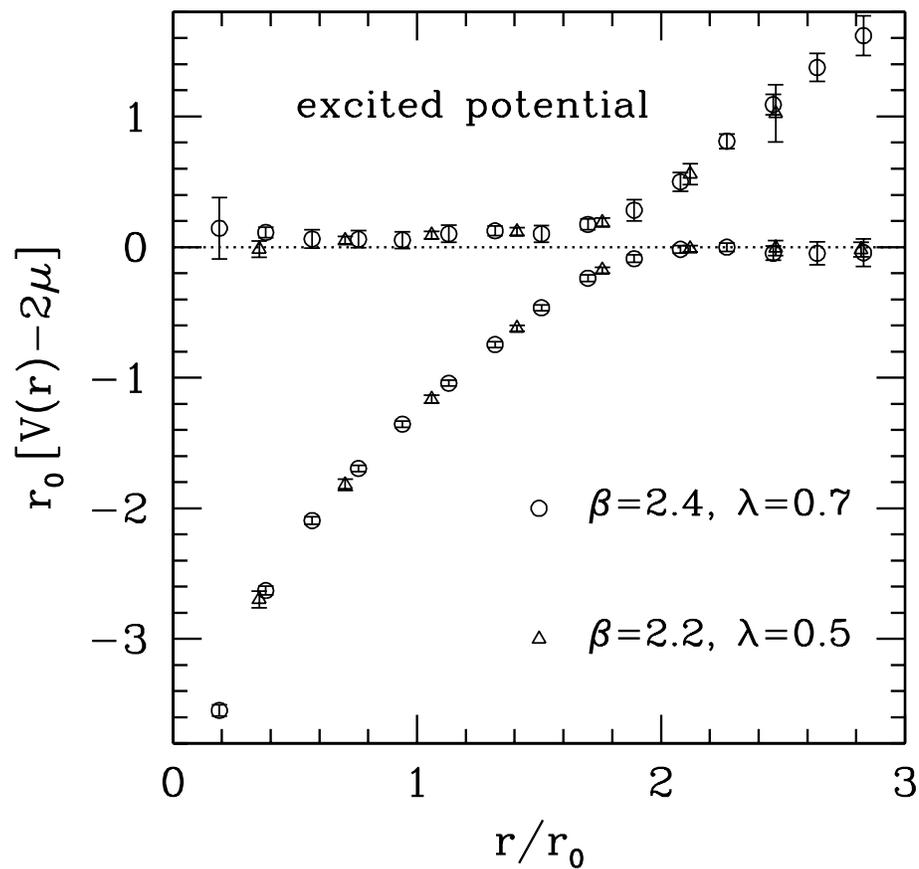}}
\vspace{-0.0cm}
\caption{Here, we show the scaling of the renormalised ground state and 
  first excited state static potentials.
 \label{f_invariant_lcp}}
\end{figure}
%%%%%%%%%%%%%%%%%%%%%%%%%%%%%%%%%%%%%%%%%%%%%%%%%%%%%%%%%%%%%%%%%%%%%%
%%%%%%%%%%%%%%%%%%%%%%%%%%%%%FIGURE%%%%%%%%%%%%%%%%%%%%%%%%%%%%%%%%%%%
\begin{figure}[tb]
\hspace{0cm}
\vspace{-1.0cm}
\centerline{\epsfig{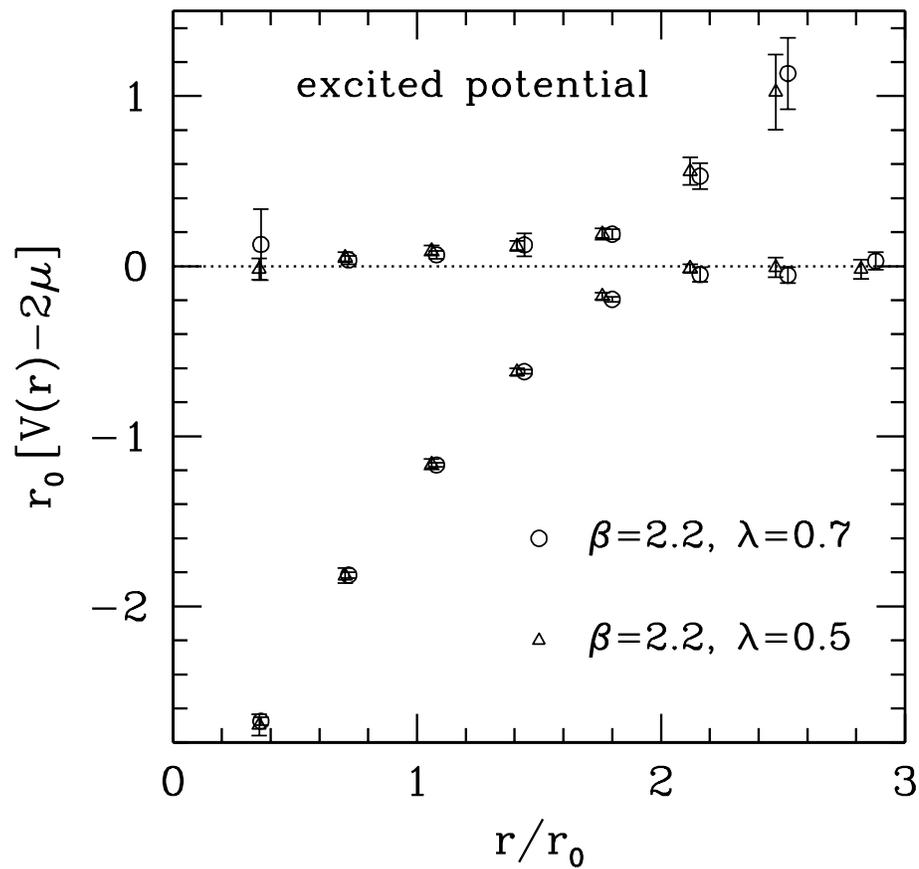}}
\vspace{-0.0cm}
\caption{Here, we show the $\lambda$-independence of the static
  potentials once the
  parameter $\kappa$ is matched by the condition $F_1=F_1^*$.
 \label{f_invariant_b22}}
\end{figure}
%%%%%%%%%%%%%%%%%%%%%%%%%%%%%%%%%%%%%%%%%%%%%%%%%%%%%%%%%%%%%%%%%%%%%%

\subsection{Static potentials}

In \fig{f_invariant_lcp}, we compare the results for the renormalised
ground state and first excited state static potentials that we
obtained at $\beta=2.4$ (the same as in \fig{f_invariant}) and at
$\beta=2.2,\;\kappa=0.2737,\;\lambda=0.5$.
The results are compatible with scaling within minute errors!
In \fig{f_invariant_b22}, we compare the static potentials for two
different values $\lambda=0.5$ and $\lambda=0.7$ at $\beta=2.2$. For
$\lambda=0.7$ we use $\kappa=0.2928$.
There is no significant difference, confirming the almost independence
on $\lambda$ that we observed
in \sect{hw_correlations} for the Higgs and W-boson mass.
These results are a strong indication for a continuum-like behavior of the
static potentials already at the small $\beta$ values that we use.

\chapter{Conclusions and Outlook \label{conclusions}}

String breaking, the flattening of the static potential at large
distances, is clearly observed in our lattice simulations of
the four-dimensional SU(2) Higgs model in the confinement ``phase''.
We are able to determine the ground state and the first excited state
static potentials with good precision. Our results confirm the
interpretation of string breaking as a level crossing phenomenon
between ``string states'' and ``two-meson states''.

In the path integral formalism the static quarks are represented by 
straight time-like Wilson lines, the string-type states are represented by 
(smeared) space-like Wilson lines and the meson-type states by 
(smeared) Higgs fields. With these ingredients, a matrix correlation
function can be constructed from which the static potentials are
extracted using a variational method.
We are able to describe the level crossing between the string-type and
meson-type states in terms of the overlaps of these states with the
true eigenstates of the Hamiltonian.
In order to understand the difficulties in observing string breaking
in recent lattice QCD simulations with dynamical fermions, we studied
the extraction of the static potential from the string-type states
(Wilson loops) alone. The overlap of the string-type states with the true
ground state at large separations drops under 10\%. The
Wilson loops have in this region a large overlap with the first
excited state. If we would extract the static potential from the
Wilson loops alone we would see the continuation of the linear rise 
of the static potential beyond the expected asymptotic value, as 
observed in the QCD simulations \cite{Burkhalter:1998wu,Aoki:1999ff}.

We addressed the question of the lattice artifacts in our
results.
The lines of constant physics (LCPs) in the parameter space of the
SU(2) Higgs model are
determined by the values of two dimensionless physical quantities
$F_1$ and $F_2$: the lattice spacing is varied by changing $\beta$ and
the bare parameters $\kappa$ and $\lambda$ are renormalised in order
to keep $F_1$ and $F_2$ constant.
The quantity $F_1$ that we choose is defined in \eq{F1}:
it is mainly sensitive to the mass of the
dynamical Higgs field and hence to $\kappa$.
We studied then the sensitivity to $\lambda$ of different
quantities. Our aim was to find a renormalised quartic Higgs coupling.
We found a candidate, the cumulant $c_4$ defined in \eq{cumulants},
but no precise conclusions could be drawn. Our results support the
conjecture that the physics in the confinement ``phase'' of the SU(2)
Higgs model is weakly dependent on $\lambda$, once a physical quantity
such as $F_1$ in \eq{F1} is kept fixed. Assuming this, we studied the
scaling behavior of different quantities between $\beta=2.2$ and
$\beta=2.4$, for a variation of the lattice spacing by almost a factor
of two.
The ground state and first excited state static potentials are
compatible with scaling within surprisingly small errors: our main
result is shown in \fig{f_invariant_lcp}. Compatibility with
scaling is also observed for the other quantities that we
investigated.

The extension of the method described in \cite{Michael:1992nc} and in
\sect{stringbreak} of our work to QCD should present ``only'' problems
related to statistical accuracy. There are two disadvantages in QCD
with respect to our situation: there is no one-link integral to reduce
the statistical variance of the correlations at large times and the
computation of quark propagators is so CPU-time consuming that one
cannot take advantage of translation invariance to reduce statistical
errors as we did. On the other hand, in QCD the quark fields are
integrated out analytically, which usually results in correlation
functions with relatively small statistical errors. As concerns the
computation of the quark propagators, the maximal variance reduction
method of reference \cite{Michael:1998sg} is a promising prospective.

The search for a method of defining LCPs in the confinement ``phase''
of the SU(2) Higgs model is still an open and interesting question: the
basic problem behind it, is to find a definition of a gauge invariant
renormalised quartic Higgs coupling.

There are other important examples in QCD of phenomena involving
mixing of states, like the computation of the glueball spectrum 
with dynamical fermions. The improvements that were made in the
observation of string breaking can inspire the research in these other
contexts.

\begin{appendix}
 \chapter{Notation conventions \label{notation}}

\section{Lattice notation}

We consider in this work a four-dimensional hyper-cubic lattice in the
Euclidean space with lattice spacing $a$
\bes\label{lattice}
 \Gamma & = & \{x/a\in\N^4\,|\,a\le x_{\mu}\le L_{\mu},\; L_0\equiv
 T\} \, .
\ees
We use Greek symbols such as $\mu$, $\nu$, to denote all directions
0,1,2,3 and Roman symbols such as $k$ to denote the space-like
directions 1,2,3. $\muh$ is the unit vector in $\mu$ direction. The
physical lattice volume is denoted by $V=TL_1L_2L_3$ and the number of
lattice points by $\Omega=V/a^4$.
The Fourier transformation of a function $f(x)$ on $\Gamma$ is defined by:
\bes\label{lattfourier}
 \tilde{f}(p) & = & a^4\sum_{x\in\Gamma}\rme^{-ip\cdot x}f(x) \, ,
\ees
where $p\cdot x\equiv p_{\mu}x_{\mu}\equiv
\sum_{\mu}p_{\mu}x_{\mu}$.
Considering periodic boundary conditions in all directions:
\bes\label{periodicbc}
 f(x+L_{\mu}\muh) & = & f(x) \, ,
\ees
the allowed lattice momenta are restricted to the first Brillouin zone
\bes\label{1stbrillouin}
 {\mathcal B} & = & \{p\,|\,p_{\mu}=\frac{2\pi}{L_{\mu}}n_{\mu},\;
 n_{\mu}=0,1,...,L_{\mu}-1\} \, .
\ees
The inverse Fourier transformation of \eq{lattfourier} is
\bes\label{inverselattfourier}
 f(x) & = & \frac{1}{a^4\Omega}\sum_{p\in{\mathcal B}} \rme^{ip\cdot
 x}\tilde{f}(p) \, .
\ees

\section{SU($N$) gauge fields}

A SU($N$) gauge field on the lattice is represented by the set of
oriented links
$U(x,\mu)_{ab}\in\SUn$ ($a,b=1,2,...,N$ are the color indices)
connecting the points $x+a\muh$ with $x$.
A gauge transformation is
defined by a field $\{\Lambda(x)\in\SUn\,|\,x\in\Gamma\}$ and
by the transformations
\bes\label{gaugetrsfu4}
 U^{\Lambda}(x,\mu) & = & \Lambda^{\dagger}(x)U(x,\mu)\Lambda(x+a\muh)
 \,.
\ees
For the pure gauge theory, we use Wilson's action which is based on the
plaquette
\bes\label{plaquette}
 P_{\mu\nu}(x) & = & U(x,\mu)U(x+a\muh,\nu)
 U^{\dagger}(x+a\nuh,\mu)U^{\dagger}(x,\nu)
\ees
and reads
\bes\label{wilsonaction}
 S_{\rmW} & = & \beta\sum_x\sum_{\mu<\nu}
 \left\{1-\frac{1}{N}\Re\tr P_{\mu\nu}(x)\right\} \,.
\ees
For vanishing lattice spacing and smooth gauge fields, 
$S_{\rmW}$ reduces to the classical
Yang-Mills action $S_{{\rm YM}}$ for gauge fields on the continuum
\bes\label{gaugecont}
 S_{\rmW} & \longrightarrow & S_{{\rm YM}} + \rmO(a^2) \,,
\ees
if the bare gauge coupling constant $g$ on the lattice satisfies
the relation
\bes\label{gaugecoupling}
 \beta & = & \frac{2N}{g^2} \,.
\ees

\section{SU(2) Higgs field \label{higgsnotation}}

A SU(2) Higgs field on the lattice is a complex scalar field
$\Phi_a(x)\;(a=1,2)$
in the fundamental representation of the gauge group SU(2): under a
gauge transformation $\{\Lambda(x)\in\SUtwo\,|\,x\in\Gamma\}$
\bes\label{gaugetrsfphi4}
 \Phi^{\Lambda}(x) & = & \Lambda^{\dagger}(x)\Phi(x) \,.
\ees
$U(x,\mu)$ is the parallel transporter for the Higgs field,
$U(x,\mu)\Phi(x+a\muh)$ transforms under gauge transformation
like $\Phi(x)$. The lattice covariant derivative is defined as
\bes\label{latcovder}
 \nabla_{\mu}\Phi(x) & = &
 \frac{1}{a}\,[U(x,\mu)\Phi(x+a\muh)-\Phi(x)] \,.
\ees

A peculiarity of the gauge group SU(2) is that the complex conjugate
representation $2^*$ is equivalent to the fundamental representation
$2$:
\bes\label{equiv}
 (i\tau_2)^{-1}U(i\tau_2) & = & U^*\,, \quad U\in\SUtwo \,,
\ees
where $\tau_r\;(r=1,2,3)$ denote the Pauli matrices. This means that
the field $\tilde{\Phi}(x)=i\tau_2\Phi^*(x)$ transforms like
$\Phi(x)$. Using this
property a representation by a $2\times 2$ matrix field $\vp(x)$ can
be constructed for the SU(2) Higgs field. We fix the
notation
\bes\label{higgsnotation1}
 \Phi(x) = \left( \bea{c} \Phi_1(x)=\phi_1(x)+i\phi_2(x) \\ 
                          \Phi_2(x)=\phi_3(x)+i\phi_4(x) \ea \right) 
 \,,
\ees
where $\phi_i(x)\;(i=1,2,3,4)$ are the four real components of
$\Phi(x)$, and we define
\bes\label{higgsnotation2}
 \vp(x) = \left( \bea{cc} \tilde{\Phi}_1(x) & \Phi_1(x) \\
 \tilde{\Phi}_2(x) & \Phi_2(x) \ea \right) =
 \sqrt{\Phi^{\dagger}(x)\Phi(x)}\cdot\alpha(x) \,,
\ees
where $\alpha(x)\in {\rm SU(2)}$. Under gauge transformation
$\vp(x)$ transforms like
\bes\label{gaugetrsf42}
 \vp^{\Lambda}(x) = \Lambda^{\dagger}(x)\vp(x) 
\ees
The scalar product of two SU(2) doublets $\Phi$ and $\Psi$ can be
written using the corresponding matrices $\vp$ and $\psi$ constructed
as in \eq{higgsnotation2}:
\bes\label{doubletmatrix}
 \Re\Phi^{\dagger}\Psi & = & \frac{1}{2}\tr(\vp^{\dagger}\psi) \,, \\
 \Im\Phi^{\dagger}\Psi & = & \frac{1}{2}\tr((i\tau_3)\vp^{\dagger}\psi)
 \,.
\ees

 \chapter{The transfer matrix in the SU(2) Higgs model \label{apptm}}

\section{Hamiltonian formalism \label{hamform}}

Along the lines of reference \cite{Luscher:TM} we construct a
canonical, Hamiltonian formalism for fields living on a timeslice. We
consider a three-dimensional lattice with sites
$\vec{x}=a\cdot(n_1,n_2,n_3),\; n_k\in\N ,\; 1\le k\le 3,\;
1\le n_k\le N$ and assume periodic boundary conditions.
The Hilbert space $\Hspace$ is the set of all complex-valued,
square-integrable wave functions $f$ with arguments
$U(\vec{x},k)\in\SUtwo$ and $\Phi(\vec{x})\in{\C}^2$
(in the following we shall denote them simply by $f(U,\Phi)$ or $|f\rangle$).
The scalar product is defined by
\bes\label{scalarprod}
 \langle f|g \rangle & = & \int\mass f(U,\Phi)^*\,g(U,\Phi) \, , \\
 \int\mass & = & \intthree \, ,
\ees
where $\rmd U$ is the Haar measure on SU(2) and $\phi_i,\;
i=1,2,3,4$ are the four real components of $\Phi$, see \eq{higgsnotation1}.
The field operators $\Phih_a(\vec{x}),\;
\Uh_{ab}(\vec{x},k)$ 
act as multiplicative operators on the wave
functions $f$:
\bes
 \left[\Uh_{ab}(\vec{x},k)\,f\right](U,\Phi) & = & 
 U_{ab}(\vec{x},k)\,f(U,\Phi) \, , \label{opu} \\
 \left[\Phih_a(\vec{x})\,f\right](U,\Phi) & = &
 \Phi_a(\vec{x})\,f(U,\Phi) \, . \label{opphi}
\ees
A gauge transformation is defined by a field of SU(2) matrices
$\Lambda(\vec{x})$:
\bes
 U^{\Lambda}(\vec{x},k) & = & \Lambda^{\dagger}(\vec{x})\,U(\vec{x},k)\,
 \Lambda(\vec{x}+a\kh) \, , \label{gtrsfu} \\
 \Phi^{\Lambda}(\vec{x}) & = & \Lambda^{\dagger}(\vec{x})\,\Phi(\vec{x}) \, .
 \label{gtrsfphi}
\ees
In the Euclidean framework this corresponds to a time independent gauge
transformation. In $\Hspace$ we define a unitary operator
representation $\Rop(\Lambda)$ such that
\bes\label{gtrsfwf}
 \left[\Rop(\Lambda)\,f\right](U,\Phi) & = &
 f(U^{\Lambda},\Phi^{\Lambda}) \, \equiv \, f^{\Lambda}(U,\Phi) \, .
\ees
Requiring gauge covariance of \eq{opu} and \eq{opphi} one has
\bes
 \Rop^{\dagger}(\Lambda)\,\Uh_{ab}(\vec{x},k)\,\Rop(\Lambda) & = &
 \Lambda^{\dagger}_{aa^{\prime}}(\vec{x})\,
 \Uh_{a^{\prime}b^{\prime}}(\vec{x},k)\,
 \Lambda_{b^{\prime}b}(\vec{x}+a\kh) \, , \label{gtrsfuop} \\
 \Rop^{\dagger}(\Lambda)\,\Phih_a(\vec{x})\,\Rop(\Lambda) & = & 
 \Lambda^{\dagger}_{aa^{\prime}}(\vec{x})\,
 \Phih_{a^{\prime}}(\vec{x}) \, . \label{gtrsfphiop}
\ees
An operator $\hat{\rmO}$ is called gauge invariant if
\bes\label{ginvop}
 \left[\,\Rop(\Lambda)\,,\,\hat{\rmO}\,\right] & = & 0 \, .
\ees
In the following we shall need operators $\hat{\rmO}$ which can
be represented with the
help of an integral kernel $K_{\rmO}(U,\Phi;\Up,\Phip)$:
\bes\label{intkernel}
 (\hat{\rmO}\,f)(U,\Phi) & = & \int\massp
 K_{\rmO}(U,\Phi;\Up,\Phip)\,f(\Up,\Phip) \, .
\ees
The gauge invariance condition \eq{ginvop} translates into
\bes\label{ginvkernel}
 K_{\rmO}(U,\Phi;\Up,\Phip) & = &
 K_{\rmO}(U^{\Lambda},\Phi^{\Lambda};U^{\prime\Lambda},\Phi^{\prime\Lambda})
 \, .
\ees
For the product of two operators $\hat{\rmO}_1\,\hat{\rmO}_2$ we have
\bes\label{prodkernel}
 K_{\rmO_1\rmO_2}(U,\Phi;\Up,\Phip) & = & \int\masspp
 K_{\rmO_1}(U,\Phi;\Upp,\Phipp)\,\cdot \\
 & & K_{\rmO_2}(\Upp,\Phipp;\Up,\Phip) \, . \nonumber
\ees
The trace\footnote{
We write $\Tr$ for the operator trace in $\Hspace$ and $\tr$ for the
color trace in the gauge group.}
of an operator is defined as
\bes\label{traceop}
 \Tr\,\hat{\rmO} & = & \int\mass K_{\rmO}(U,\Phi;U,\Phi) \, .
\ees

\section{The transfer matrix in the temporal gauge \label{transfmat}}

We define an operator $\trans_{\rm temp}$ through the following kernel:
\bes\label{transtemp}
 K_{\rmT}^{\rm temp}(U,\Phi;\Up,\Phip) & = & T_{\rmH}(U,\Phi)\cdot
 T_{\rmG}(U)\cdot S(U,\Phi;\Up,\Phip)\cdot \\
 & & T_{\rmG}(\Up)\cdot T_{\rmH}(\Up,\Phip) \, , \nonumber
\ees
where
\bes
 & & T_{\rmG}(U) = \exp\left\{-\frac{\beta}{2}\sum_{\vec{x}}\sum_{1\le
     i<j\le3} \left[1-\frac{1}{2}\tr P_{ij}(\vec{x})\right]\right\} \,
     \label{tmgker} \\
 & & T_{\rmH}(U,\Phi) = \exp\left\{-\frac{1}{2}\sum_{\vec{x}}\Bigg[ 
     \lambda\left(\Phid(\vec{x})\Phi(\vec{x})-1\right)^2
     \right. \label{tmhker} \\
 & & \left. \left. -\kappa\sum_{k=1}^3\left(
     \Phid(\vec{x})U(\vec{x},k)\Phi(\vec{x}+a\kh) + 
     \Phid(\vec{x}+a\kh)\Ud(\vec{x},k)\Phi(\vec{x})
     \right) \right] \right\} \, , \nonumber \\
 & & S(U,\Phi;\Up,\Phip) = \exp\left\{\sum_{\vec{x}}
     \Bigg[\kappa\left(\Phid(\vec{x})\Phip(\vec{x}) + 
     \Phipd(\vec{x})\Phi(\vec{x})\right) 
     \right. \label{tmsker} \\
 & & \left. \left. -\frac{1}{2}\left(\Phid(\vec{x})\Phi(\vec{x}) + 
     \Phipd(\vec{x})\Phip(\vec{x})\right)
     -\beta\sum_{k=1}^3\left(
     1-\frac{1}{2}\tr(U(\vec{x},k)\Upd(\vec{x},k))
     \right) \right] \right\} \, . \nonumber
\ees
We included the terms quadratic in $\Phi$ and $\Phip$ separately, in
$S$ \eq{tmsker} instead of $T_{\rmH}$ \eq{tmhker} for later convenience.

\subsection{Positivity \label{tmpos}}

{\bf Proposition 1.}
\begin{itemize}
 \item[a)]
  $\trans_{\rm temp}$ is a selfadjoint, bounded and gauge invariant
  operator in $\Hspace$.
 \item[b)]
  $\trans_{\rm temp}$ is strictly positive for $\kappa>0$ and $\lambda>0$ \\
  (if $\lambda=0$ strict positivity holds for $0<\kappa<1/6$).
\end{itemize}
Selfadjointness of $K_{\rmT}^{\rm temp}$ means
$K_{\rmT}^{\rm temp}(\Up,\Phip;U,\Phi)^*=
K_{\rmT}^{\rm temp}(U,\Phi;\Up,\Phip)$.
This follows from the fact that
the kernel $K_{\rmT}^{\rm temp}$ is symmetric under interchange
of $\{U,\Phi\}$ with $\{\Up,\Phip\}$ and that the substitutions
$U_{ab}(\vec{x})\rightarrow U_{ab}^*(\vec{x}),\;
\Phi_a(\vec{x})\rightarrow\Phi_a^*(\vec{x})$ leave the
integration measure and the action invariant. \\
In order for $K_{\rmT}^{\rm temp}$ to be bounded, one has to discuss the cases
$\lambda>0$ and $\lambda=0$.
If $\lambda>0$ the kernel is bounded for all values of $\kappa$
because of the term quartic in the Higgs field.
If $\lambda=0$ one has to show that the matrix
\bes
 B_{\vec{x}a,\vec{y}b} & = & \delta_{\vec{x},\vec{y}}\delta_{ab}
 - \kappa\sum_{k=1}^3\left[U_{ab}(\vec{x},k)\delta_{\vec{y},\vec{x}+a\kh}
 + \Ud_{ab}(\vec{y},k)\delta_{\vec{x},\vec{y}+a\kh}\right]
\ees
is strictly positive. This is the same as for fermionic fields
\cite{Luscher:TM}: the restriction $|\kappa|<1/6$ has to be imposed. \\
Gauge invariance of $K_{\rmT}^{\rm temp}$ (\eq{ginvkernel}) is obvious.

To prove strict positivity of $\trans_{\rm temp}$ one has to verify that
\bes\label{positivity}
 & & \int\mass\massp
 f(U,\Phi)^*\,K_{\rmT}^{\rm temp}(U,\Phi;\Up,\Phip)\,f(\Up,\Phip)
 \nonumber \\
 & & \equiv\;\langle f|\trans_{\rm temp}|f \rangle\;>\;0 
 \qquad \mbox{for all $f\neq 0$ in $\Hspace$.}
\ees
If $K_{\rmT}^{\rm temp}$ is bounded it is enough to show strict positivity for
the kernel $S(U,\Phi;\Up,\Phip)$ because the kernels $T_{\rmG}$ and $T_{\rmH}$
can be absorbed in a new wave function
$g=T_{\rmG}\,T_{\rmH}\,f\in\Hspace$.
Furthermore, one has
\bes
 S(U,\Phi;\Up,\Phip) & = &
 \prod_{\vec{x}}S_{\rmH}(\Phi(\vec{x});\Phip(\vec{x})) \cdot
 \prod_{\vec{x},k}S_{\rmG}(U(\vec{x},k);\Up(\vec{x},k)) \, , \nonumber \\
 S_{\rmH}(\Phi;\Phip) & = & \exp\left\{\kappa\left[\Phid\Phip + \Phipd\Phi
 \right] - \frac{1}{2}(\Phid\Phi+\Phipd\Phip) \right\} \, , \nonumber \\
 S_{\rmG}(U;\Up) & = & \exp\left\{-\beta\left[1 - \frac{1}{2}\tr(U\Upd)
 \right]\right\} \, . \nonumber
\ees
and we are left to prove that
\bes\label{positivity2}
 \int\rmd U\,\rmd\Up\,\rmd^4\phi\,\rmd^4\phip\,h(U,\Phi)^*\,
 S_{\rmH}(\Phi;\Phip)\,S_{\rmG}(U;\Up)\,h(\Up,\Phip) \; > \; 0 \,
\ees
for all square integrable nonvanishing functions $h(U,\Phi)$ depending
on one link variable $U$ and one field variable $\Phi$.
Changing to the matrix notation for the Higgs field
$\vp=\rho\alpha,\; \rho\ge 0,\; \alpha\in\SUtwo$ one has
\bes
 \Phid\Phip + \Phipd\Phi & = & \tr(\vp^{\dagger}\vp^{\prime}) \, , \\
 \int_{\blackboardrrm^4}\rmd \phi_1(x)\cdot\cdot\cdot\rmd \phi_4(x)
 & \propto & \int_0^{\infty}\rho^3\rmd\rho\int_{\SUtwo}\rmd\alpha \, .
\ees
Expanding
\bes
 \exp\{\kappa\tr(\vp^{\dagger}\vp^{\prime})\} & = &
 \sum_{n=0}^{\infty}\frac{(\kappa\rho\rho^{\prime})^n}{n!}
 \left[\tr(\alpha^{\dagger}\alpha^{\prime})\right]^n \, , \nonumber
\ees
we recognise that
$\left[\tr(\alpha^{\dagger}\alpha^{\prime})\right]^n$ is the trace of
the tensor product representation of SU(2) which is composed of n
quarks\footnote{A quark is a vector transforming
according to the fundamental representation of
the gauge group, an anti-quark to its complex conjugate.
In the case of SU(2) they are equivalent.}\cite{Luscher:TM}.
Reducing out the tensor product we end up with
\bes\label{characterexp}
 S_{\rmH}(\rho,\alpha;\rho^{\prime},\alpha^{\prime}) & = &
 \sum_{n=0}^{\infty}\frac{(\kappa\rho\rho^{\prime})^n}{n!}
 \sum_{\nu}b_{\nu}(n)\chi^{(\nu)}(\alpha^{\dagger}\alpha^{\prime})\cdot
 \rme^{(\rho^2+\rho^{\prime 2})/2} \, .
\ees
$\nu$ labels the set of
all inequivalent unitary irreducible representations of SU(2) and
$\chi^{(\nu)}$ is the character of the representation $\nu$.
The coefficient $b_{\nu}(n)$ is the number of times the representation
$\nu$ occurs when reducing out the tensor product.
Since all irreducible representations can be obtained in this way,
for every $\nu$ exists a value of $n$ for which $b_{\nu}(n)\neq 0$.
Similarly, the character expansion for $S_{\rmG}$ is \cite{Luscher:TM}
\bes
 S_{\rmG}(U;\Up) & = & \rme^{-\beta}\cdot
 \sum_{\nu}c_{\nu}\chi^{(\nu)}(U\Upd) \, , \quad c_{\nu}>0 \, .
\ees
Defining
\bes\label{hnfunc}
 h_n(U,\alpha) & = & \int_0^{\infty}\rmd\rho\,\rho^{3+n}\,h(U,\rho\alpha)\,
 \rme^{-\rho^2/2}
\ees
and using the relation
\bes\label{charrel}
 \int\rmd U\,\chi^{(\nu)}(VU)\,\chi^{(\nu^{\prime})}(\Ud W) & = &
 \delta_{\nu\nu^{\prime}}\frac{1}{d_{\nu}}\chi^{(\nu)}(VW) \, ,
\ees
where $d_{\nu}=\chi^{(\nu)}(\one)$ is the dimension of the
representation $\nu$, the integral in \eq{positivity2} becomes
\bes\label{positivity3}
 \sum_{\nu,\nu^{\prime}}\sum_n C(\nu,\nu^{\prime},n)
 \int\rmd V\,\rmd W\,
 \left|\int\rmd U\rmd\alpha\, h_n^*(U,\alpha)\, \chi^{(\nu)}(UV)
 \chi^{(\nu^{\prime})}(\alpha^{\dagger}W) \right|^2 & &
\ees
where $C(\nu,\nu^{\prime},n)=d_{\nu}d_{\nu^{\prime}}c_{\nu}\kappa^n
b_{\nu^{\prime}}(n)\rme^{-\beta}$.
We are finally left with a sum of strictly positive terms provided
$\kappa>0$.

One can show that the condition $\kappa>0$ is also necessary. It is
possible to construct a function
$h(U,\rho\alpha)=h(\rho)$ such that $h_n$ \eq{hnfunc}
vanishes for all even $n$ and is different from zero for at least
one odd $n_0$. If $\kappa$ is negative, for this function the whole sum
\eq{positivity3} is then negative.

\section{Charge sectors \label{chargesect}}

\subsection{Minkowski space continuum \label{chargesmink}}

The classical Lagrange density for the SU(2) Higgs model in Minkowski
space is
\bes\label{lagrangeden}
 {\cal L} & = & -\frac{1}{4}F_a^{\mu\nu}F_{a\mu\nu} +
 (\rmD^{\mu}\Phi)^{\dagger}(\rmD_{\mu}\Phi) - V(\Phi) \, , \\
 F_a^{\mu\nu}(x) & = & \pd^{\mu}A^{\nu}_a(x) - \pd^{\nu}A_a^{\mu}(x) -
 g\veps_{abc}A_b^{\mu}(x)A_c^{\nu}(x) \, , \nonumber \\
 \rmD^{\mu}\Phi(x) & = &
 \left[\pd^{\mu}+igA_a^{\mu}(x)\pauli{a}\right]\Phi(x) \, , \nonumber \\
 V(\Phi(x)) & = & m_0^2\Phi^{\dagger}(x)\Phi(x) +
 \lambda_0[\Phi^{\dagger}(x)\Phi(x)]^2 \, , \nonumber
\ees
where the indices $a,b,c$ label the three generators of SU(2)
$\pauli{a}$, $\tau_a$ being the Pauli matrices, and
$g,\,m_0,\;\mbox{and}\;\lambda_0$ are the bare couplings.
The real vector fields $A_a^{\mu}$ carry a
Lorentz index $\mu$ and a group index $a$. $\Phi$ is the complex Higgs
SU(2) doublet field.
The canonical conjugate momenta to the fields $A_a^{\mu}$
and $\Phi,\; \Phid$ are
\bes
 \Pi_a^{\mu}(x) = \frac{\pd{\cal L}}{\pd(\pd_0A_{a\mu}(x))} & = &
 -F_a^{\mu 0}(x) = \left\{\begin{array}{cl} -E_a^i & \mu=i=1,2,3 \\
 0 & \mu=0 \end{array} \right. \label{cmoma1} \\
 \pi = \frac{\pd{\cal L}}{\pd(\pd_0\Phi(x))} & = &
 [\rmD^0\Phi(x)]^{\dagger} \, , \label{cmomphi1} \\
 \pi^{\dagger} = \frac{\pd{\cal L}}{\pd(\pd_0\Phi^{\dagger}(x))} & = &
 \rmD^0\Phi(x) \label{cmomphid1} \, .
\ees
$E_a^i$ is the non-Abelian electric field strength.
The generalisation of Gauss' law follows from the
equations of motion in the Lagrange formalism \cite{KHuang:QLGF}:
\bes\label{gauss1}
 \rmD_iE^i_a \equiv \pd_iE^i_a+g\veps_{abc}A_b^iE_c^i & = & j_a^0 \, .
\ees
$j_a^{\mu}$ is the matter-field current and in particular
\bes\label{matcur0}
 j_a^0 & = &
 -ig\left[\pi\pauli{a}\Phi-\Phid\pauli{a}\pi^{\dagger}\right] \, .
\ees 

Performing the Legendre transformation to the Hamilton density one has
to face the problem that the canonical conjugate momentum to $A^0_a$ is
zero. One can choose a gauge in which the fields $A^0_a$ vanish
identically, the temporal gauge \cite{KHuang:QLGF}. There is
still some freedom left in the choice of the gauge, namely the
time independent gauge transformations.
The canonical quantisation of the theory proceed considering wave
functionals
\bes\label{wfunccont}
 \Psi[A_a^i(\vec{x}),\Phi(\vec{x})] & &
\ees
where the fields are taken at fixed time. 
In this representation of the states the field operators
$\hat{A}_a^i(\vec{x})$ and $\hat{\Phi}(\vec{x})$ act as multiplicative
operators and the conjugate momenta are represented by functional
derivatives
\bes
 \hat{E}_a^i(\vec{x}) & = & i\frac{\delta}{\delta A_{ai}(\vec{x})}
 \, , \label{cmoma2} \\
 \hat{\pi}(\vec{x}) & = & -i\frac{\delta}{\delta \Phi(\vec{x})}
 \, , \label{cmomphi2} \\
 \hat{\pi}^{\dagger}(\vec{x}) & = & -i\frac{\delta}{\delta \Phid(\vec{x})}
 \label{cmomphid2} \, .
\ees
Under a time independent gauge transformation
$\Lambda(\vec{x})=\one-i\omega_a(\vec{x})\pauli{a}$ with infinitesimal
parameters $\omega_a(\vec{x})$ the fields transform like:
\bes
 \dg A_a^i(\vec{x}) & = & \frac{1}{g}\pd^i\omega_a(\vec{x}) +
 \veps_{abc}\omega_b(\vec{x})A_c^i(\vec{x}) \, , \label{deltaA} \\
 \dg \Phi(\vec{x}) & = & -i\omega_a(\vec{x})\pauli{a}\Phi(\vec{x})
 \, , \label{deltaphi} \\
 \dg \Phid(\vec{x}) & = & i\omega_a(\vec{x})\Phid(\vec{x})\pauli{a}
 \, . \label{deltaphid}
\ees
Now, we would like to find the generators $\hat{\rmG}_a$ of the gauge
transformations on the wave functionals $\Psi$: 
\bes\label{ggen}
 \dg \Psi & = & -\frac{i}{g}\int\rmd^3x\, 
 \omega_a(\vec{x})\,\hat{\rmG}_a(\vec{x})\Psi
\ees
To this aim we write:
\bes\label{Psivariation}
 & & \dg \Psi = \int\rmd^3x\left\{ \frac{\delta\Psi}{\delta
 A_{ai}(\vec{x})}\dg A_a^i(\vec{x}) +
 \frac{\delta\Psi}{\delta\Phi(\vec{x})}\dg\Phi(\vec{x}) +
 \dg\Phid(\vec{x})\frac{\delta\Psi}{\delta\Phid(\vec{x})} \right\}
 \nonumber \\
 & & \qquad = -\frac{i}{g}\int\rmd^3x\,
 \omega_a(\vec{x})\bigg\{\left[\pd_i\hat{E}_a^i(\vec{x}) + 
 g\veps_{abc}A_b^i(\vec{x})\hat{E}_c^i(\vec{x})\right]\Psi + \nonumber \\ 
 & & \qquad\quad ig\left[(\hat{\pi}(\vec{x})\Psi)\pauli{a}\Phi(\vec{x}) -
 \Phid(\vec{x})\pauli{a}(\hat{\pi}^{\dagger}(\vec{x})\Psi)\right]
 \bigg\} \, .
\ees
In the second line we used integration by part. From this we
conclude that
\bes\label{gaugegen}
 \hat{\rmG}_a(\vec{x}) & = &
 \rmD_i\hat{E}^i_a(\vec{x})-\hat{j}^0_a(\vec{x})
 \, ,
\ees
where the operators on the right hand side are defined by
\bes
 \rmD_i\hat{E}^i_a(\vec{x}) & = & \pd_i\hat{E}_a^i(\vec{x}) + 
 g\veps_{abc}\hat{A}_b^i(\vec{x})\hat{E}_c^i(\vec{x}) \, , \nonumber \\
 \hat{j}_a^0(\vec{x}) & = &
 -ig\left[\hat{\Phi}_m(\vec{x})\left(\pauli{a}\right)_{lm}\hat{\pi}_l(\vec{x}) - 
 \hat{\Phi}^{\dagger}(\vec{x})\pauli{a}\hat{\pi}^{\dagger}(\vec{x})
 \right] \, . \nonumber
\ees
We can classify the wave functionals $\Psi$ according to the action of
$\hat{\rmG}_a$.
Gauss' law
\bes\label{gauss2}
 \hat{\rmG}_a\Psi & = & 0
\ees
is equivalent to the gauge invariance of the state $\Psi$.
On states transforming like
\bes\label{varpsirep}
 \dg\Psi & = & -i\omega_a(\vec{x})T_a^{(\nu)}\Psi \, ,
\ees
where $T_a^{(\nu)}$ are the generators of the irreducible representation
$\nu$ of SU(2), the operator $\hat{G}$ acts as
\bes\label{statcharge}
 \hat{\rmG}_a(\vec{x}^{\prime}) & = &
 gT_a^{(\nu)}\delta^{(3)}(\vec{x}-\vec{x}^{\prime}) \, .
\ees
Thus wave functionals transforming under gauge transformation
according to the representation $\nu$ of $\Lambda(\vec{x})$ for some
fixed spacial point $\vec{x}$ express
the presence of a static external charge at position $\vec{x}$
in the representation $\nu$.

\subsection{Lattice formulation \label{chargelatt}}

Strict positivity of $\trans_{\rm temp}$ allows the definition of an
Hamiltonian $\ham$:
\bes\label{hamiltonian}
 \ham & = & -\frac{1}{a}\ln\trans_{\rm temp}
\ees
From the properties of $\trans_{\rm temp}$ it follows that
$\ham$ is selfadjoint (has only real eigenvalues), bounded from below
and gauge invariant.

The gauge symmetry of $\ham$ allows to choose its eigenstates with a
well defined transformation property under the gauge group. The
Hilbert space $\Hspace$ can then be classified according to the
irreducible representations of SU(2) on each point of the lattice
\cite{Rainer:PhD}. As we saw inspecting Gauss' law in the continuum
these representations define the charge sectors of
$\Hspace$. The transfer matrix is restricted to a specific charge
sector by multiplying it with the projector $\projector$ onto this
sector. We give two examples of charge sectors which are of interest
for us.

The zero charge sector (vacuum sector) corresponds to the
gauge invariant wave
functions. We denote by $|n_0\rangle$ the eigenstates of $\ham$ with
eigenvalues $E_n^{(0)}$ which form a basis of this sector. The
corresponding projection operator
$\projector_0=\sum_{n_0}|n_0\rangle \langle n_0|$ is
\bes\label{proj0}
 (\projector_0\,f)(U,\Phi) & = & \int\massw f^W(U,\Phi) \, .
\ees
The transfer matrix operator in the zero charge sector is
$\trans_0=\trans_{\rm temp}\projector_0$ and has the kernel
\bes\label{kernelT0}
 K_{\rmT}^0(U,\Phi;\Up,\Phip) & = & \int\massw
 K_{\rmT}^{\rm temp}(U,\Phi;U^{\prime W^{\dagger}},\Phi^{\prime W^{\dagger}})
\ees

A wave function $f$ in the sector of $\Hspace$ with a static quark
at position $\vec{x}$ and a static anti-quark at position $\vec{y}$
transforms as
\bes\label{qqbarsector}
 f^{\Lambda}_{ab} & = & \sum_{a^{\prime},b^{\prime}=1}^2
 \Lambda_{aa^{\prime}}(\vec{x})\, f_{a^{\prime}b^{\prime}}\,
 \Lambda^{\dagger}_{b^{\prime}b}(\vec{y})
\ees
We choose a basis $|n_{\qqb},a,b\rangle$ with eigenvalues $E_n^{(\qqb)}$
of $\ham$ independent of $a,b$. We drop in the following the color
indices $a,b$ of the states when an implicit sum over them is meant.
The projection operator $\projector(\vec{x},\vec{y})=
\sum_{n_{\qqb}} |n_{\qqb}\rangle \langle n_{\qqb}|$ onto this sector is
\bes\label{projqqbar}
 (\projector(\vec{x},\vec{y})\,f)(U,\Phi) \, = \, \int\massw 
 4\,\tr W^{\dagger}(\vec{x})\,\tr W(\vec{y})\,f^W(U,\Phi) \, . & &
\ees
The transfer matrix operator in this sector is
$\trans_{\qqbxy}=\trans_{\rm temp}\projector(\vec{x},\vec{y})$
and has the kernel
\bes\label{kernelTqqbar}
 K_{\rmT}^{\qqb}(U,\Phi;\Up,\Phip) & = & \int\massw
 4\,\tr W^{\dagger}(\vec{x})\,\tr W(\vec{y})\,\cdot \\
 & & K_{\rmT}^{\rm temp}
 (U,\Phi;U^{\prime W^{\dagger}},\Phi^{\prime W^{\dagger}}) \, .
\ees

\section{Reconstruction of the Euclidean expectation values \label{reco}}

We consider operators $\Oop(\vec{x},\vec{y})_{ab}$ which transform
under gauge transformation according to
\bes\label{corrop}
 \Rop^{\dagger}(\Lambda) \Oop(\vec{x},\vec{y})_{ab} \Rop(\Lambda)
 & = & \Lambda^{\dagger}_{aa^{\prime}}(\vec{x})
 \Oop(\vec{x},\vec{y})_{a^{\prime}b^{\prime}}
 \Lambda_{b^{\prime}b}(\vec{y}) \, .
\ees
These operators create out of
the vacuum states belonging to the sector with a static pair of
quarks. Examples are:
\bes\label{corropex}
 \hat{\Phi}_a(\vec{x})\,\hat{\Phi}_b^{\dagger}(\vec{y}) \quad
 \mbox{and} \quad
 \hat{U}(\vec{x},\vec{y})_{ab}
\ees
$U(\vec{x},\vec{y})$ denotes the product of links along a path
connecting $\vec{y}$ with $\vec{x}$. Because of \eq{opu} and
\eq{opphi} we can directly associate the observables
$O(\vec{x},\vec{y})_{ab}=\Phi_a(\vec{x})\Phi_b^*(\vec{y})$ and
$U(\vec{x},\vec{y})_{ab}$.

\subsection{Reconstruction theorem \label{recotheo}}

{\bf Proposition 2.}
\begin{itemize}
 \item[a)] \[ \Tr(\trans_0^M)=\intfour \exp(-S) \equiv Z \] is
   the partition function on a four-dimensional lattice with time
   coordinate $x_0/a=1,2,...,M$ and periodic boundary conditions.
 \item[b)] \[ \Tr\left(\sum_{b,c}
   \hat{\rmB}(\vec{x},\vec{y})^{\dagger}_{cb} \trans_{\qqbxy}^{t/a}
   \hat{\rmA}(\vec{x},\vec{y})_{bc} \trans_0^{M-t/a} \right) \,/\,
   \Tr(\trans_0^M) \] \\
   \[ = \langle \tr\left( A(x,y) U(y,y+t\hat{0})
   B(x+t\hat{0},y+t\hat{0})^{\dagger}
   U(x,x+t\hat{0})^{\dagger} \right) \rangle \equiv C_{{\rm A}{\rm B}} \] \\
   where $x_0=y_0$ and $\hat{\rmA},\hat{\rmB}$ are operators of the type of
   \eq{corrop}. This is the amplitude for the transition from the 
   state $A$ to the state $B$ over a time interval $t$.
\end{itemize}

We first inspect the kernel of $\trans_0$ in
\eq{kernelT0}. Identifying $W(\vec{z})$ with the time-like links
$U(z,0),\; z=(z_0,\vec{z})$, renaming
$\Phip(\vec{z})\rightarrow\Phi(z+a\hat{0})$ and
$\Up(\vec{z},k)\rightarrow U(z+a\hat{0},k)$ we see that \eq{tmsker},
excluding the terms quadratic in $\Phi$ and $\Phip$ separately,
becomes
\bes\label{timegauge}
 S(U,\Phi;U^{\prime W^{\dagger}},\Phi^{\prime W^{\dagger}}) & = &
 \exp\{-\Delta S(z_0,z_0+a)\} \, .
\ees
$\Delta S(z_0,z_0+a)$ are the terms in the action which couples
variables on the neighboring
timeslices $t=z_0$ and $t=z_0+a$. From this we see that the integration over
the time-like links in the path integral is equivalent to the
projection onto the gauge invariant sector of
$\Hspace$. \eq{timegauge} is also the reason why we called 
$\trans_{\rm temp}$ the transfer matrix in the temporal gauge.
The other factors in $K_{\rmT}^{\rm temp}$ contain the
pieces of the action depending only on the fields in the timeslices
and the trace is equivalent to the periodic boundary conditions in the
time direction.

The numerator of the left hand side of b) can be written as
\bes\label{reconumlhs}
 & & \prod_{z,\mu}\int_{\SUtwo}\rmd U(z,\mu) \cdot
 \prod_z\int_{\blackboardrrm^4}\rmd^4 \phi(z)\,
 \tr\left(A(x,y)B(x+t\hat{0},y+t\hat{0})^{\dagger}\right)\,\cdot \nonumber \\
 & & 2^{t/a}\, \tr\Ud(x+(t-a)\hat{0},0)\, \tr\Ud(x+(t-2a)\hat{0},0)\,
 \cdot\cdot\cdot\, \tr\Ud(x,0)\,\cdot \nonumber \\
 & & 2^{t/a}\, \tr U(y,0)\, \tr U(y+a\hat{0},0)\, \cdot\cdot\cdot
 \tr U(y+(t-a)\hat{0},0)\,\cdot \nonumber \\
 & & \exp(-S) \, .
\ees
Because the action and the measure are gauge invariant, 
for any observable $O$ one has
$\langle O^{\Lambda}\rangle=\langle O\rangle$, where 
$O^{\Lambda}$ is the gauge transformed observable. Gauge
transforming \eq{reconumlhs}, integrating over
$\prod_{w_0=x_0}^{x_0+t}\int\prod_{\vec{w}}\rmd\,\Lambda(w_0,\vec{w})$
and applying \eq{charrel} one sees that the product of traces in
\eq{reconumlhs} is glued into one trace, the right hand side of b).

\subsection{The static potential}

Using the spectral decomposition
$\trans_0=\sum_{n_0} |n_0\rangle \exp(-aE_n^{(0)}) \langle n_0|$
the partition function can be written as
\bes\label{partfct}
 Z & = & \sum_{n_0} \exp\left\{-MaE_n^{(0)}\right\} \, .
\ees
Inserting $\trans_{\qqbxy}=\sum_{n_{\qqb}} |n_{\qqb}\rangle
\exp(-aE_n^{(\qqb)}) \langle n_{\qqb}|$ in the quantum mechanical
expression for the correlation $C_{{\rm A}{\rm B}}$ we have:
\bes\label{corrAB}
 C_{{\rm A}{\rm B}} & = & \frac{1}{Z}\sum_{n_0,n_{\qqb}} \left(
 \sum_{b,c} \langle n_0|\hat{\rmB}^{\dagger}_{cb}|n_{\qqb} \rangle
 \langle n_{\qqb}|\hat{\rmA}_{bc}|n_0 \rangle \right) \cdot \\
 & & \exp\left\{-t(E_n^{(\qqb)}-E_n^{(0)})\right\} \cdot
 \exp\left\{-MaE_n^{(0)}\right\}
 \nonumber
\ees
In the limit of $Ma-t \gg 1/m_1^{(0)}$, where $m_1^{(0)}$ 
is the mass gap in the gauge invariant sector, and 
$t \gg (E_1^{(\qqb)}-E_0^{(\qqb)})^{-1}$ we have the asymptotic
behavior
\bes\label{corrABasympt}
 C_{{\rm A}{\rm B}} & \sim & \alpha_{\rm B}^*\,\alpha_{\rm A}\,
 \exp\left\{-t(E_0^{(\qqb)}-E_0^{(0)})\right\} \, ,
\ees
where $\alpha_{\rm A}=\langle 0_{\qqb}|\hat{\rmA}|0\rangle$ and the trace
over the color indices of the states and operators is
implicit. The quantity
\bes\label{statpot}
 V & = & E_0^{(\qqb)}-E_0^{(0)}
\ees
is called the static potential.

 \chapter{Updating algorithms \label{mcsu2higgs}}

In this appendix, we give a detailed description of the algorithms that
we used for the Monte Carlo simulation of the SU(2) Higgs model on the
lattice. Before describing the algorithms, we want to discuss the
generation of random numbers which are an essential part of the updating.

\section{Random Numbers \label{random}}

Because the updating process is stochastic, one needs to generate a
large number of independent random numbers. We use
the high-quality random number generator of M. {L\"uscher}
\cite{Luscher:ranfloat,Luscher:1994dy}, derived from an algorithm
originally proposed by Marsaglia and Zaman \cite{MarZam}.
It generates random floating point numbers distributed uniformly 
in the range $[0,1)$. From these flat random numbers we have to
construct random numbers with probability distributions needed 
in the updating of the gauge and Higgs field variables.

To be clear with the terminology, one speaks in general of random
numbers $\eta$ distributed in an interval $[a,b]\subset\blackboardrrm$
with probability density $p(\eta)$. The normalisation of the density
is such that $\int_a^b p(\eta)\,\rmd\eta=1$.
A uniform (or flat) distribution
of a random number $\eta$ in the interval $[a,b]$ corresponds to
$p(\eta)=1/(b-a)$. The probability distribution
$P(\eta)$ is related to the probability density by
\bes\label{probdistrib}
 \frac{\rmd P(\eta)}{\rmd\eta} & = & p(\eta) \,.
\ees
If one considers a transformation of variable $\tilde{\eta}=b(\eta)$
the probability density $\tilde{p}(\tilde{\eta})$ of the transformed
variable is related to $p(\eta)$ by the requirement
\bes\label{probvartrsf}
 \rmd P(\eta) & = & \rmd \tilde{P}(\tilde{\eta}) \,.
\ees

First of all, we consider the generation of random numbers
$\tilde{\eta}$ distributed according to the 
Gaussian density
\bes\label{gaussian}
 \tilde{p}(\tilde{\eta}) & = &
 \frac{1}{\sqrt{\pi}}\rme^{-\tilde{\eta}^2} \,.
\ees
It is better to consider the generation of 
a couple $(\tilde{\eta}_1,\tilde{\eta}_2)$ of independent Gaussian random
numbers. Independent means that
\bes\label{indepran}
 \rmd \tilde{P}(\tilde{\eta}_1,\tilde{\eta}_2) & = &
 \tilde{p}(\tilde{\eta}_1)\rmd\tilde{\eta}_1\,
 \tilde{p}(\tilde{\eta}_2)\rmd\tilde{\eta}_2 \,.
\ees
With the variable transformations $\tilde{\eta}_1=\rho\cos\theta$ and
$\tilde{\eta}_2=\rho\sin\theta$, where $\rho\in[0,\infty)$ and
$\theta\in[0,2\pi)$ we can rewrite \eq{indepran} using
\eq{probvartrsf} as
\bes
 \rmd P(\rho,\theta) & = & 
 \rmd \tilde{P}(\tilde{\eta}_1,\tilde{\eta}_2) \;=\;
 \frac{1}{\pi}\rho\rme^{-\rho^2}\rmd\rho\rmd\theta \nonumber \\
 & = &
 \rmd\left(1-\rme^{-\rho^2}\right)\,\rmd\left(\frac{\theta}{2\pi}\right) \,.
\ees
This is equivalent to the generation of a couple $(\eta_1,\eta_2)$ of
independent flat random numbers in the range $[0,1)$ together with the
transformations $\eta_1=1-\exp(-\rho^2)$ and
$\eta_2=\theta/(2\pi)$. The final result is then
\bes\label{gaussian2}
 \tilde{\eta}_1 & = & \sqrt{-\ln(1-\eta_1)}\cos(2\pi\eta_2) \quad
 \mbox{and} \nonumber \\
 \tilde{\eta}_2 & = & \sqrt{-\ln(1-\eta_1)}\sin(2\pi\eta_2) \,.
\ees
Gaussian random numbers are needed in the updating of the Higgs field
variables, see \sect{hbhiggs}.

Secondly, we consider the generation of random numbers distributed 
in $[0,\infty)$ according to the density
\bes\label{ranhblink}
 p(y) & = & \frac{2}{\sqrt{\pi}}\sqrt{y}\exp(-y)\Theta(y) \,,
\ees 
where
\bes\label{heaviside}
 \Theta(x-x_0) & = & \left\{\bea{ll} 1 & \mbox{for}\quad x>x_0 \\ 
                           0 & \mbox{for}\quad x<x_0 \ea\right.
\ees
is the Heaviside function.
This can be done by generating a number
$a\in[0,\infty)$ according to the Gaussian density
\bes\label{vara}
 p(a) & = & \frac{2}{\sqrt{\pi}}\rme^{-a^2}\Theta(a)
\ees
and independently a number $b\in[0,\infty)$ with the exponential density
\bes\label{varb}
 p(b) & = & \rme^{-b}\Theta(b) \,. 
\ees
To generate $a$ one uses \eq{gaussian2} replacing $2\pi\eta_2$ with 
$\pi\eta_2/2$ as the argument for the trigonometric functions.
To generate $b$ one simply takes a flat distributed random number
$\eta$ in the range $[0,1)$ and sets $b=-\ln(1-\eta)$.
We define the variables
\bes\label{vary}
 \bar{y}\;=\;a & \mbox{and} & y\;=\;a^2+b 
\ees
with ranges $y\in[0,\infty)$ and $\bar{y}\in[0,\sqrt{y})$. From
\eq{probvartrsf} we get the relation
\bes
 \rmd P(\bar{y},y) & = & \rmd P(a,b) \;=\;
 \frac{2}{\sqrt{\pi}}\rme^{-y}\Theta(\bar{y})\Theta(\sqrt{y}-\bar{y}) 
 \rmd\bar{y}\rmd y \,,
\ees
which means that the variables $\bar{y}$ and $y$ are distributed with
probability density
$p(\bar{y},y)=2/\sqrt{\pi}\exp(-y)\Theta(\bar{y})\Theta(\sqrt{y}-\bar{y})$.
The probability density for $y$ alone is obtained by integrating
$p(\bar{y},y)$ over all possible values of $\bar{y}$:
\bes\label{distry2}
 p(y) & = & \int_0^{\sqrt{y}} 
 \frac{2}{\sqrt{\pi}}\rme^{-y}\Theta(y)\rmd\bar{y} \;=\;
 \frac{2}{\sqrt{\pi}}\sqrt{y}\rme^{-y}\Theta(y) \,.
\ees
The random numbers $y$ from \eq{vary}
are needed in the updating of the gauge link 
variables, see \sect{hblink}.

Now, we are ready to describe the different parts of the HOR algorithm
introduced in \sect{simulation}.

\section{Updating of the gauge field}

To discuss the updating of the link variable $U(x,\mu)$, 
we split the action \eq{action3} into
\bes
 S & = & -\frac{\beta}{2}\tr\{U(x,\mu)W^{\dagger}(x,\mu)\} +
 \nonumber
 \\ & & \mbox{terms independent of $U(x,\mu)$} \, , \\
 W(x,\mu) & = & V(x,\mu) +
 \frac{2\kappa}{\beta}\vp(x)\vp^{\dagger}(x+a\muh) \, , \label{linkaction}
\ees
where $V(x,\mu)$ is the sum of the products of links over the six
``staples'' around the link $U(x,\mu)$
\bes\label{staples}
 V(x,\mu) & = & \sum_{\nu\neq\mu}
 \{U(x,\nu)U(x+a\nuh,\mu)U^{\dagger}(x+a\muh,\nu) + \nonumber \\
 & & \quad U^{\dagger}(x-a\nuh,\nu)U(x-a\nuh,\mu)U(x-a\nuh+a\muh,\nu)\} \, .
\ees
We denote the part of the action depending on $U(x,\mu)$ in
\eq{linkaction} by $S(U(x,\mu))$.

\subsection{Heatbath \label{hblink}}

The heatbath algorithm for the gauge field in
the SU(2) Higgs model is a simple modification of the algorithm
for the pure SU(2) gauge theory
\cite{Fabricius:1984wp,Kennedy:1985nu,Weisz:puresu2}.
We drop the arguments $(x,\mu)$ of the link to be updated.
The new link $\Up$ is chosen independently of $U$ according
to the Boltzmann distribution
\bes\label{gheatbath1}
 \rmd P(\Up) & \sim & 
 \rme^{\frac{\beta}{2}\tr(\Up W^{\dagger})}\,\rmd \Up \,.
\ees
The matrix $W$ defined in \eq{linkaction} can be written as
$W=\sqrt{\det W}\,\hat{W}$, where $\hat{W}\in\SUtwo$.
Using the invariance of the Haar measure we obtain from \eq{gheatbath1}
\bes\label{gheatbath2} 
 \rmd P(U^{\prime}\hat{W}) & \sim &
 \rme^{\frac{1}{2}\rho\tr(U^{\prime})}\,\rmd U^{\prime} \,,
\ees
where $\rho=\beta\sqrt{\det W}$.
Writing $\Up$ in the quaternionic representation
\bes\label{quaternions}
 U^{\prime} = a_0+ia_j\tau_j \quad,\quad a_{\mu}\in\blackboardrrm
 \quad,\quad a^2 = a_{\mu}a_{\mu} = 1 \,,
\ees
where $\tau_j\;(j=1,2,3)$ are the Pauli matrices, the Haar measure for
SU(2) takes the form \cite{MontMuen}
$\rmd\Up=\frac{1}{\pi^2}\delta(a^2-1)\rmd^4a$ and
\eq{gheatbath2} becomes
\bes\label{gheatbath3} 
 \rmd P(U^{\prime}\hat{W}) & \sim &
 \frac{1}{\pi^2}\delta(a^2-1)\rme^{\rho a_0}\,da_0d^3a \,.
\ees
The heatbath algorithm consists then in the following:
\begin{enumerate}
\item Generation of $a_{\mu}$ according to the distribution
\bes\label{fabhaan}
 \rmd P(a_{\mu}) & \sim &
 \sqrt{1-a_0^2}\rme^{\rho a_0}\rmd a_0\,\rmd^3n\delta(n^2-1) \,,  
\ees
where $a_j=n_j\sqrt{1-a_0^2}$. 
This is done using the method described in
\cite{Weisz:puresu2} which is a slight variation of that of Fabricius
and Haan \cite{Fabricius:1984wp}.
In order to generate $a_0\in[-1,1]$ with the distribution
\bes\label{distra0}
 \rmd P(a_0) & \sim & \sqrt{1-a_0^2}\rme^{\rho a_0}\rmd a_0
\ees
we perform the change of variable
\bes\label{yhblink}
 y & = &\rho(1-a_0)\in[0,2\rho] \,.
\ees 
According to \eq{probvartrsf} we rewrite \eq{distra0} as
\bes\label{distry}
 \rmd P(y) & \sim &
 (2-\frac{y}{\rho})^{1/2}\sqrt{y}\rme^{-y}\rmd y \,.
\ees
The generation of random numbers $y$ according to \eq{distry} cannot
be done exactly. Instead, one generates $y$ according to the
probability density \eq{ranhblink} as explained in \sect{random}.
An accept-reject step takes into account
the omitted factor $(2-\frac{y}{\rho})^{1/2}$.
A flat random number $\eta\in[0,1)$ is generated: if
\bes\label{accrej}
 2\eta^2 & \le & 2-\frac{y}{\rho}
\ees
the change $a_0=1-y/\rho$ is accepted. We notice that the values
$y>2\rho$ are automatically rejected in \eq{accrej}.
The components $n_j$ of the vector $n$ in \eq{fabhaan} are
uniformly distributed on the surface of the unit sphere in three
dimensions. The generation of $n_j$ requires two flat distributed
random numbers $\eta_1$ and $\eta_2$ in the interval [0,1):
\bes
 n_1 & = & 1-2\eta_1 \,, \\
 n_2 & = & \sqrt{1-n_1^2}\cos(2\pi\eta_2) \,, \\
 n_3 & = & \sqrt{1-n_1^2}\sin(2\pi\eta_2) \,.
\ees

\item If \eq{accrej} is fulfilled, $U$ is substituted by 
\bes\label{hbu} 
 U^{\prime}\hat{W}\;=\;\frac{1}{\sqrt{\det W}}U^{\prime}W \,.
\ees
\end{enumerate}
Simulating the SU(2) Higgs model on a $12^4$ lattice with 
parameters $\beta=2.0$, $\kappa=0.25$ and $\lambda=0.5$ 
(this point is in the confinement ``phase'')
the change \eq{hbu} is accepted in 95\% of the cases.

\subsection{Overrelaxation \label{orlink}}

The new link $U$ proposed is 
\bes\label{goverrel} 
 U^{\prime} & = &
 \hat{W}U^{\dagger}\hat{W}\;=\;\frac{1}{\det W}W\Ud W 
\ees
This change is microcanonical,
i.e. $S(U)=-\rho/2\tr(U\hat{W}^{\dagger})=S(\Up)$ because of the
reality of the trace of SU(2) elements. Since the inverse relation
of \eq{goverrel} is $U=\hat{W}\Upd\hat{W}$, \eq{goverrel} is a
reflection of $U$ and $\Up$ and therefore the factors $p_{\rm C}$
cancel in the acceptance probability \eq{metropolis}. The change
\eq{goverrel} is always accepted.

\section{Updating of the Higgs field}

For the updating of the Higgs field we use the algorithms proposed in
\cite{Bunk:1994xs}. In particular the over-relaxation algorithm has
been proved to be very efficient for the reduction of
autocorrelations.

The part of the SU(2) Higgs model action \eq{action3} 
depending on the Higgs field reads
\bes\label{phiaction1} 
 S_{\phi} &  = & \sum_x\left\{
\frac{1}{2}\tr(\vp^{\dagger}(x)\vp(x)) + \lambda\left[
\frac{1}{2}\tr(\vp^{\dagger}(x)\vp(x))-1 \right]^2 \right. \nonumber\\
& & \left. -\kappa\sum_{\mu}\tr(\vp^{\dagger}(x)U(x,\mu)\vp(x+\hat{\mu}))
\right\} \,.
\ees
Combining \eq{higgsnotation1} and \eq{higgsnotation2}
the $2\times2$ matrix $\vp$ can be written in terms of the four real
components $\phi_i(x)\;(i=1,2,3,4)$ as
\bes\label{higgsnotation12} 
 \vp(x) & = & \left( \bea{cc} \phi_3(x)-i\phi_4(x) &
 \phi_1(x)+i\phi_2(x) \\ -\phi_1(x)+i\phi_2(x) & \phi_3(x)+i\phi_4(x)
 \ea \right) \,.
\ees
If we do the same for the matrix
\bes\label{matrixB}
 B(x) & = & \kappa\sum_{\mu}\left[ 
 U(x,\mu)\vp(x+\hat{\mu})+U^{\dagger}(x-\hat{\mu},\mu)\vp(x-\hat{\mu})
 \right] \nonumber \\ 
 & = & \left( \bea{cc} b_3(x)-ib_4(x) & b_1(x)+ib_2(x) \\ 
 -b_1(x)+ib_2(x) & b_3(x)+ib_4(x) \ea \right) \quad
 b_i\in\blackboardrrm \,,
\ees
we can write the action \eq{phiaction1} in the four component notation
of reference \cite{Bunk:1994xs}
\bes\label{phiaction2}
 S_{\phi} = \sum_x\left\{ (\phi(x)-b(x))^2 +
 \lambda\left[ \phi^2(x)-1 \right]^2 - b^2(x) \right\} \,,
\ees
where the squares mean the scalar product in $\blackboardrrm^4$.

\subsection{Heatbath \label{hbhiggs}}

We drop the index $i$ and the argument $x$ of the Higgs field variable
$\phi_i(x)\;(i=1,2,3,4)$ to be updated.
A new Higgs variable $\phi^{\prime}$ has to be generated according to
the distribution
\bes\label{phipotential} 
 \rmd P(\phi^{\prime}) \;\sim\;
 \rme^{-V(\phi^{\prime})}\,\rmd^4\phi^{\prime} \quad,\quad 
 V(\phi) = (\phi-b)^2+\lambda(\phi^2-1)^2 \,.
\ees
This is best achieved by introducing a free parameter $\alpha$ which
parametrises the splitting of the potential $V(\phi)$ in a quadratic and in
a quartic part
\bes\label{splitting} 
 V(\phi) & = & \alpha(\phi-\alpha^{-1}b)^2 +
 \lambda(\phi^2-v_{\alpha}^2)^2 - c_{\alpha} \,,
\ees
where 
\bes 
 v_{\alpha}^2 & = & 1+\frac{\alpha-1}{2\lambda} \,, \label{valpha} \\
 c_{\alpha} & = & \lambda(v_{\alpha}^4-1)+(\alpha^{-1}-1)b^2 \,. 
\ees
A trial Higgs variable is generated according to the Gaussian
density (see \eq{gaussian2})
\bes\label{gaussran}
 p_{\rm trial}(\phi^{\prime}) & = & 
 \left(\frac{\alpha}{\pi}\right)^2
 \rme^{-\alpha(\phi^{\prime}-\alpha^{-1}b)^2} 
\ees
and is accepted with probability
\bes\label{probhbphi}
 w_{\rm acpt}(\phi^{\prime}) & = &
 \rme^{-\lambda(\phi^{\prime 2}-v_{\alpha}^2)^2} \,.
\ees
The parameter $\alpha$ is chosen so that it maximises the acceptance
rate defined as
\bes
 A(\alpha) & = & \int \rmd^4\phi\, p_{\rm trial}(\phi)w_{\rm
   acpt}(\phi) \,,
\ees
which can be rewritten using \eq{splitting} as
\bes\label{acchbphi}
 A(\alpha) & = & \alpha^2\rme^{-c_{\alpha}}A(1) \,.
\ees
Differentiating with respect to $\alpha$ yields the cubic equation 
\bes\label{acccubic}
 f(\alpha)\equiv \alpha^3 - (1-2\lambda)\alpha^2 - 4\lambda\alpha & =
 & 2\lambda b^2 \,.
\ees
The exact solution of \eq{acccubic} is inconvenient because the
dependence on $b^2$ forces to solve it separately for each update.
We use the approximate solution proposed in \cite{Bunk:su2higgs}.
The function $f(\alpha)$ in \eq{acccubic} vanishes at
\bes\label{alpha0}
 \alpha_0 & = & 
 \frac{1}{2}-\lambda+\left[(\frac{1}{2}-\lambda)^2+4\lambda\right]^{1/2} \,.
\ees
Expanding $f(\alpha)$ in powers of $(\alpha-\alpha_0)$ and omitting
the cubic term, \eq{acccubic} becomes
\bes\label{alpha2}
 (\alpha_0^2+4\lambda)(\alpha-\alpha_0) +
 (6\alpha_0+4\lambda-2)\frac{1}{2}(\alpha-\alpha_0)^2 & = & 2\lambda
 b^2 \,.
\ees
The positive solution of \eq{alpha2} is
\bes\label{alphasol}
 \alpha & = & h_0 + [h_1+h_2\,b^2]^{1/2} \,,
\ees
with
\bes
 h_0 & = & \alpha_0-\frac{\alpha_0^2+4\lambda}{6\alpha_0+4\lambda-2} \,,\\
 h_1 & = & \left(\frac{\alpha_0^2+4\lambda}{6\alpha_0+4\lambda-2}\right)^2
 \,,\\
 h_2 & = & \frac{4\lambda}{6\alpha_0+4\lambda-2} \,.
\ees
The constants $\alpha_0$ and $h_i\;(i=0,1,2)$ can be computed in
advance because they do not depend on $b$. For the update of $\phi$ it
is only necessary to determine $b$, find $\alpha$ according to
\eq{alphasol} and the corresponding $v_{\alpha}^2$ from \eq{valpha}.
Simulating the SU(2) Higgs model for relatively large value of
$\lambda$, the choice $\alpha=1$, corresponding to a naive splitting of
the potential $V(\phi)$ in \eq{splitting}, would lead to a low
acceptance $A(1)$ in \eq{acchbphi}. Using \eq{alphasol} we get values
$A=70\%$ in the confinement ``phase'' ($\beta=2.0$, $\kappa=0.25$,
$\lambda=0.5$) and $A=59\%$ in the Higgs ``phase'' ($\beta=2.3$,
$\kappa=0.32$, $\lambda=1.0$), both the simulations performed on a
$12^4$ lattice.

\subsection{Overrelaxation \label{orhiggs}}

The following change of the Higgs variable is proposed \cite{Bunk:1994xs}:
\bes\label{bunkoverrel}
 \phi & \longrightarrow & \phi^{\prime}=2\alpha^{-1}b-\phi \,.
\ees
This change is not microcanonical, but it is a reflection of $\phi$ and
$\phi^{\prime}$, since the inverse relation of \eq{bunkoverrel} is 
$\phi=2\alpha^{-1}b-\phi^{\prime}$. According to \eq{metropolis}
the new Higgs variable is accepted with a probability
\bes 
 p_{\rm A} & = & \min\left\{ 1,\exp\left[V(\phi)-V(\phi^{\prime})\right]
\right\} \nonumber\\ & = & \min\left\{
1,\exp\left[\lambda(\phi^2-\phi^{\prime 2}) 
(\phi^2+\phi^{\prime 2}-2v_{\alpha}^2)\right] \right\} \,.
\ees
For the acceptance $A$ of the change \eq{bunkoverrel} we get values
$A=79\%$ in the confinement ``phase'' ($\beta=2.0$, $\kappa=0.25$,
$\lambda=0.5$) and $A=68\%$ in the Higgs ``phase'' ($\beta=2.3$,
$\kappa=0.32$, $\lambda=1.0$), both the simulations performed on a
$12^4$ lattice.

 \chapter{One-link integral and approximation of the modified 
 Bessel functions \label{bessel}}

In \sect{onelink}, we presented a method for reducing the
statistical variance of correlation functions involving static
charges which are represented in the path integral formalism 
by straight time-like Wilson lines. The
time-like links can be replaced (with some restriction) by their 
expectation value (called one-link integral) 
in the fixed configuration of the other field variables.

First of all, we present the derivation of the result
\eq{onelinkintegral}. We omit writing the $(x,0)$ dependence of $U$
and $W$, for the rest the notation is the same as in \sect{onelink}. 
We denote by $\hat{W}$ the projection
$W/\sqrt{\det(W)}$ of $W$ into SU(2). The one-link
integral can be written after a change of integration variable as
\bes\label{olint}
 \overline{U} & = & \frac{\int\rmd U\, 
 U\,\exp\{(\rho/2)\tr(U)\}}{\int\rmd U\, 
 \exp\{(\rho/2)\tr(U)\}}\,\hat{W} \, ,
\ees
where $\rho=\beta\sqrt{\det(W)}$.
To solve analytically these integrals, we use 
the following character expansion for SU(2) \cite{MontMuen}
\bes\label{charexp}
 \exp\{(\rho/2)\tr(U)\} & = & \frac{2}{\rho}\sum_{\nu}
 (2\nu+1)\,I_{2\nu+1}(\rho)\,\chi^{(\nu)}(U) \, ,
\ees
where $\nu=0,1/2,1,3/2,...$ labels the irreducible representations of
SU(2) with characters $\chi^{(\nu)}$ and $I_n$ is the modified Bessel
function of integer order $n$. With help of \eq{charexp} and 
the orthogonality relations for the characters, it is easy to show that
$\int\rmd U\, \exp\{(\rho/2)\tr(U)\}=(2/\rho)I_1(\rho)$. The integral
in the numerator of \eq{olint}
\bes
 F & = & \int\rmd U\, U\,\exp\{(\rho/2)\tr(U)\} \,
\ees
defines a linear transformation in $\C^2$ with the property that
\bes
 A\,F & = & F\,A\,, \quad A\in\SUtwo \, .
\ees
From the Schur's Lemma of representation theory, it follows that F is
equal to a complex constant times the identity matrix $\one$. 
Writing the SU(2) link $U$ in the quaternionic representation
$U=1/2\tr(U)\one+i\vec{u}\cdot\vec{\tau}$, where
$\tau_i\;(i=1,2,3)$ are the Pauli matrices and $u_i\in\blackboardrrm$,
we see that only the component $1/2\tr(U)\one$ gives a contribution to
$F$. By noting that $\tr(U)\equiv\chi^{(1/2)}(U)$ and using again
\eq{charexp} together with the
orthogonality property of the characters, we conclude that $F=(2/\rho)
I_2(\rho)\one$, giving the desired result
\bes\label{beres}
 \overline{U} & = & \frac{I_2(\rho)}{I_1(\rho)}\,\hat{W} \,.
\ees
%%%%%%%%%%%%%%%%%%%%%%%%%%%%%FIGURE%%%%%%%%%%%%%%%%%%%%%%%%%%%%%%%%%%%
\begin{figure}[tb]
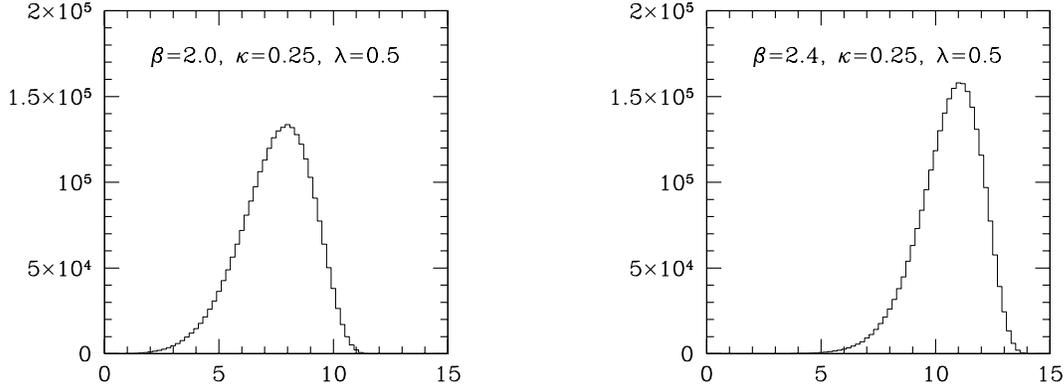

\hspace{0cm}
\vspace{-1.0cm}
\parbox{6.2cm}{
\centerline{\epsfig{file=plots/rhob20.epsi,width=6.0cm}}}
\hfill
\parbox{6.2cm}{
\centerline{\epsfig{file=plots/rhob24.epsi,width=6.0cm}}}
\vspace{0.5cm}
\caption{Here, we show a histogram plot for the distribution of
  $\rho=\beta\sqrt{\det(W(x,0)))}$ measured on each point $x$ of
  a $6^4$ lattice for 2000 field configurations. The simulations are
  performed  for two representative parameter sets.
  \label{rhodist}}
\end{figure}
%%%%%%%%%%%%%%%%%%%%%%%%%%%%%%%%%%%%%%%%%%%%%%%%%%%%%%%%%%%%%%%%%%%%%%

The modified Bessel functions of integer order $I_n(x)$ have the
following asymptotic behavior for large $x$
\bes\label{bessellarge}
 I_n(x) & = & \frac{\rme^x}{\sqrt{2\pi x}}\,\left[
 1+O\left(\frac{1}{x}\right) \right] \, .
\ees
For small $x$ we get from the Taylor expansion of $I_n(x)$
\bes\label{besselsmall}
 I_n(x) & = & \frac{1}{n!}\left(\frac{x}{2}\right)^n\,\left[
 1+O\left(x^2\right) \right] \, .
\ees
In order to choose a good numerical approximation for the ratio
$R(\rho)=I_2(\rho)/I_1(\rho)$
we measured the distribution
of the argument $\rho$ in Monte Carlo simulations on a $6^4$ lattice
for two values $\beta=2.0,\;2.4$. The results are shown in
\fig{rhodist}.
For $\rho$ lying in the interval $[0,13]$, we use two different
Chebyshev approximations \cite{NumRec}, 
for the function $R(\rho)/\rho$ (this choice is motivated by
\eq{besselsmall}) in the range $\rho\in[0,5]$ 
and for $R(\rho)$ in the range $\rho\in[5,13]$.
The Chebyshev approximation of a function $f(x)$ in the interval 
$x\in[a,b]$ is defined as
\bes\label{chebyappr}
 f(x) & \approx & \sum_{k=0}^m c_k\,T_k(y) \,, \quad
 y\equiv\frac{x-\frac{1}{2}(b+a)}{\frac{1}{2}(b-a)}\in[-1,1] \, ,
\ees
where $T_k(y)=\cos(k\,\arccos(y))$ is the Chebyshev polynomial of
degree k. The Chebyshev coefficients $c_k,\;k=1,2,...,m$ 
are rapidly decreasing with $k$ so that the error of the
approximation is dominated by $c_{m+1}T_{m+1}(y)$ and is
bounded by $|c_{m+1}|$. We computed the
coefficients $c_k$ using the program MAPLE: the value of $m$ is
chosen such that we reach a precision of $10^{-6}$
in the considered ranges. We need $m=10$ for the approximation of
$R(\rho)/\rho$ in the range $\rho\in[0,5]$
and $m=8$ for the approximation of $R(\rho)$ in the range $\rho\in[5,13]$.
The Chebyshev polynomials satisfy the recurrence relation
$T_{k+1}(y)=2yT_k(y)-T_{k-1}(y),\; k\ge1$. An elegant and efficient
way to evaluate the sum in \eq{chebyappr} is the
Clenshaw's recurrence formula described in reference \cite{NumRec}.

When $\rho>13$, which is rarely the case as can be seen from
\fig{rhodist}, we write
\bes\label{ratiopolyapprox}
  R(\rho) & = & \frac{\sqrt{\rho}\,\rme^{-\rho}\,I_0(\rho) -
    2\,\sqrt{\rho}\,\rme^{-\rho}\,I_1(\rho)/\rho} 
    {\sqrt{\rho}\,\rme^{-\rho}\,I_1(\rho)}
\ees
and use the polynomial approximations for
$\sqrt{\rho}\,\rme^{-\rho}\,I_n(\rho)\;(n=0,1)$
in negative powers of $t=\rho/3.75$
given in reference \cite{AbraSteg} for the range $\rho>=3.75$
(these approximations are motivated by \eq{bessellarge}).
For $\sqrt{\rho}\,\rme^{-\rho}\,I_1(\rho)$ we take the polynomial
approximation up to the power $t^{-7}$, which guarantees a
precision of $10^{-6}$ for $\rho>13$. Because of the extra power $\rho^{-1}$
in the numerator of \eq{ratiopolyapprox}, we take
the polynomial approximation for
$\sqrt{\rho}\,\rme^{-\rho}\,I_0(\rho)$ up to the power $t^{-8}$, which
also guarantees a precision of $10^{-6}$.

 \chapter{Parallelisation \label{parallel}}

In this appendix, we describe the parallelisation of the program for
the simulation of the SU(2) Higgs model. Geometrically, the
parallelisation consists in the partition of the lattice
on different processor elements (PEs).
This is necessary for two reasons: the simulation is faster and
the memory space
needed on a $32^4$ lattice to store the field variables does not fit
into the $128\,\MB$ memory of one PE. 
We adopt a two-dimensional partitioning of the lattice. The $xy$-plane
is distributed among the PEs and the $t$ and $z$ directions,
for fixed $x$ and $y$ coordinates, are entirely contained on each PE. 
This choice was motivated by the
measurement of the correlation function for the static potential:
apart from smearing, it
can be performed in the $tz$-plane on each PE independently.

The programming aspects of the parallelisation are related to the
communication between PEs.
The program is written in the language FORTRAN 90 for a CRAY T3E
machine.  For the communication between PEs we used the CRAY SHMEM
(logically shared, distributed memory access)
routines. In order that a local PE can pass (receive) a data object to
(from) a remote PE there must be a known relationship between 
the local and remote address of the data object. 
This is realised by declaring the data object with the 
{\em save} attribute or in a {\em common} block.
On CRAY systems, this means that the address of the data object
in the memory of the local and remote PE is the same.
These data objects are called {\em symmetric}.

In this appendix, we set for simplicity $a=1$. We {\em emphasise}
all the names of variables, FORTRAN 90 statements and routines 
that we used in the program.
%%%%%%%%%%%%%%%%%%%%%%%%%%%%%FIGURE%%%%%%%%%%%%%%%%%%%%%%%%%%%%%%%%%%%
\begin{figure}[tb]
\hspace{0cm}
\vspace{-1.0cm}
\centerline{\epsfig{file=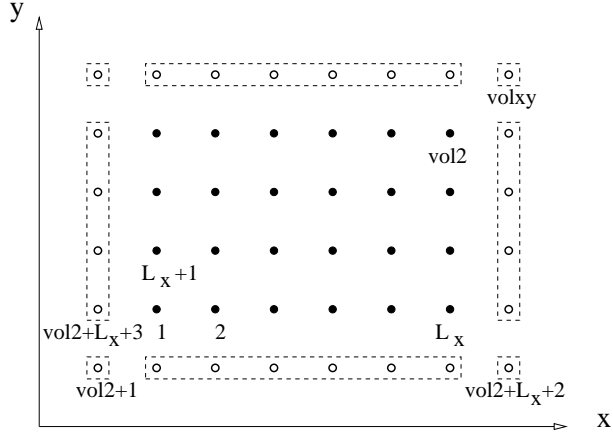,width=8cm}}
\vspace{0.5cm}
\caption{Here, we show the $xy$-partition on each PE. The filled
  circles are the local points, i.e. the points on which the updating 
  is performed. The empty circles are the remote points, i.e. the
  boundary points copied from the neighboring PEs. 
  The dashed lines enclose points belonging to the same neighboring PE.
  The labels under the points are the values of the lexicographic
  index $xs$.
\label{f_geometry}}
\end{figure}
%%%%%%%%%%%%%%%%%%%%%%%%%%%%%%%%%%%%%%%%%%%%%%%%%%%%%%%%%%%%%%%%%%%%%%

\section{Geometry of the two-dimensional partition \label{geometry}}

The lattice points in the $xy$-plane are partitioned among 
$N_x\times N_y$ PEs. For fixed $x$ and $y$ coordinates, the points in
$t$ and $z$ directions are entirely contained on the PE.
Therefore, each PE works on a sub-lattice with $L_x\times L_y\times
L\times T$ points, 
where $L_x=L/N_x$ and $L_y=L/N_y$. During the updating of the
fields on a sub-lattice, we need field variables from
sub-lattices having a common boundary in the $xy$-plane.
Therefore, we store a
copy of these boundary field variables. The number of points on each
PE is then $(L_x+2)\times(L_y+2)\times L\times T$.

The PEs are labelled with a number $npe$ that can be queried with the
SHMEM routine $my\_pe()$ and goes from 0 to $N_xN_y-1$. To give an
``identity'' to the PEs (a relation with the geometrical sub-lattice they
work on), we map the number $npe$ into a ``processor lattice'' in the
$xy$-plane by introducing the arrays 
\bes\label{pearray}
 peup(npe,i) \quad \mbox{and} \quad pedn(npe,i)\,, \quad (0\le npe\le
 N_xN_y-1;i=1,2)\,,
\ees
where $i=1(2)$ labels the $x$($y$) direction. The value of the array
element $peup(npe,i)$
($pedn(npe,i)$) is the number of the nearest-neighbor PE in positive
(negative) $i$ direction of the PE labelled by $npe$. 
This is also the way periodic boundary conditions are implemented
in the $x$ and $y$ directions.

The $xy$-partition on a
PE is represented in \fig{f_geometry}. We decided to label the points
on a sub-lattice with three coordinates
\bes\label{subpoints}
  (z,t,xs)\,, \quad 1\le z\le L\,, \quad 1\le t\le T\,, \quad
  1\le xs\le volxy\,.
\ees
The index $xs$ is the lexicographic index in the $xy$-plane and is
constructed labelling the local points (filled
circles in \fig{f_geometry}),
i.e. the points on which the updating is performed,
from 1 to $vol2=L_xL_y$. 
The remote points (empty
circles), i.e. the boundary points
copied from neighboring PEs, are labelled
from $vol2+1$ to $volxy=(L_x+2)(L_y+2)$.
The order of the labelling is shown in
\fig{f_geometry}: the $x$ coordinate runs fastest.

The space-time movements to the nearest neighbor points 
on the sub-lattice (``hopping'') are given by the arrays
\bes
 & & zup(z) \quad \mbox{and} \quad zdn(z)\,, 
 \quad (1\le z\le L)\,, \label{zarray} \\
 & & tup(t) \quad \mbox{and} \quad tdn(t)\,,
 \quad (1\le t\le T)\,, \label{tarray} \\  
 & & iup(xs,i) \quad \mbox{and} \quad idn(xs,i)\,, 
 \quad (1\le xs\le volxy;i=1,2) \,, \label{xyarray}
\ees
where $i=1(2)$ again labels the $x$($y$) direction. The ending ``$up$''
(``$dn$'') refers to the positive (negative) direction.

For each local point with index $1\le xs\le vol2$,
we define an array $boundary(xs)$ whose components are of
the derived data type \\[0.2cm]
{\em
   type, public :: boundary\_type\\
   \hspace*{.5cm}logical :: tf\\
   \hspace*{.5cm}integer :: ndir\\
   \hspace*{.5cm}integer, dimension(3) :: xtg\\
   \hspace*{.5cm}integer, dimension(3) :: pe\\
   end type boundary\_type}\\[0.2cm]
The component $tf$ is $.true.$ if the point $xs$ 
is on the boundary of the $xy$-partition
(default value is $.false.$). The component $ndir$ is the
number of neighboring PEs that have a copy of the point $xs$: we have
$ndir=3$ for the points in the corners of the $xy$-partition and
$ndir=1$ for the other boundary points (default value is 0).
The component $xtg(i)\;(i=1,2,3)$ (default value 0) is the
lexicographic index of the copy of the point $xs$
on the neighboring (target) PE with number $pe(i)$ (default value -1).

The arrays (\ref{pearray}), (\ref{zarray}), (\ref{tarray}) and
(\ref{xyarray})
together with $boundary(xs)$ are defined in the module {\em geometry}. 
Once the geometry is initialised, it will be used throughout 
the program without modification.
%%%%%%%%%%%%%%%%%%%%%%%%%%%%%FIGURE%%%%%%%%%%%%%%%%%%%%%%%%%%%%%%%%%%%
\begin{figure}[tb]
\hspace{0cm}
\vspace{-1.0cm}
\centerline{\epsfig{file=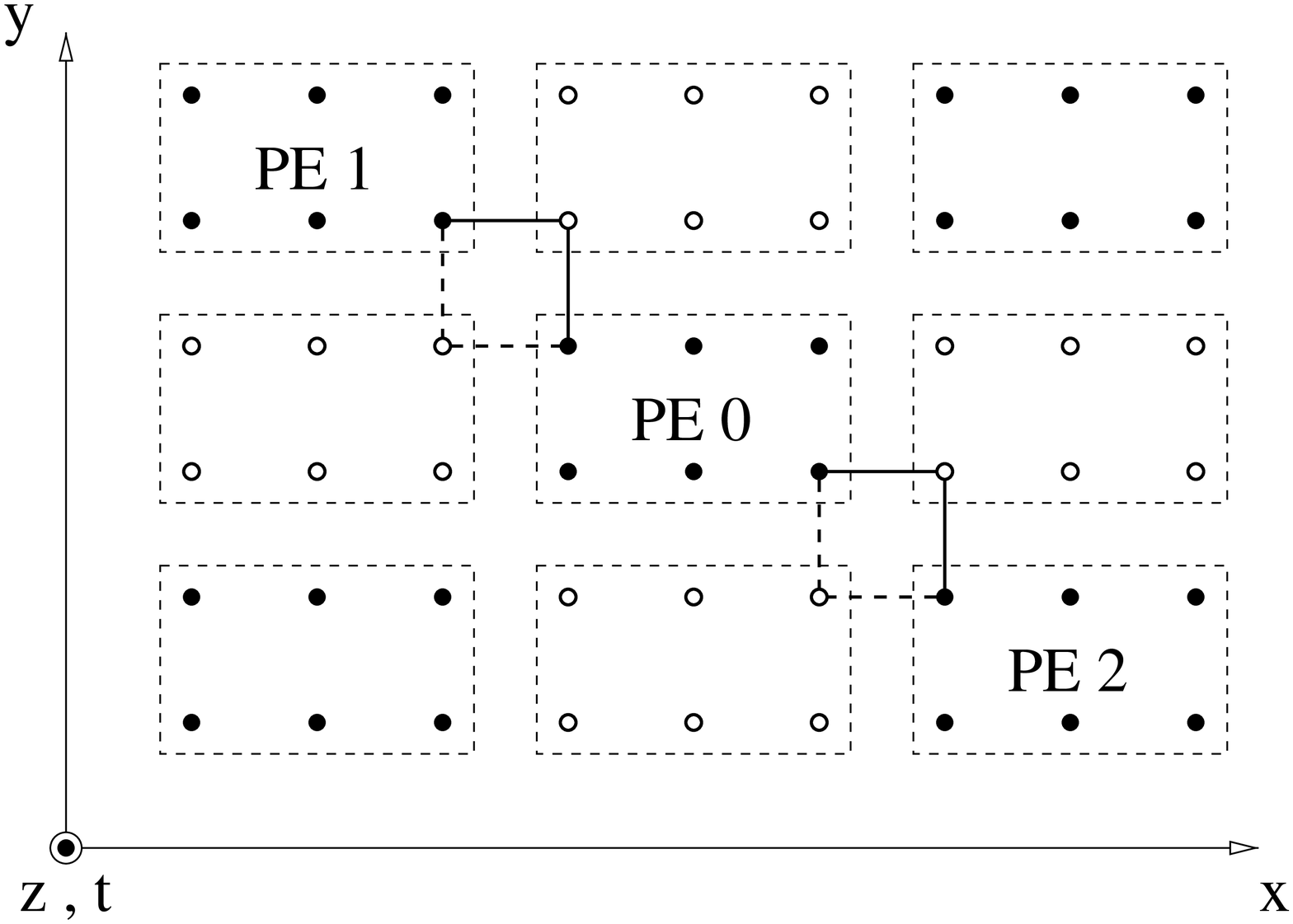,width=8cm}}
\vspace{0.5cm}
\caption{Here, we illustrate the
  updating of the gauge links. The filled circles represent
  points on which the links (full lines) are going to be updated.
  The empty circles represent points where the links (dashed lines) do not
  change because of the different parity of their PEs.
  The dashed boxes
  delimit the local points on the PEs. When PE0 updates the links
  $U(x,\mu)$ with $\mu=1$ ($\mu=2$) it needs links with $\mu=2$
  ($\mu=1$) from PE2 (PE1). If all PEs update links with the same
  direction $\mu$ and pass the links at the boundary of the
  $xy$-partition before changing the
  direction, there is no danger of using links from other PEs which
  are not up to date.
\label{f_updating}}
\end{figure}
%%%%%%%%%%%%%%%%%%%%%%%%%%%%%%%%%%%%%%%%%%%%%%%%%%%%%%%%%%%%%%%%%%%%%%

\section{Communication between PEs \label{communication}}

The updating of the gauge and Higgs field requires the exchange of
field variables between the PEs. This is the reason why we have a copy
of the field variables at the boundary of the
$xy$-partition belonging
to neighboring PEs. It is essential that these boundary
values are ``up to date'', i.e. they don't get
modified simultaneously on their local PEs. In order to avoid the
use of field variables that do not have the current value, we associate
to each PE a parity $par$ which can assume the values 0 or
1. According to the value of $par$, the updating is processed on points
belonging to one of the two time sections
\bes\label{timeboxes}
 1\le t\le \frac{T}{2} \quad (par=0) \quad \mbox{or} \quad 
 \frac{T}{2}+1\le t\le T \quad (par=1)
 \,.
\ees
Before the updating starts, the value of $par$ is initialised on the PEs.
The PE with index 0 has $par=0$ and the parity of two nearest-neighbor
PEs is different.
Before changing the value of $par$, the PEs pass the
field variables at the boundary of the $xy$-partition (in the time
section which has been updated).
This gives the correct algorithm for the
Higgs field, which has only nearest-neighbor interactions.
The updating of a gauge link on a point is more complicated since the
plaquettes containing the link extend to next-to-nearest-neighbor points.
When updating the gauge field in a given time section,
the PEs must all work on the links
$U(x,\mu)$ with the direction $\mu$ and they pass the boundary
links before changing the direction.
This procedure is necessary for two reasons. Firstly,
when updating time-like links, gauge links from nearest-neighbor PEs at a
different time coordinate are needed, but they are space-like. Secondly,
the updating of links in the $x$($y$) direction requires the use of
gauge links at the same time coordinate on next-to-nearest neighbor
PEs (which have also the same parity),
but they are directed in the $y$($x$) direction. This second
situation is illustrated in \fig{f_updating}.

The gauge field is represented by SU(2) matrices and the Higgs field
can be represented according to \eq{higgsnotation2} as a real constant
times a SU(2) matrix. We can parametrise both of these types of 
$2\times2$ matrices as
\bes\label{matrices2x2}
 \left(\bea{cc} \alpha & \beta \\ -\beta^* & \alpha^* \ea\right)\,,
 \quad
 \alpha=(a\_re)+i(a\_im),\beta=(b\_re)+i(b\_im)\in\C \,,
\ees
and we define for them the derived data type\\[0.2cm]
{\em
      type, public :: su2\\
      \hspace*{.5cm}sequence\\
      \hspace*{.5cm}real(kind=double)  ::  a\_re,a\_im,b\_re,b\_im\\
      end type su2}\\[0.2cm]
The attribute {\em sequence} means that the components of the data object
are stored in the sequence that they are declared.
The name of the data type is $su2$ but it can be used for a larger
class of matrices than only SU(2), as discussed above.
The $kind=double$ of the components means that they are double
precision real numbers with a storage size of $8\,\rmB$.
In the module $su2\_type$ we extend the intrinsic operators *,+,--
to act between data objects of the type $su2$ with the usual meaning of
matrix operations.
We also extend the assignment = such that
$A=c$ sets the components of a $su2$ data object $A$
corresponding to the identity matrix times the real number $c$.
We define new operators
between $su2$ data objects $.doth.$ and $.hdot.$, where
$A.doth.B$ and $A.hdot.B$ correspond
to the matrix multiplications $AB^{\dagger}$ and $A^{\dagger}B$
respectively. Moreover, we define the function $det$, which has a
$su2$ data object as argument and returns the value of the determinant
of the corresponding matrix.
There is one point concerning the precedence of the
operators which is worthwhile to mention. An extended intrinsic operator
maintains its precedence, a defined binary operator has the lowest
precedence. For example, the expressions
\bes\label{opprecedence}
 A.doth.B+C \quad \mbox{and} \quad (A.doth.B)+C
\ees
correspond to the matrix operations $A(B+C)^{\dagger}$ and
$AB^{\dagger}+C$ respectively, which are clearly different.

The gauge and Higgs field on the lattice are represented by arrays $u$
and $phi$ of data objects of the type $su2$ with indices
\bes\label{uphifields}
 u(z,t,xs,mu) \quad \mbox{and} \quad phi(z,t,xs) \,.
\ees
The ordering of the indices is chosen carefully.
The language FORTRAN stores
the array $A(i_1,i_2,...,i_n)$ in the computer memory by columns. This
means that the index $i_1$ is the fastest and the index $i_n$ the
slowest. When the array elements are used in the program
they are copied into the
memory cash of the machine. A better performance of the code is
reached when the sequence
of the indices $i_1,...,i_n$ in the definition of the array $A$ 
corresponds to the inverse sequence
of $do$ loops over the indices in the program (the loop over $i_n$ is
the outermost and the loop over $i_1$ the innermost). Why the sequence
of indices in (\ref{uphifields}) is the best for our purposes
will become clear soon.

The communication between PEs on a CRAY T3E machine is
implemented with the SHMEM routines $shmem\_put$ and $shmem\_get$.
The routine $shmem\_put$ has the following arguments:\\[0.2cm]
{\em
\hspace*{.5cm}shmem\_put(target,source,size,target {\rm PE}).}\\[0.2cm]
The first argument $target$ is the target array on a remote PE
to which we put the data.
Actually, what is passed in FORTRAN is the address
in memory of the first element of the array $target$. 
The array $target$ must be a {\em symmetric} array, i.e. its components
have the same address in memory on each PE.
The second argument $source$ is the array on the local PE from which
the data are transferred. The third argument $size$ is the number of
words w ($1\,{\rm w}\;=\;8\,\rmB\;=\;64\,{\rm bits}$)
that are transferred. The last
argument is the number of the remote PE to which the data are passed.
The arguments of the routine $shmem\_get$ are\\[0.2cm]
{\em
\hspace*{.5cm}shmem\_get(target,source,size,source {\rm PE}).}\\[0.2cm]
The interpretation is similar but the $target$ array, to which the
data are transferred, resides on the
local PE and the source array, from which we get the data, is
on a remote PE, whose number is given by the last argument. The array
$source$ must be {\em symmetric}.

As concerns the time duration of the communication between PEs, a
useful quantity to know is
the time interval between the start of the data transfer
and the availability of the data on the target PE, called data latency
$\tau_{\rm L}$. For the CRAY T3E we have
\bes\label{latency}
 \tau_{\rm L} & = & 0.3\times10^{-6}\,\sec \,.
\ees
In a simplified but rather realistic model,
we can parametrise the total time per byte
needed for transferring a block of data of size $P$ (in bytes B) as
\bes\label{transfertime}
 t_{\rm B} & = & a + \frac{b}{P} \,,
\ees
where $a$ is the asymptotic ($P\to\infty$) transfer rate (bandwidth)
and has the value
$1\,\sec/300\,\MB$. Identifying the data latency as the time for
transferring $1\,\rmB$,
we can set the parameter $b\approx\tau_{\rm L}$. 
If we define the critical size $P_c$ for the block of data to be transferred
from the condition
\bes\label{criticalsize}
 \frac{b}{P_c} & = & \frac{a}{10} \,,
\ees
which means a 10\% deviation from the asymptotic rate of transfer, we
get $P_c=1000\,\rmB$. For this size of the data block transferred, the
time losses due to communication are only 10\%.

The fundamental operation of synchronisation of the PEs is achieved by
calling the SHMEM routine\\[0.2cm]
{\em
\hspace*{.5cm}barrier().}\\[0.2cm]
The call of the $barrier()$ routine ``announces'' the arrival of a PE
at that line of the code.
The PE suspends the execution of commands until all PEs have
called $barrier()$. A barrier ensures that a PE,
prior to synchronising with other PEs,
has completed all previously issued local memory stores and remote
memory updates issued via SHMEM routine calls such as
$shmem\_put$.

In the following example of parallelised code, we give
the structure of the updating of the gauge field
contained in the module $update\_gaugefield$.
The outermost loop changes the parity $par$ of the PE which in turn 
determines through (\ref{timeboxes}) the time section $tmin(par)\le t\le
tmax(par)$ in which the update is performed. The second loop is over
the directions $mu$ of the links, for the reasons we explained at the
beginning of this section and also because of the integrated autocorrelation
times as discussed in \sect{simulation}.
The third loop runs over the lexicographic
index $xs$ in the local $xy$-plane. For fixed $par$, $mu$ and $xs$, the
links $u(z,t,xs,mu)$ in the specified range of $t$ and for all $z$ are
updated: if $xs$ is at the boundary of the $xy$-partition,
the package of field variables
$u(1\le z\le L,tmin(par)\le t\le tmax(par),xs,mu)$ is sent to the
neighboring PEs that need a copy of it. The information on 
which PE needs a copy and where the copy has to be stored is contained
in the components of the array element $boundary(xs)$.
The parameter $package$ is the
size in words of the data block to be sent
and is equal to $4\times L\times T/2$.
For $L=32=T$ this corresponds to $16\,\KB$, which is well above the
critical size $P_c=1000\,\rmB$ in \eq{criticalsize}.
Now, it becomes clear why we choose the order given in (\ref{uphifields}) 
for the arguments of the fields:
the $z$ coordinate is always sent entirely in
the package, the $t$ coordinate between a lower and an upper
bound. This package must lie in one memory sequence, in fact
what is passed to the target PE is the address of the first component
of the sequence (as can be seen in the code example) and the
length in words of the sequence.
The argument $mu$ of the gauge field in (\ref{uphifields}) is the last
because of the structure of the updating.
After having updated the links in a time section for fixed direction $mu$,
there is a barrier. The copies of the boundary links must be up to date
before processing the next direction. Once the updating in the time section
is terminated, the parity of the PE is changed.
The updating of the Higgs field in the module $update\_higgsfield$ has
the same structure without loop over $mu$.

Actually, not all the boundary field variables that are copied are
needed during the updating. The simplified structure and the relatively
large size of data blocks transferred justify
these small losses in performance. The performance of the
whole program has been established with the tool APPRENTICE to be
\bes\label{performance}
 93.4\,{\rm Mflop}/\sec/{\rm PE} \,.
\ees
For comparison, the theoretical performance of the CRAY T3E
machine is $900\,{\rm Mflop}/\sec/{\rm PE}$ but (\ref{performance}) is
a respectable number for a program written in a high level language.\\
\begin{centerline}{
\begin{minipage}{12.5cm}
\vspace*{.5cm}
Example of parallelised code: updating of the gauge field.\\[0.2cm]
{\em
do pp=1,2\\[0.2cm]
\hspace*{.5cm}do mu=0,3\\[0.2cm]
\hspace*{1.cm}do xs=1,vol2\\[0.2cm]
\hspace*{1.5cm}do t=tmin(par),tmax(par)\\
\hspace*{2.cm}do z=1,L\\
\hspace*{2.5cm}{\rm ! updating of {\em u(z,t,xs,mu)}}\\
\hspace*{2.cm}end do\\
\hspace*{1.5cm}end do\\
\hspace*{1.5cm}{\rm ! transfer of fields on boundary points}\\
\hspace*{1.5cm}if (boundary(xs)\%tf.eqv..true.) then\\
\hspace*{2.cm}do i=1,boundary(xs)\%ndir\\
\hspace*{2.5cm}call shmem\_put(u(1,tmin(par),boundary(xs)\%xtg(i),mu), \&\\
\hspace*{5.2cm}u(1,tmin(par),xs,mu), \&\\
\hspace*{5.2cm}package, \&\\
\hspace*{5.2cm}boundary(xs)\%pe(i))\\
\hspace*{2.cm}end do\\
\hspace*{1.5cm}end if\\[0.2cm]
\hspace*{1.cm}end do\\
\hspace*{1.cm}{\rm ! synchronisation}\\
\hspace*{1.cm}call barrier()\\[0.2cm]
\hspace*{.5cm}end do\\
\hspace*{.5cm}{\rm ! parity switch}\\
\hspace*{.5cm}par=modulo(par+1,2)\\[0.2cm]
end do}
\vspace*{.5cm}
\end{minipage}}
\end{centerline}

\section{Program overview}

Our program for the simulation of the SU(2) Higgs model on the lattice
is composed of the following modules:\\
\begin{centerline}{
\begin{minipage}{6cm}
\vspace*{.5cm}
{\em 
module su2\_type\\
module global\\
module geometry\\
module ranfloat\\
module initial\_config\\
module run\_updating\\
module update\_gaugefield\\
module update\_higgsfield\\
module smearing\\
module observables\\
module kinds\\
module timing\\
module utilities\\
module gauge\_transformation\\
program main}
\vspace*{.5cm}
\end{minipage}}
\end{centerline}
Some of the modules have been already described. Here, we give a short
description of the other modules in which important definitions or 
subroutines of the program can be found. We refer to the comments in
the code for a more detailed information.

The module {\em global} contains the global variables and
parameters. The global parameters for which the user can choose 
a value (with some restrictions) are:
the numbers of PEs in $x$ and $y$ 
directions ($N_x\equiv npe\_x$ and $N_y\equiv npe\_y$: must be either even
numbers or 1),
the lattice sizes $L\equiv space\_size$ (must be divisible by $N_x$
and $N_y$) and 
$T\equiv time\_size$ (must be even),
the numbers of smeared fields ($smlevel\_phi$ and $smlevel\_link$ for the
Higgs and gauge field respectively) and corresponding smearing levels
(arrays $philevels$ and $ulevels$), 
the smearing strength
$\epsilon\equiv omega\_link$ for the gauge field (see
\eq{smearopgauge}, the smearing procedure for the Higgs field
in \eq{smearophiggs2} has no free parameters) and
the maximal space and time extensions of the
correlations for the measurement of the static potentials and the
static-light meson spectrum ($rmax$ and $tmax$).

The module {\em ranfloat} contains the random number generator as proposed
by M. {L\"uscher} \cite{Luscher:ranfloat,Luscher:1994dy} and
subroutines for the generation of random numbers with distributions
needed in the updating algorithms, see \sect{random}.

The module {\em run\_updating} organises the updating of the fields, the
measurements of the observables and writes the measurements into
files.

The module {\em smearing} contains the computation of the one-link
integrals (see \App{bessel}) and the construction of the smeared fields.
The subroutine $ini\_eigphi()$ contains the coefficients describing the
approximate wave functions for the static-light meson ground, first and
second excited states.

The module {\em observables} contains the subroutines for the measurements
of the plaquette, the Higgs length squared and its square, the
gauge invariant links, the
matrix correlation functions for the static potentials (in the
$tz$-plane) and for the static-light meson spectrum. The translation
invariance and the isotropy of the space are used to average the
measured quantities over the lattice points and the directions.

In order that the program {\em main} is executed, 
several run parameters are read during
execution from a file {\em parameter.d}.
The run parameters needed are listed at the end of this section.
For the continuation of an old run ($start=1$) the
parameters from 4. to 9. must be omitted, for the start from a given
configuration ($start=2$) they must be substituted by $4.\;confpath$
(name of the path where the start configuration is stored) and $5.\;confname$
(basic name of the start configuration files). The parameter $bitseed$
is the initialisation of a random bit generator
\cite{Luscher:ranfloat} 
which produces for each PE different initialisation seeds 
of the Marsaglia-Zaman random number generator.
The random number sequences on the PEs are different
for different seeds \cite{Luscher:ranfloat}.
We recall that one iteration
is composed of one heatbath sweep for both gauge and Higgs field
and $N_{\rm OR}$ times the combination of one
over-relaxation sweep for the gauge field plus three over-relaxation
sweeps for the Higgs field. The measurements of the matrix correlation
functions for the static potentials and for the
static-light meson spectrum are performed only each 
$iter\_to\_meas$ iterations.

\begin{centerline}{
\begin{minipage}{12.5cm}
\vspace*{.5cm}
Parameters to be written in the file $parameter.d$.\\[0.2cm]
{\em
1. path\\
{\rm full path where to write the output files}\\
2. name\\
{\rm basic name for the output files}\\
3. start=0,1,2\\
{\rm 0: new run, 1: continuation of old run,
2: start from a thermalised configuration}\\
4. $\beta$ ({\rm or} confpath) \\
5. $\kappa$ ({\rm or} confname) \\
6. $\lambda$\\
7. $1\le bitseed\le2^{23}-1$\\
{\rm seed for the initialisation of the random number generators}\\
8. order\\
{\rm 0: hot start, $\neq0$ cold start}\\
9. therm\\
{\rm number of thermalisation iterations}\\
10. N\_iter\\
{\rm number of updating iterations}\\
11. cpu\_time\_max\\
{\rm maximal CPU time in seconds}\\
12. N\_OR\\
{\rm ratio (over-relaxation sweeps)/(heatbath sweeps) for the gauge field}\\
13. iter\_to\_meas\\
{\rm ratio (iterations updating)/measurements}\\
14. N\_save\\
{\rm number of iterations after which the data are written into
  files}}
\vspace*{.5cm}
\end{minipage}}
\end{centerline}

\end{appendix}
   \bibliography{lattice}        %or whatever your .bib file is
   \bibliographystyle{h-elsevier}   %if you use h-elsevier.bst
\chapter*{Acknowledgements}

\begin{itemize}

\item[$\star$] I would like to thank Rainer Sommer for having proposed
  me such a beautiful argument for my Ph.D. thesis. He shared with me
  the difficulties and the nice moments. His professional and human
  support was fundamental.
\item[$\star$] I am indebted also to other members of the 
  ALPHA collaboration for their help: 
  to J. Heitger, who introduced me to the study of
  the SU(2) Higgs model, to S. Capitani, whose F-programs were
  the basis from which I constructed my own ones and to B. Bunk, 
  who suggested the study of the cumulants.
\item[$\star$] I would like to thank H. {St\"uben} from
  the Konrad-Zuse-Zentrum {f\"ur} Informationstechnik Berlin (ZIB)
  togheter with D. Pleiter for their explanations of the basics 
  of the parallel programming on a CRAY T3E machine and their
  assistance.
  I thank the ZIB for having given me the possibility to perform
  simulations on the CRAY T3E.
\item[$\star$] I would like to thank T. Kleinwort from the
  Rechenzentrum at DESY Zeuthen for having solved many of my ``small''
  computing problems.
\item[$\star$] I had always the strong support of my parents, Milena
  and Giacomo.
\item[$\star$] For this special moment of my life, my last words
  are for Silke.

\end{itemize}

\end{document}